\documentclass[12pt, draftclsnofoot, onecolumn]{IEEEtran}
\usepackage{amsmath,amsthm}
\usepackage{graphics} 
\usepackage{epsfig,epstopdf} 
\usepackage[export]{adjustbox}
\usepackage[dvipsnames]{xcolor}
\usepackage{amsmath,bigints,bbm} 
\usepackage{amssymb}  
\usepackage[T1]{fontenc}
\usepackage[utf8]{inputenc}
\usepackage{authblk}
\usepackage{cite}
\usepackage{cleveref}
\usepackage{caption}
\usepackage{subcaption}
\usepackage{mathtools}
\usepackage{enumitem}
\usepackage{lettrine}

\newtheorem{theorem}{\textbf{\text{Theorem}}}
\newtheorem{lemma}{\textbf{\text{Lemma}}}

\newtheorem{corollary}{Corollary}

\crefrangelabelformat{equation}{(#3#1#4--#5#2#6)}
\crefname{equation}{Eq.}{Eqs.}
\Crefname{equation}{Equation}{Equations}

\newtheoremstyle{problemstyle}  
{3pt}                                               
{3pt}                                               
{\normalfont}                               
{}                                                  
{\bfseries\itshape}                 
{\normalfont\bfseries:}         
{.5em}                                          
{}                                                  
\theoremstyle{problemstyle}

\usepackage{algorithm}
\usepackage{algorithmic}
\usepackage{tikz}

\usepackage[a4paper, total={7in, 9.4in}]{geometry}
\usepackage{setspace}	

\begin{document}

\title{\huge Optimal Deployment of Tethered Drones for Maximum Cellular Coverage in User Clusters}

\author{  
\IEEEauthorblockN{
Osama M. Bushnaq, {\em Student Member, IEEE}, Mustafa A. Kishk, {\em Member, IEEE},
Abdulkadir Çelik, {\em Senior Member, IEEE}, Mohamed-Slim Alouini, {\em Fellow, IEEE}, 
and Tareq Y. Al-Naffouri, {\em Senior Member, IEEE}.
} 
\thanks{The authors are with Computer, Electrical, and Mathematical Sciences and Engineering (CEMSE) Division at King Abdullah University of Science and Technology (KAUST), Thuwal,  23955-6900, KSA.\\
This manuscript was accepted in the IEEE Transaction on Wireless Communications on Nov. 2, 2020.}
}
	
	\maketitle
	
		\begin{abstract}		
			Unmanned aerial vehicles (UAVs) have recently received a significant interest to assist terrestrial wireless networks thanks to their strong line-of-sight links and flexible/instant deployment. However, UAVs' assistance is limited by their battery lifetime and wireless backhaul link capacity. At the expense of limited mobility, tethered UAVs (T-UAVs) can be a viable alternative to provide seamless service over a cable that simultaneously supplies power and data from a ground station (GS). Accordingly, this paper presents a comparative performance analysis of T-UAV and regular/untethered UAV (U-UAV)-assisted cellular traffic offloading from a geographical area that undergoes heavy traffic conditions. By using stochastic geometry tools, we first derive joint distance distributions between the hot-spot users, the terrestrial base station (TBS), and the UAV. To maximize the end-to-end signal-to-noise ratio, a user association policy is developed, and corresponding association regions are analytically identified. Then, the overall coverage probability of the U-UAV/T-UAV-assisted system is derived for given locations of the TBS and the U-UAV/T-UAV. Moreover, we analytically prove that optimal UAV location falls within a partial surface of the spherical cone centered at the GS. 
			Numerical results show that T-UAV outperforms U-UAV given that sufficient GS locations accessibility and tether length are provided.    
		\end{abstract}
	

	\section{Introduction}\label{sec:intro}

		\lettrine{U}{nmanned} aerial vehicles (UAVs) have rapidly gained a tremendous interest to be used in numerous emerging commercial and military applications such as aerial surveillance, border protection, traffic control, transportation, logistics, precision agriculture, search \& rescue missions, disaster recovery, \dots, etc. In particular, UAV-based airborne communications bring a major paradigm shift to the information and communication technology (ICT) sector, which primarily depends upon a terrestrial communication and networking infrastructure \cite{halim, Fotouhi_2019}. Indeed, UAVs can offer salient attributes to today's fixed telecom infrastructure, including strong line-of-sight backhaul/access links, flexible/instant deployment, and extra degrees of freedom for the controlled mobility \cite{Mozaffari_2019}. 
		
		In the context of wireless communications, the ambitious quality-of-service demands (i.e., high-rate, ultra-reliable, and low-latency) of the next-generation networks can be fulfilled by UAV-assisted cellular communications, whereby UAVs are integrated with the terrestrial cellular infrastructure for various applications \cite{Zeng2016}. In this regard, UAVs have been recently envisioned as aerial base stations \cite{Zhang_2020,  Mozaffari2016}, relays \cite{Zhang_2020,Gesbert_2019,kouzayha2019stochastic}, user equipments (UE) \cite{Sofie_2019}, and data fusion access points \cite{Bushnaq_2018, Bushnaq_2019, Gong_2018,Zhang2020_sense}. Thanks to UAV's instant and cost-efficient deployment, UAV-assisted cellular communication is especially suitable for providing extra coverage to geographical regions that experience heavy traffic conditions, which are also referred to as \textit{hot-spots}. Unless this heavy traffic is caused by an extraordinary event (e.g., natural disasters), hot-spots generally follow a spatio-temporal pattern that is caused by mass events such as sports matches, concerts, conferences, exhibitions, demonstrations, \dots, etc. Unlike the high cost of deploying fixed terrestrial base station (TBS) to serve these occasional or periodic events, UAVs can hover over the hot-spot and assist the existing TBSs to provide ground users with better coverage.

		Nonetheless, utilizing UAVs as aerial base stations has two main drawbacks: Firstly, the limited capacity of state-of-the-art batteries poses a daunting challenge for the operational lifetime of UAVs. Therefore, a UAV cannot be available throughout the entire mission duration as it is required to return to a charging/docking station, charge/replace its battery, and return back to the hot-spot region. Secondly, the service quality offered to the hot-spots is restricted by the capacity of the backhaul link between the UAV and TBS. Although UAV is fully flexible to be deployed anywhere, the backhaul link capacity restrains its deployment region to a space around the TBS. Tethered UAVs (T-UAVs) can be a viable alternative to supply both power and data over a cable from a ground station (GS), which can be located on a rooftop or a mobile station \cite{kishk2019capacity}. Given a set of accessible GS locations, T-UAVs can also fly between GSs to serve hot-spots that do not overlap in the temporal domain. Nevertheless, the T-UAVs are also susceptible to the following limitations \cite{kishk2019TUAV}: Firstly, the optimal GS location may not be readily available. Therefore, the number of GS location (e.g., building density) and their accessibility (i.e., the permission of the residents) has an impact on the optimal deployment strategies. Secondly, the tether length and inclination angle of the T-UAV restrain the freedom of mobility around the GS. At this point, it is worth noting that the backhaul link capacity of a regular/untethered UAV (U-UAV) plays the role of tether by limiting the distance from the TBS. Considering that both systems have virtues and drawbacks, the main objective of this paper is to provide a comparative performance analysis of U-UAV and T-UAV-assisted cellular traffic offloading under the practical challenges\footnote{Throughout the paper, the term `UAV' is used to refer both U-UAVs and T-UAVs.}.
		
		\subsection{Related Work}
		
		The limited energy supply at the UAV forms a critical challenge for the deployment of the aerial BSs. Energy-efficient UAV communication is studied in \cite{Zhang2017Energy}. However, it is shown in \cite{Fotouhi_2019, Mozaffari_2019, Alzenad2017} that the communication power is negligible compared to the mechanical power consumed during hovering and traveling. Therefore, improving the communication power efficiency has a negligible impact on the overall UAV energy efficiency. The propulsion power consumption can be reduced by controlling the UAV speed and hovering height \cite{Zorbas2016, Zeng2019Energy}. In \cite{Zeng2019Energy}, the UAV propulsion energy and communication related energy are minimized while satisfying a throughput constraint for the served users. Battery replacement/recharging approaches are proposed in \cite{Li_2017, ERDELJ2019101612, SHARMA201794}, where solutions can significantly improve the U-UAV availability at the expense of extra cost and complexity. The limited UAV battery lifetime issue is addressed in \cite{Galkin2019Magazine}, where UAV swapping, battery swapping, and laser wireless charging are discussed.

		Deployment of aerial BSs is studied in \cite{Lyu_2017, Yanikomeroglu_2016, Chen2018}. In \cite{Lyu_2017}, the UAVs are deployed to guarantee the coverage of a group of ground users. Similarly in \cite{Yanikomeroglu_2016}, the UAV is placed to serve the maximum number of users with the maximum possible A2G link quality. In \cite{Chen2018}, a relaying UAV is placed optimally to minimize the overall outage and bit error rate. Using stochastic geometry tools, \cite{Galkin2019} characterizes the coverage probability of UAVs hovering over a hot spot and uses that to optimally place them. In order to improve the overall user QoS, the backhaul link and the association policy must be carefully studied.  In \cite{kouzayha2019stochastic}, the UAV-assisted network is assessed, assuming a mmWave backhauling for random ground BS and UAV locations. In \cite{Alzenad2018FSO}, point-to-point free-space optics (FSO) links are proposed for UAV backhaul/fronthaul connection. In \cite{cicek2018backhaulaware}, the UAV placement problem is solved to maximize the data rate while considering limited backhaul and radio access capacity.	In \cite{Wang2020}, 
		a joint precoding optimization scheme is proposed for secure UAV-aided NOMA network. In \cite{Cheng2018}, the trajectory of the UAV is optimized for data offloading from the edge of multiple cells. In \cite{Cheng2018}, UE either associates with the UAV or a close TBS and experience interference from the close non serving TBSs and/or UAV. Unlike the above works dealing with the deployment of U-UAVs, we consider T-UAV deployment and compare its performance with U-UAV under practical scenarios.
	
		The use of tethered UAVs in cellular communication has attracted attention recently. As discussed in~\cite{kishk2019capacity}, T-UAVs have two main advantages: (i) having a stable power supply through the tether connecting the UAV to the GS and (ii) having a reliable wired data-link connecting the UAV to the GS. In~\cite{kishk2019TUAV}, the average path-loss for a point-to-point link between a T-UAV and a ground user is derived and optimized. In~\cite{8644135}, the authors propose a novel UAV-based communication system for a post-disaster setup. In particular, U-UAVs are used for providing cellular service for disaster areas, while T-UAVs are used to provide backhaul links for the U-UAVs. Unlike existing literature, this paper focuses on optimizing the T-UAV placement to provide cellular service for multiple ground users. To achieve this, we use tools from stochastic geometry to model the locations of ground users. This is motivated by the tractability of stochastic geometry tools and their ability to provide closed-form expressions for various performance metrics~\cite{HH1,HH2,Haenggi_2009}. More details on the contributions of this paper are provided next.
		
		
	\subsection{Main Contributions}

	This paper provides a comparative performance analysis between U-UAVs and T-UAVs, which are deployed to maximize the coverage of high QoS demanding users' within a hot-spot region. To the best of authors' knowledge, this is the first work to consider tethered UAV-assisted communication in a multi-user scenario. Although U-UAV and T-UAV systems have different virtues and drawbacks, the comparison is still valid and meaningful since both systems' performance is evaluated under identical network setups. We believe such a comparison will help network operators decide whether deploying a U-UAV or T-UAV is suitable based on hardware specifications and environmental parameters. The technical contributions of the paper can be summarized as follows:

	\begin{itemize} 
		
		\item 
		A stochastic geometry-based analysis is provided for coverage performance of U-UAVs and T-UAVs over a circular hot-spot region where UEs are uniformly distributed. While U-UAVs are limited by being available for a given duty cycle period, T-UAVs are restrained by a maximum tether length, inclination angle, and GS location accessibility. 
		
		\item  
		The paper derives the joint probability density function (PDF) of distances between TBS and UAV to a reference user. In general, these derivations are useful for systems where two nodes (regardless of their locations) interact with a uniformly distributed node within a circular cluster. The derived PDF is especially helpful for cellular networks where neither the TBS nor UAV is located at the center of the geographical region of interest.
		
		\item
		To obtain the overall system coverage probability, a user association policy is developed, and the association regions are identified. The end-to-end coverage probability is analyzed based on the aerial access and backhaul links for users associated with the UAV. 
		
		\item 
		Since the search space of the deployment area is very large, we analytically prove that optimal UAV location falls within the surface of the spherical cone centered at the GS. 
	\end{itemize}  
		Extensive simulation results are presented to validate analytical results and compare U-UAV and T-UAV-assisted systems' performance.  
		
	\subsection{Paper Notations and Organization}
		The remainder of the paper is organized as follows: Section \ref{sec:sys_model} describes the system model and characterizes access and backhaul links. Section \ref{sec:coverage} derives the joint distance PDFs and the coverage probabilities. Section \ref{sec:deployment} analytically characterizes the optimal hovering space. Then, Section \ref{sec:results} presents the numerical results. Lastly, Section \ref{sec:conc} concludes the paper with a few remarks. Table \ref{table:notation} details the notation convention used in the paper.

\section{System Model}
\label{sec:sys_model}

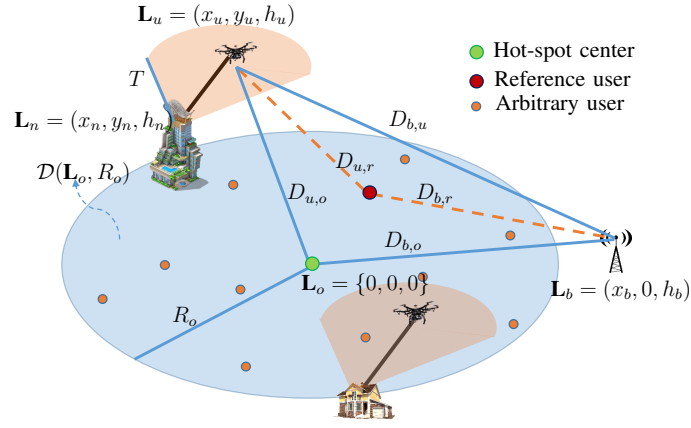
\begin{figure}
		\centering
		\begin{tikzpicture}[thick,scale=0.7, every node/.style={scale=0.7}]
		\draw (0, .7) node[inner sep=0] {	\includegraphics[trim={6cm 4cm  11cm 3cm},clip, width=.65 \linewidth]{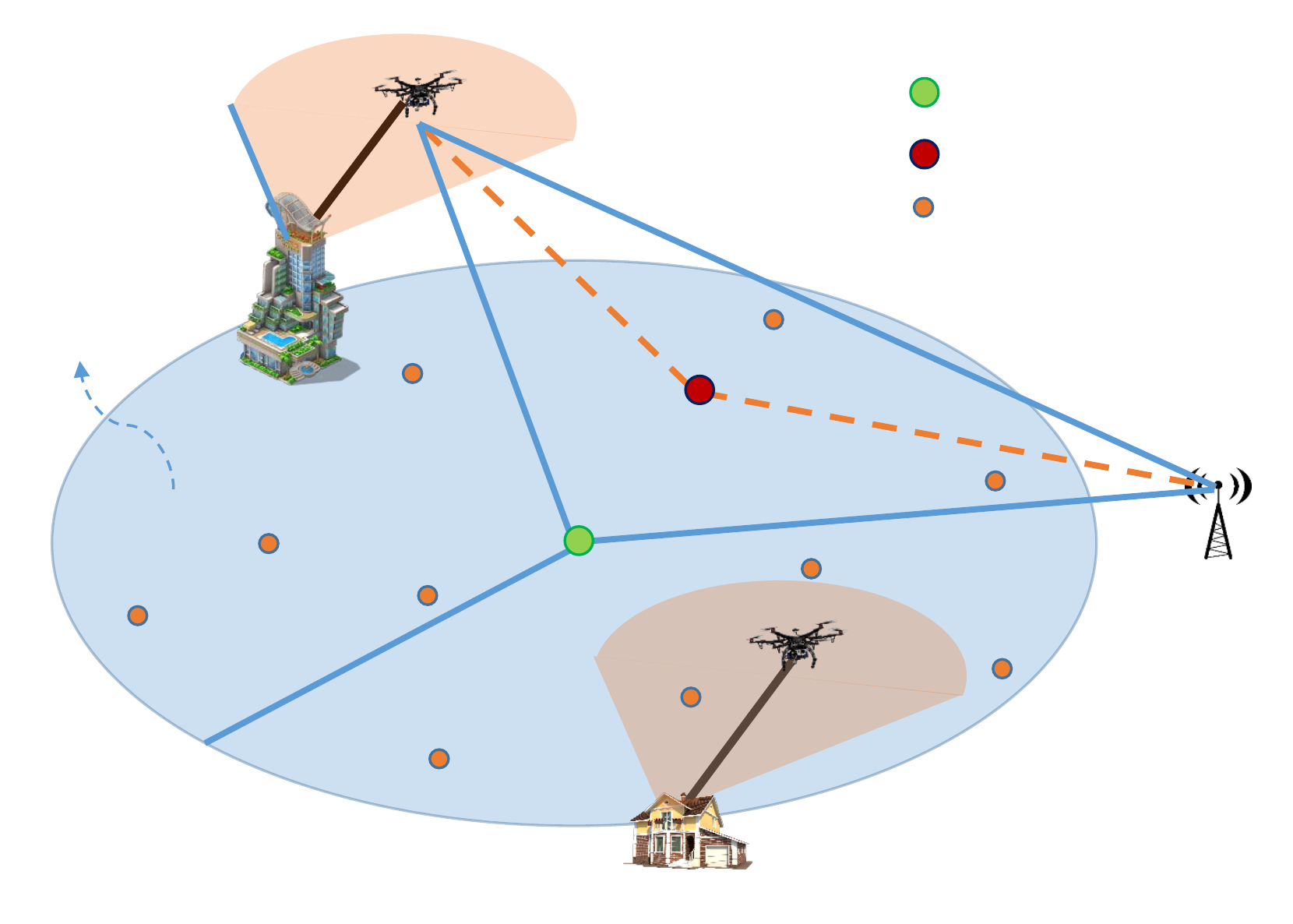}};
		\draw (4.13,4.) node {{Hot-spot center}};	
		\draw (4.1,3.5) node {{Reference user}};	
		\draw (4.05,3) node {{Arbitrary user}};
		\draw (0.4,-0.4) node {{$\textbf{L}_o = \{0,0,0\}$}};
		\draw (-.7, 1.3) node {{$D_{u,o}$}};	
		\draw (.2, 1.9) node {{$D_{u,r}$}};	
		\draw (-4.7, 2.7) node {{$\textbf{L}_n =(x_n,y_n,h_n)$}};	
		\draw (-2.4, 4.7) node {{$\textbf{L}_u =(x_u,y_u,h_u)$}};
		\draw (1.2, 2.7) node {{$D_{b,u}$}};
		\draw (-3.9, 3.5) node {{$T$}};	
		\draw (5.2, -.5) node {{$\textbf{L}_b =(x_b,0,h_b)$}};
		\draw (1.1, .4) node {{$D_{b,o}$}};
		\draw (1.7, 1.3) node {{$D_{b,r}$}};
		\draw (-3, -1) node {{$R_o$}};
		
		\draw (-4.9, 1.7) node {{$\mathcal{D}(\textbf{L}_o,R_o)$}};	
		\end{tikzpicture}
		\caption{Data offloading through T-UAV system model.} \label{fig:sys_model2}
	\end{figure}

	We consider improving downlink wireless coverage in highly crowded areas with heavy traffic conditions, which is referred to as \textit{hot-spots} throughout the paper. The hot-spot region is modeled as a disk $\mathcal{D}(\textbf{L}_o,R_o) \subset\mathbb{R}^2$ centered at the origin $\textbf{L}_o$ with radius $R_o$, (similar to \cite{Galkin2019, Dhillon2017Oct}), where UEs are assumed to be uniformly distributed.
Without loss of generality, the TBS location $\textbf{L}_b = \{x_b,0,h_b\}$ is assumed to be at the $x$-axis for convenience. We aim to offload downlink traffic from a TBS to an U-UAV or a T-UAV. The U-UAV is a regular UAV that can freely hover at any location in $\mathbb{R}^3$. However, it has a defined flight duration due to the limited lifetime of battery. Therefore, its service availability is modeled by a duty-cycle parameter $A \in [0,1]$ which is determined based on charging and serving durations of the U-UAV. Another restraint on the U-UAV is the limited capacity of the backhaul link between the TBS and the U-UAV, which has a critical impact on deployment and user association strategies as overall coverage probability is jointly determined by access (UAV--UE) and backhaul (TBS--UAV) links. On the other hand, the T-UAV is connected to a ground station (GS) which uninterruptedly supplies both power and data through a tether. The GSs can be installed on $N$ potential rooftops whose locations are denoted by $\textbf{L}_n=\{x_n,y_n,h_n\}$, $n\in[1,N]$. On the negative side, the mobility of T-UAV is restrained by its maximum tether length $T$ and minimum inclination angle $\phi$. Note that, in practice, the tether cannot be completely stretched due to wind and gravity. We assume that the maximum distance between the UAV and the GS is $T$.	As a result, the reachability of T-UAV is restricted to the following spherical cone
	\begin{align}\label{sperical_cone}
	\mathcal{M}_n = \Bigg\{  x_u,y_u,h_u  : \ ||\textbf{L}_n - \textbf{L}_u|| \leq T,\ \arcsin\left( \dfrac{h_u - h_n}{||\textbf{L}_n - \textbf{L}_u|| } \right) \geq \phi   \Bigg\},
	\end{align}
	where the GS location $\textbf{L}_n$ is the center of spherical cone and $\textbf{L}_u=\{x_u,y_u,h_u\}$ is the location of the T-UAV. The system model is illustrated in Fig. \ref{fig:sys_model2}. 
	
	It is worth noting that interference plays an essential role in the performance analysis of both the U-UAV and the T-UAV systems. While neglecting interference is acceptable in rural environments, a thorough investigation is needed to illustrate the coverage performance in urban environment where interference is significant. At this early T-UAV assisted cellular network analysis stage, we neglect the interference and leave it for future work.

	In the rest of the paper, we will focus our analysis on a reference UE (RUE), which is randomly selected from the disk $\mathcal{D}(\textbf{L}_o,R_o)$ and located at $\textbf{L}_r$. In the following subsections, we characterize the terrestrial access link between the TBS and the RUE, the ground-to-air (G2A) aerial backhaul link between the TBS and the U-UAV, and the air-to-ground (A2G) aerial access link between the UAV and the RUE.
	\begin{table}
		\caption{Summary of the notations.}	\label{table:notation} 
		\footnotesize
		\centering
		\begin{tabular}
			{| p{4.5cm} | p{11cm} |} 
			\hline
			\textbf{Notation} & \textbf{Description}  \\ [0.5ex] 
			\hline\hline
			$\{\cdot\}_r / \{\cdot\}_b / \{\cdot\}_u$ & Subscripts refer to the RUE/TBS/UAV \\
			\hline
			$\textbf{L}_i$ & Location of an arbitrary point $i$ in 3D\\
			\hline
			$D_{i,j}$ & Euclidean distance between $\textbf{L}_i$ and $\textbf{L}_j$, $D_{i,j} \triangleq \| \textbf{L}_i - \textbf{L}_j\|$\\
			\hline
			$\{ \cdot \}'$ & Ground projection of a point or a distance\\
			\hline
			$\mathcal{L}(\textbf{L}_i,\textbf{L}_j)$ & The line segment formed by connecting the points at $\textbf{L}_i$ and $\textbf{L}_i$ \\
			\hline
			$\angle(\textbf{L}_i,\textbf{L}_j,\textbf{L}_k)$ & The angle at $\textbf{L}_j$ formed by moving from $\mathcal{L}(\textbf{L}_i,\textbf{L}_j)$ to $\mathcal{L}(\textbf{L}_j,\textbf{L}_k)$ counterclockwise  \\
			\hline
			${\textbf{L}}_x^+$ & A point in the positive $x$ direction, i.e., ${\textbf{L}}_x^+ = \{ \infty,0,0\}$ \\
			\hline
			$\mathcal{C}(\textbf{L}_i,R_i)$ & Circle centered at $\textbf{L}_i$ with radius $R_i$ \\
			\hline
			$\mathcal{D}(\textbf{L}_i,R_i)$ & Disk centered at $\textbf{L}_i$ with radius $R_i$ \\	
			\hline
			$\mathcal{A}(\textbf{L}_j,R_j,\textbf{L}_i,R_i) \subseteq \mathcal{C}(\textbf{L}_j,R_j)$ & Arc centered at $\textbf{L}_j$ with radius $R_j$ within $\mathcal{C}(\textbf{L}_i,R_i)$ \\
			\hline
			$|| \cdot ||$ & $\ell_2$ norm \\
			\hline
			$| \cdot |$ & Absolute value for scalars or Lebesgue measure for sets \\
			\hline
		\end{tabular}
	\end{table}

	\subsection{Terrestrial Access Links (TBS--RUE)}
	
	The TBS$\rightarrow$RUE access link is assumed to experience free-space path-loss as well as Rayleigh fading. As a result of path-loss, the transmitted signal power decays with distance, i.e., $D_{b,r}^{-\alpha_b}$ where $D_{b,r}=\vert\vert \textbf{L}_b-\textbf{L}_r \vert\vert$ is the three dimensional (3D) Euclidean distance between the TBS and RUE, and $\alpha_b$ is the path-loss decay exponent. Accordingly, the signal to noise ratio (SNR) at the RUE is expressed as
	
	\begin{align}
	\label{eq:SNR_b-r}
	\text{SNR}_{b,r} &= \dfrac{\rho_b G_{b,r} D_{b,r}^{-\alpha_b}  }{\sigma_n^2},
	\end{align}
	where $G_{b,r}$ is the channel gain, $\rho_b$ is the TBS transmission power, and $\sigma_n^2$ is the noise variance. Following the Rayleigh fading assumption, $G_{b,r}$ is exponentially distributed with the probability density function (PDF) $f_{G_{b,r}}(g) =  \mu e^{-{g}{\mu}}$, where $\mu$ is the fading parameter.
	
	\subsection{Aerial Access (UAV--RUE) and Backhaul (TBS--U-UAV) Links}
	Both aerial access and backhaul links are assumed to experience free-space line-of-sight (LoS) and non-line-of-sight (NLoS) attenuation path-loss as well as Nakagami-$m$ fading. The probability of having a LoS transmission between a UAV and an arbitrary location is given by
	\begin{align} \label{eq:p_los_exact}
	\kappa_{u,i}^{\text{LoS}} = \prod_{k=0}^{K} \left( 1 - \exp \left( -  \dfrac{ \left( h_u - \dfrac{(k+0.5)(h_u - h_i)}{K+1} \right) ^2 }{2\gamma_1^2} \right) \right),
	\end{align}
	where $K= \lfloor D_{b,i} \sqrt{\gamma_2 \gamma_3}-1\rfloor $ and $\gamma_i$'s are environmental parameters. Specifically, $\gamma_1$, $\gamma_2$ and $\gamma_3$ represent the building heights distribution, the ratio of built up land to the total land area, and the average number of buildings per km$^2$, respectively \cite{ITU_R_2012}. If the TBS height is fixed to $h_b$, \eqref{eq:p_los_exact} can be approximated for the TBS--U-UAV backhaul link as \cite{Akram2014Dec2},
	\begin{align}
	\kappa_{b,u}^{\text{LoS}} = \left( 1+a_b \exp \left[-b_b \left( \arcsin \left(\dfrac{h_u -h_b}{D_{b,u}} \right)-a_b \right) \right] \right)^{-1},
	\end{align}  
	where $a_b$ and $b_b$ are approximation parameters depending on $h_b$, $\gamma_1$, $\gamma_2$ and $\gamma_3$. Similarly, the LoS probability between the UAV and the RUE, which is assumed at height $h_r=0$, can be approximated as,
	\begin{align}
	\kappa_{u,r}^{\text{LoS}} = \left( 1+a_r \exp \left[-b_r \left( \arcsin \left(\dfrac{h_u}{D_{u,r}} \right)-a_r \right) \right] \right)^{-1},
	\end{align}  
	where $a_r$ and $b_r$ are functions of $h_r$, $\gamma_1$, $\gamma_2$ and $\gamma_3$. 
	
	
	Following the Nakagami-$m$ fading assumption, the G2A/A2G channel gain $G_{i,j}$ between two arbitrary points $\textbf{L}_i$ and $\textbf{L}_j$ is Gamma distributed with the PDF
	\begin{align}
	f_{G_{i,j}}(g) = \dfrac{m^m g_u^{m-1}}{\Gamma(m)} \exp(-m),
	\end{align}
	where $\Gamma(\cdot)$ is the gamma function. This reduces to Rayleigh fading for $m=1$ and approximates Rician fading for $m>1$ \cite{simon_2005digital}. Given the aforementioned G2A/A2G channel characteristics, the signal to noise ratio (SNR) for the aerial access link is expressed as
	\begin{align}
	\label{eq:SNR_ur}
	\text{SNR}_{u,r} &= \dfrac{\rho_u G_{u,r} D_{u,r}^{-\alpha_u} }{\sigma_n^2  \eta_k},
	\end{align}
	where $\rho_u$ is the UAV transmission power and $\eta_k, \forall k \in \{ \text{LoS, NLoS}\}$, are the attenuation coefficients for the LoS/NLoS links. Likewise, the SNR of the aerial backhaul link between the TBS and U-UAV is expressed as
	\begin{align}
	\label{eq:SNR_bu}
	\text{SNR}_{b,u} &= \dfrac{\rho_u G_{b,u} D_{b,u}^{-\alpha_u} }{\sigma_n^2  \eta_k}.
	\end{align}
	While the U-UAV acts as a relay between the RUE and the TBS, the T-UAV is directly connected to the core network via a fiber optics packed high-speed ultra-reliable link. Therefore, we assume that for the T-UAV $\text{SNR}_{b,u} \gg \text{SNR}_{u,r}$ holds all the time.
	
	\subsection{Association Policy}
	\label{sec:Association}
	The RUE associates with the TBS or the UAV based on the one that provides a higher average access link SNR\footnote{Here, we assume that the RUE is agnostic to the backhaul link conditions.}\cite{Alzenad2019, Andrews2011}. Accordingly, in case of LoS and NLoS aerial access links, the RUE respectively associates with the UAV if it is located within the following areas,
	\begin{align}
	\mathcal{B}_{{\text{LoS}}}^u &= \Bigg \{ x_r, y_r  : \overline{\text{SNR}}_{b,r}  < \overline{\text{SNR} }_{u,r}^{\text{LoS}} \Bigg\}  
	= \left\{x_r,y_r :   D_{u,r}  \leq \left(\dfrac{D_{b,r}^{\alpha_b}}{\eta_{\text{LoS}}} \right)^{\frac{1}{\alpha_u}} \right\}, \label{B_los}  \\
	\mathcal{B}_{{\text{NLoS}}}^u &= \Bigg\{x_r, y_r :   \overline{\text{SNR}}_{b,r}  < \overline{\text{SNR}}_{u,r}^{\text{NLoS}} \Bigg\}  
	= \left\{x_r,y_r  :  D_{u,r} \leq \left(\dfrac{D_{b,r}^{\alpha_b}}{\eta_{\text{NLoS}}} \right)^{\frac{1}{\alpha_u}} \right\},  \label{B_nlos}
	\end{align}
	where $\overline{\text{SNR}}_{b,r}$ and $\overline{\text{SNR}}_{u,r}^k, k \in \{\text{LoS, NLoS}\},$ are the average SNRs for terrestrial and aerial access links, respectively. Notice in \eqref{B_los} and \eqref{B_nlos} that we always have $\mathcal{B}_{{\text{NLoS}}}^u \subset \mathcal{B}_{{\text{LoS}}}^u$ due to the fact that $\eta_{\text{LoS}} < \eta_{\text{NLoS}}$. 
 
	Even though it is better to employ an instantaneous end-to-end SNR based association policy, we rather consider the average end-to-end SNRs for two practical reasons. First, considering an association policy based on the access link does not require overhead communication to feedback the backhaul link state to the RUE. Second, the average SNR is rather useful to avoid frequent unnecessary handovers caused by channel gain fluctuations. Since the average SNRs are location dependent, the hot-spot is divided into regions that associate with the UAV and regions that associate with the TBS [c.f. Fig. \ref{fig:uni_asc}].

\section{Coverage Performance Analysis}
\label{sec:coverage}
	Throughout this section, we focus our attention on a randomly located RUE within the hot-spot region, i.e., $\textbf{L}_r \in \mathcal{B}_o$. As a result of randomness, we first derive necessary distance distributions between the TBS/UAV and the RUE. Then, coverage performance of access and backhaul links are analyzed by using these distance distributions as building blocks. 

\subsection{Distance Distributions}

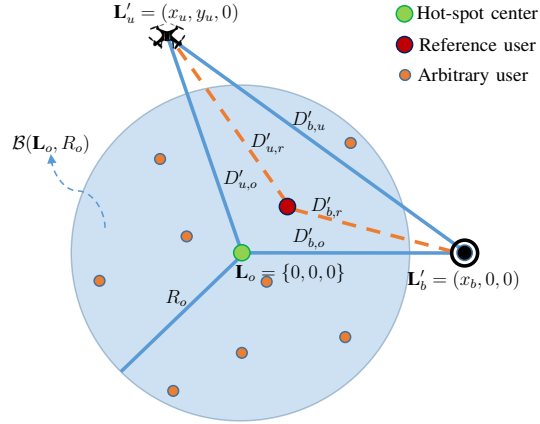
\begin{figure}
		\centering
		\begin{tikzpicture}[thick,scale=0.6, every node/.style={scale=0.6}]
		\draw (0, 0) node[inner sep=0] {	\includegraphics[trim={8cm 5cm  14cm 3cm},clip, width=.65 \linewidth]{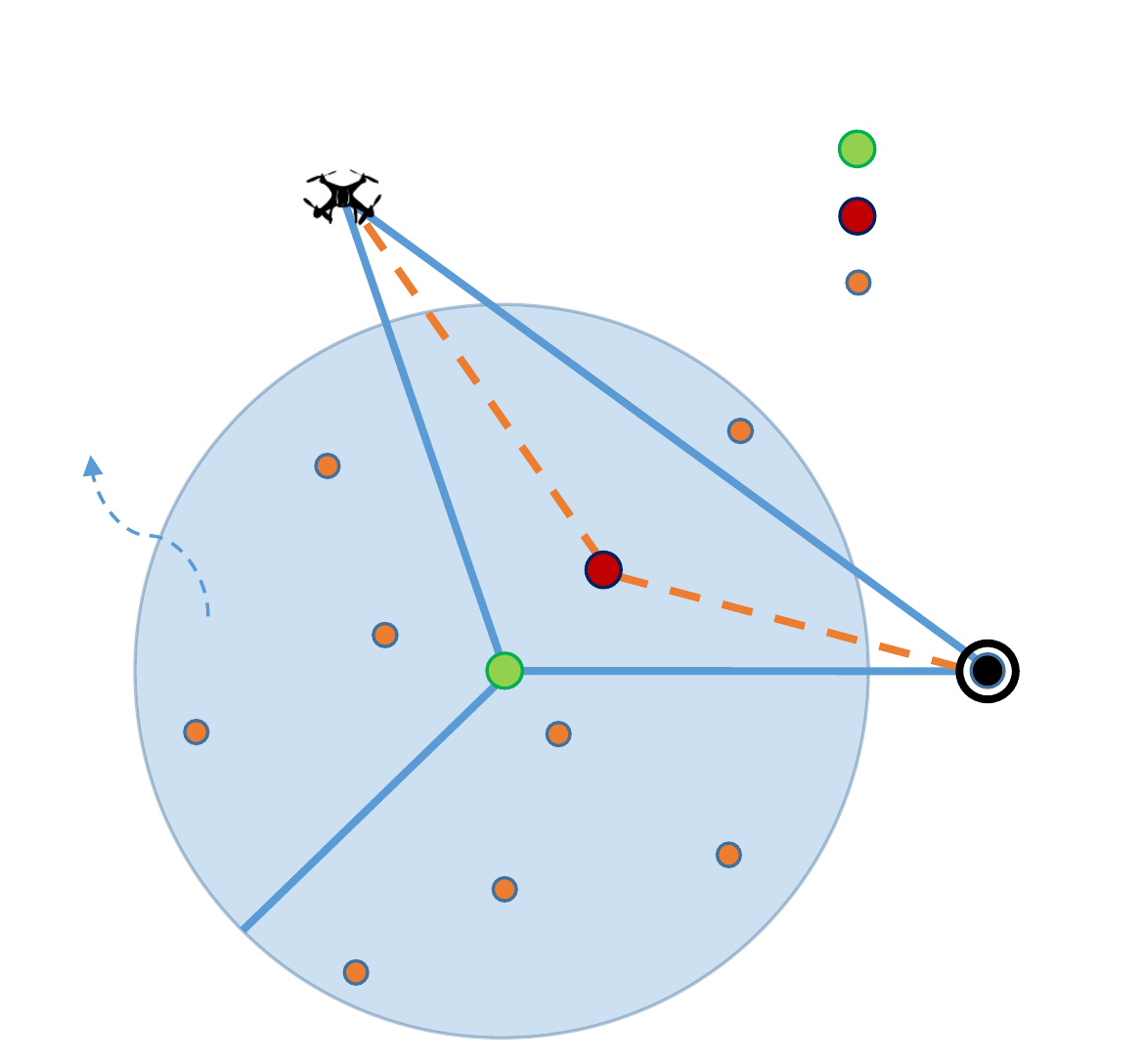}};
		\draw (4.5,3.85) node {{Hot-spot center}};	
		\draw (4.5,3.2) node {{Reference user}};	
		\draw (4.4,2.5) node {{Arbitrary user}};
		\draw (0.4,-1.85) node {{$\textbf{L}_o = \{0,0,0\}$}};
		\draw (-.1, 1) node {{$D_{u,r}'$}};	
		\draw (-2.1, 3.9) node {{$\textbf{L}_u' =(x_u,y_u,0)$}};
		\draw (.8, 1.5) node {{$D_{b,u}'$}};
		\draw (.8, -1.1) node {{$D_{b,o}'$}};
		\draw (-.7, .2) node {{$D_{u,o}'$}};
		\draw (4.2, -2) node {{$\textbf{L}_b' =(x_b,0,0)$}};
		\draw (1.2, -.4) node {{$D_{b,r}'$}};
		\draw (-2.1, -2.4) node {{$R_o$}};
		\draw (-4.85, 1.) node {{$\mathcal{B}(\textbf{L}_o,R_o)$}};	
		\end{tikzpicture}
		\caption{Aerial view of the system model in Fig. \ref{fig:sys_model2}.} 
		\label{fig:sys_model3}
	\end{figure}

	Coverage performance is primarily determined by two joint factors: SNR levels of access/backhaul links and user association resulting from the SNR levels. It is obvious from \eqref{eq:SNR_b-r}, \eqref{eq:SNR_ur}, and \eqref{eq:SNR_bu} that the SNR levels are highly dependent on the RUE's random location and thus its random distance to the TBS and the UAV, i.e., $D_{b,r}$ and $D_{u,r}$, respectively. In what follows, we consider projected distances over the $x-y$ plane for the sake of a better presentation. As illustrated in Fig. \ref{fig:sys_model3}, projected distances are defined as $D_{b,r}' \triangleq \sqrt{D_{b,r}^2-h_b^2}$ and $D_{u,r}' \triangleq \sqrt{D_{u,r}^2-h_u^2}$. Likewise, a projected location of an arbitrary point is denoted by $\textbf{L}_i' \triangleq  \{x_i,y_i,0\}$. To derive the coverage probability, one first needs to compute the joint PDF of $D_{b,r}'$ and $D_{b,r}'$ as well as their marginal PDFs. To this end, we provide formal definitions of a line segment, circle and arc as follows.
	
	\noindent \textbf{Definitions:} The line segment connecting the points $\textbf{L}_i$ and $\textbf{L}_j$ is defined as $\mathcal{L}(\textbf{L}_i,\textbf{L}_j) \triangleq \overline{\textbf{L}_i\textbf{L}}_j$. Likewise, the circle centered at $\textbf{L}_i'$ with radius $R_i$ is defined as 
	\begin{align}
	\mathcal{C}(\textbf{L}_i',R_i) = \left\{x,y : (x-x_i)^2 + (y-y_i)^2 = R_i^2 \right\}.
	\end{align}
	 For any two intersecting circles, we define the arc of $\mathcal{C}({\bf{L}}_j',R_j)$ located inside $\mathcal{C}({\bf{L}}_i',R_i)$ as 
	\begin{align}
	\mathcal{A}({\bf{L}}_j',R_j,{\bf{L}}_i',R_i) = \left\{x,y : (x-x_j)^2 + (y-y_j)^2 = R_j^2, \;  (x-x_i)^2 + (y-y_i)^2 \leq R_i^2  \right\}.
	\end{align}
	
	In the following Lemma, we derive the PDF of the distance between a uniformly distributed point within $\mathcal{D}(\textbf{L}_o,R_o)$ and any arbitrary point on the $x-y$ plane.  
	\lemma The PDF of the distance between a uniformly distributed point $\textbf{L}_j'$ within $\mathcal{D}(\textbf{L}_o,R_o)$ and any arbitrary fixed point $\textbf{L}_i'$ is given by
	\begin{align}
	\label{eq:f_R_i}
	f_{D_{i,j}'} (r_i) =   
	\begin{cases}
	\dfrac{2r_i}{R_o^2} & 0 \leq r_i \leq \max(0, R_o -D_{i,o}'),  \\
	\dfrac{2r_i}{\pi R_o^2} \arccos\left (\dfrac{(D_{i,o}')^2 +r_i^2-R_o^2} {2D_{i,o}' r_i}\right) & |R_o -D_{i,o}'| \leq r_i \leq  R_o + D_{i,o}' ,
	\end{cases}  
	\end{align}
	where $D_{i,o}' = \sqrt{x_i^2 +y_i^2}$ is the distance between $\textbf{L}_i'$ and $\textbf{L}_o$.
	\proof See Appendix \ref{app:f_D} \label{lemma:f_D}.  \hfill $\blacksquare$
	
	Accordingly, the PDF of distance between the RUE and the ground projection of the TBS and the UAV locations can be directly obtained by replacing $\textbf{L}_b'$ and $\textbf{L}_u'$ with $\textbf{L}_i'$ in \eqref{eq:f_R_i}.
	
	\lemma For a given distance between the RUE and the projected TBS location, $D_{b,r}'$, the conditional PDF of distance between a uniformly distributed RUE location $\textbf{L}_r \in \mathcal{B}(\textbf{L}_o,R_o)$ and the projected UAV location at $\textbf{L}_u'$ is given by
	\begin{align}\label{eq:cond_pdf}
	&f_{D_{u,r}'|D_{b,r}'}(r_u|r_b) =\\
	\nonumber & \quad \begin{cases}
	\dfrac{w}{2\pi r_b}  & 0 \leq r_b \leq \max(0, R_o -D_{b,o}'), \quad    D_{b,u}'-r_b \leq r_u \leq D_{b,u}'+r_b ,\\
	\dfrac{w  \mathbbm{1}_{ \{ \check{\theta}_b \leq \check{\theta}_u \leq \hat{\theta}_b \} }  }{| \mathcal{A}(\textbf{L}_b',r_b,\textbf{L}_o,R_o) | }& |R_o -D_{b,o}'| \leq r_b \leq  R_o +D_{b,o}', \quad  D_{b,u}'-r_b \leq r_u \leq  \|{\bf \check{L} }_b-\textbf{L}_u'\|, \\
	\dfrac{w}{2| \mathcal{A}(\textbf{L}_b',r_b,\textbf{L}_o,R_o) | } & |R_o -D_{b,o}'| \leq r_b \leq  R_o +D_{b,o}', \quad    \|{\bf \check{L} }_b-\textbf{L}_u'\| \leq r_u \leq  \|{\bf \hat{L} }_b-\textbf{L}_u'\|, \\
	\dfrac{w \mathbbm{1}_{ \{ \check{\theta}_b \leq \hat{\theta}_u \leq \hat{\theta}_b \} }  }{| \mathcal{A}(\textbf{L}_b',r_b,\textbf{L}_o,R_o) | } & |R_o -D_{b,o}'| \leq r_b \leq  R_o +D_{b,o}', \quad   \|{\bf \hat{L} }_b-\textbf{L}_u'\| \leq r_u \leq D_{b,u}'+ r_b,
	\end{cases} 
	\end{align}
	where $D_{b,o}'=\vert\vert \textbf{L}_b' \vert\vert$, $D_{b,u}'=\vert\vert \textbf{L}_b' - \textbf{L}_u' \vert\vert$, $\mathbbm{1}_{ \{\cdot\} }$ is the indicator function, 
	\begin{align}
	w &= \dfrac{2 r_u}{D_{b,u}'} \dfrac{1}{ \sqrt{ 1 - \left(  \dfrac{ (D_{b,u}')^2 +r_b^2 - r_u^2}{2D_{b,u}' r_b }  \right)^2 } },  \text{ and}
	\end{align}
	\begin{align}
	| \mathcal{A}(\textbf{L}_b',r_b,\textbf{L}_o,R_o) | &=	2 r_b \arccos\left( \dfrac{(D_{b,o}')^2 +r_b^2-R_o^2}{2 D_{b,o}' r_b} \right).
	\end{align}
	The locations ${\bf \check{L} }_b = \{\check{x}_b,\check{y}_b,0 \}$ and ${\bf \hat{L} }_b= \{\hat{x}_b,\hat{y}_b,0 \}$ are the points of intersection between $\mathcal{C}(\textbf{L}_o,R_o)$ and $\mathcal{C}(\textbf{L}_b',r_b)$ expressed as
	\begin{alignat}{2}
	\check{x}_{b} &= \check{x}_{b} & &= \dfrac{R_o^2 - r_b^2 +(D_{b,o}')^2}{2D_{b,o}'} \\
	\check{y}_{b} &=- \hat{y}_{b} & &= \sqrt{ R_o^2 - (\hat{x}_{b})^2} .
	\end{alignat}
	Denoting ${\textbf{L}}_x^+ = \{ \infty,0,0\}$ as a point in the positive $x$ direction, $\check{\theta}_b = \angle ({\textbf{L}}_x^+,\textbf{L}_b,{\bf \check{L} }_b) $ and $\hat{\theta}_b= \angle ({\textbf{L}}_x^+,\textbf{L}_b,{\bf \hat{L} }_b)$ are the angles at ${\bf L }_b'$ formed by moving from the line $\mathcal{L}( {\textbf{L}}_x^+,\textbf{L}_b )$ to $ \mathcal{L} (\textbf{L}_b, {\bf \check{L} }_b) $ and $ \mathcal{L} (\textbf{L}_b, {\bf \hat{L} }_b) $ counter clockwise. Similarly, $\check{\theta}_u = \angle ({\textbf{L}}_x^+,\textbf{L}_b,{\bf{L} }_u') $ and $\hat{\theta}_u=( \pi + \check{\theta}_u) \mod 2\pi$.	
	\proof See Appendix \ref{app:f_Du|Db} \label{lemma:f_Du|Db_general}. \hfill $\blacksquare$

\subsection{Coverage Probability}

	The coverage probability is defined as the probability that the received SNR is greater than a threshold $\beta$. In this subsection, we derive the coverage probability of access and backhaul links for given TBS and UAV locations. 
	\lemma For a given SNR threshold $\beta$, the coverage probability of the Rayleigh fading terrestrial access link (TBS--RUE) is defined as $P_{b,r}(\beta) \triangleq \mathbb{P} \left[ \text{SNR}_{b,r} > \beta \right]$ and given by
	\begin{align}
	P_{b,r}(\beta) =  \int_{-\infty}^{\infty}  P_{b,r\vert r_b}(\beta) f_{D_{b,r}'}(r_b) \; d r_b,
	\end{align}
	where $P_{b,r \vert r_b}(\beta)=\exp\left( - \bar{\beta}_b (r_b^2+h_b^2)^{\alpha_b/2} \right)$ is the coverage probability for a given distance to the TBS ($r_b$), $\bar{\beta}_b = \dfrac{\sigma_n^2 \beta}{\rho_b}$, and $f_{D_{b,r}'}(r_b)$ is the PDF of distance between $\textbf{L}_b$ and $\textbf{L}_r$ [c.f. Lemma \ref{lemma:f_D}].
	\label{theorem:p_c_b}
	\proof See Appendix \ref{app:p_c_b}.  \hfill $\blacksquare$
	
	\lemma For a given SNR threshold $\beta$, the coverage probability of the G2A/A2G Nakagami-$m$ fading aerial access link (UAV--RUE) is defined as $P_{u,r}(\beta) \triangleq \mathbb{P} \left[ \text{SNR}_{u,r} > \beta \right]$ and given by
	\begin{align}\label{eq:p_u-r}
	P_{u,r}(\beta) = \int_{-\infty}^{\infty}   \sum_{i\in \{\text{LoS},\text{NLoS}\}} \kappa_{u,r}^{i}  P_{u,r \vert r_u }^i(\beta) f_{D_{u,r}'}(r_u) \; dr_u,
	\end{align}
	where $P_{u,r \vert r_u }^i(\beta)= \sum_{k=0}^{m-1} \dfrac{ (m  \bar{\beta}_u (r_u^2+h_u^2)^{\alpha_u/2} {\eta_i})^k }{k!} \exp(-m  \bar{\beta}_u (r_u^2+h_u^2)^{\alpha_u/2} {\eta_i}), \ \forall i \in \{\text{LoS},\text{NLoS}\}$, is the LoS/NLoS coverage probability for a given distance to the UAV ($r_u$), $\bar{\beta}_u = \dfrac{\sigma_n^2 \beta}{\rho_u}$, and $f_{D_{b,r}'}(r_b)$ is the PDF of distance between $\textbf{L}_u$ and $\textbf{L}_r$ [c.f. Lemma \ref{lemma:f_D}].
	\label{theorem:p_c_u} 
	\proof See Appendix \ref{app:p_c_u}. \hfill $\blacksquare$

	\corollary For a given SNR threshold $\beta$, the coverage probability of the G2A/A2G Nakagami-$m$ fading aerial backhaul link (TBS--UAV) is defined as $P_{b,u}(\beta) \triangleq \mathbb{P} \left[ \text{SNR}_{b,u} > \beta \right]$ and given by
	\begin{align}\label{eq:p_b-u}
	P_{b,u}(\beta) = \sum_{i\in \{\text{LoS},\text{NLoS}\}} \kappa_{b,u}^{i} \sum_{k=0}^{m-1} \dfrac{ (m  \bar{\beta}_u D_{b,u}^{\alpha_u} {\eta_i})^k }{k!} \exp \left(-m  \bar{\beta}_u D_{b,u}^{\alpha_u} {\eta_i} \right) ,
	\end{align}
	where $D_{b,u}$ is the distance between the TBS and the UAV. 
	\proof This corollary follows by substituting the random RUE location into the deterministic TBS location in Lemma \ref{theorem:p_c_u}.  \hfill $\blacksquare$
	
	For a given SNR threshold $\beta$, the end-to-end coverage probability of the RUE associated with the UAV is defined as $P_{b,u,r}(\beta) \triangleq \mathbb{P}\left[ \min  \left( \text{SNR}_{u,r},\text{SNR}_{b,u} \right)  >\beta   \right] $ and given by
	\begin{align}
	\label{eq:P_bur}
	P_{b,u,r}(\beta) & = \mathbb{P}\left( (\text{SNR}_{u,r}>\beta) \quad {\cap} \quad (\text{SNR}_{b,u}>\beta) \right) \nonumber  \\
	&= \mathbb{P}\left( \text{SNR}_{u,r}  >\beta   \right) \mathbb{P}\left( \text{SNR}_{b,u} >\beta   \right) = P_{b,u}(\beta)P_{u,r}(\beta),
	\end{align}
	which follows from Lemma \ref{theorem:p_c_u}, Corollary 1, and independent out-of-band backhaul and access links assumption. For the T-UAV, \eqref{eq:P_bur} reduces to $P_{b,u,r}(\beta) = P_{u,r}(\beta)$ because a high capacity fiber link is assumed to reliably connect the T-UAV to the core network, i.e.,  $P_{b,u}(\beta)=1$ for the T-UAV. Based on the above coverage performance analyses and the association policy given in \eqref{B_los} and \eqref{B_nlos}, the overall T-UAV/U-UAV-assisted system coverage probabilities are given in the following theorems. 
	
	{\theorem Given the association policy in \eqref{B_los} and \eqref{B_nlos}, the T-UAV-assisted system coverage probability of UEs within the hot-spot $\mathcal{D}(\textbf{L}_o,R_o)$ is given by
	\begin{align}\label{eq:p^t}
	P^t (\beta)&=  \int_{-\infty}^{\infty}\left(\int_{-\infty}^{\lambda_{\text{LoS}} } E_{u,r}^{LoS} \; d r_u + \int_{\lambda_{\text{LoS}}}^{\infty} E_{b,r}^{LoS} \; d r_u  + \int_{-\infty}^{\lambda_{\text{NLoS}} } E_{u,r}^{NLoS} \; d r_u + \int_{\lambda_{\text{NLoS}} }^{\infty} E_{b,r}^{NLoS} \; d r_u  \right) d r_b ,
	\end{align}
	where the terms are given by $\lambda_{\text{LoS}}=\left(\dfrac{r_b^{\alpha_b}}{\eta_{\text{LoS}}} \right)^{\frac{1}{\alpha_u}}$, $\lambda_{\text{NLoS}}=\left(\dfrac{r_b^{\alpha_b}}{\eta_{\text{NLoS}}} \right)^{\frac{1}{\alpha_u}}$, 
	\begin{align}
	E_{b,r}^{\text{LoS}} &= \kappa_{u,r}^{\text{LoS}} P_{b,r \vert r_b}(\beta)  f_{D_{b,r}'} (r_b ) f_{D_{u,r}'|D_{b,r}'} (r_u|r_b ) ,\\
	E_{b,r}^{\text{NLoS}} &= \kappa_{u,r}^{\text{NLoS}} P_{b,r \vert r_b}(\beta)  f_{D_{b,r}'}(r_b ) f_{D_{u,r}'|D_{b,r}'} (r_u|r_b) , \\
	E_{u,r}^{\text{LoS}} &= \kappa_{u,r}^{LoS}  P_{u,r \vert r_u}^{\text{LoS}}(\beta) f_{D_{b,r}'} (r_b ) f_{D_{u,r}'|D_{b,r}'} (r_u|r_b ) , \text{ and}\\
	E_{u,r}^{\text{NLoS}} &= \kappa_{u,r}^{NLoS}  P_{u,r\vert r_u}^{\text{NLoS}}(\beta) f_{D_{b,r}'} (r_b ) f_{D_{u,r}'|D_{b,r}'}(r_u|r_b ) .
	\end{align} 
	\proof The proof follows directly from Lemmas \ref{lemma:f_D}- \ref{theorem:p_c_u} and the association policy in \eqref{B_los} and \eqref{B_nlos}. \hfill $\blacksquare$
}

	In \eqref{eq:p^t}, the first and third terms correspond to the coverage probability under LoS and NLoS aerial access links, while the second and forth terms correspond to the coverage probability for the terrestrial access links. Also notice that  \eqref{eq:p^t} does not consider the backhaul link since $P_{b,u}(\beta)=1$.

	{\theorem Given the association policy in \eqref{B_los} and \eqref{B_nlos}, the U-UAV-assisted system coverage probability of UEs within the hot-spot $\mathcal{D}(\textbf{L}_o,R_o)$ is given by
	\begin{align}\label{eq:p^u}
	\nonumber P^u (\beta) &= A \int_{-\infty}^{\infty}\left(\int_{-\infty}^{\lambda_{\text{LoS}} } E_{b,u,r}^{LoS} \; d r_u + \int_{\lambda_{\text{LoS}}}^{\infty} E_{b,r}^{LoS} \; d r_u  + \int_{-\infty}^{\lambda_{\text{NLoS}} } E_{b,u,r}^{NLoS} \; d r_u + \int_{\lambda_{\text{NLoS}} }^{\infty} E_{b,r}^{NLoS} \; d r_u  \right) d r_b \\
	&+ (1-A) P_{b,r}(\beta),
	\end{align}
	where
	$E_{{b,u,r}}^{LoS} = P_{b,u} E_{u,r}^{LoS}$ and $E_{{b,u,r}}^{NLoS} = P_{b,u} E_{u,r}^{NLoS}$.
	\proof The proof follows directly from Lemmas \ref{lemma:f_D}- \ref{theorem:p_c_u} and the association policy in \eqref{B_los} and \eqref{B_nlos}. \hfill $\blacksquare$}

	In \eqref{eq:p^u}, the first term is the coverage probability given that U-UAV is available while the second term is the coverage probability over the TBS due to the unavailability of the U-UAV. For given TBS location ($\textbf{L}_b$), UAV  location ($\textbf{L}_u$), and SNR threshold ($\beta$), Theorem 1 and Theorem 2 derive the coverage probability provided by a T-UAV and U-UAV for a random user within the hot-spot $\mathcal{D}(\textbf{L}_o,R_o)$, respectively.
	 
\section{Optimal UAV hovering location}
\label{sec:deployment}

	The UAV deployment plays a critical role in maximizing the overall system performance. In the previous sections, the U-UAV and the T-UAV system performances are analyzed for a given UAV location $\textbf{L}_u$. Therefore, it is necessary to find the optimal UAV location for the maximum system coverage. Accordingly, the UAV deployment problem can be formulated for U-UAV and T-UAV as 
	\begin{align}
	\mathsf{P}_\mathsf{u}: & \hspace{.7cm} \underset{\textbf{L}_u \in \mathbb{R}^3}{\max} \hspace{.3cm} P^u(\beta), \label{eq:P_u} \\ 
	\mathsf{P}_\mathsf{t}: & \quad \underset{\textbf{L}_u  \in \, \underset{n}{\bigcup} \, \mathcal{M}_n, \, \forall n}{\max} P^t(\beta), \label{eq:P_t}
	\end{align}
	respectively. Note how in \eqref{eq:P_u} the U-UAV can be located anywhere in $\mathbb{R}^3$ while in \eqref{eq:P_t} the T-UAV mobility is restricted to the $n$-th spherical cone centered at the $n$-th GS location. Considering the highly non-linear nature and large search space of these problems, we first narrow down the problem search space by proving that the optimal deployment location falls within a specific subspace. 
	
	\begin{itemize}[leftmargin=*]
		\item 
		Given that the hot-spot is centered at the origin and the TBS is located at $\text{L}_b = \{x_b,0,h_b\}$, one can observe the symmetry of the UAV locations around the $x$-axis as shown in Fig. \ref{fig:symmetry_around_x}. For the U-UAV, we therefore only study the half space $\{y \geq 0\}$ and generalize the result for the other half without loss of generality. For the T-UAV, we also study the case $y_n \geq 0$ only and generalize the findings to the other half space. We note that some part of the spherical cone $\mathcal{M}_n$ may belong to the half space $\{y \leq 0\}$  if $\textbf{L}_n$ is near the $x$-axis [c.f Fig. \ref{fig:symmetry_around_x}]. In this case, the cropped spherical cone is denoted by $\bar{\mathcal{M}}_n$. We ignore the cropped part of the spherical cone since it is symmetric to a subset of the spherical cone within $\{y \geq 0\}$.
		
		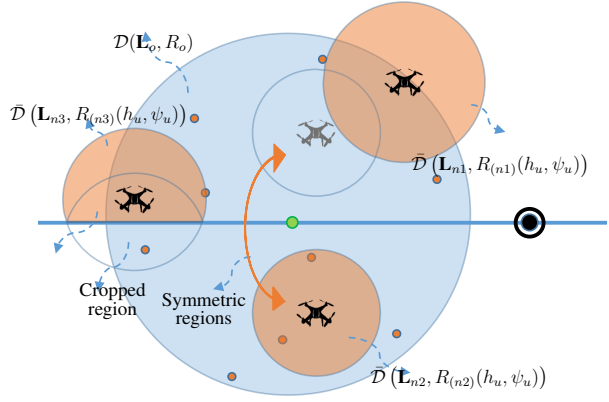
\begin{figure}
			\centering
			\begin{tikzpicture}[thick,scale=0.6, every node/.style={scale=0.6}]
			\draw (0, 0) node[inner sep=0] {	\includegraphics[trim={8cm 5cm  14cm 5cm},clip, width=.7 \linewidth]{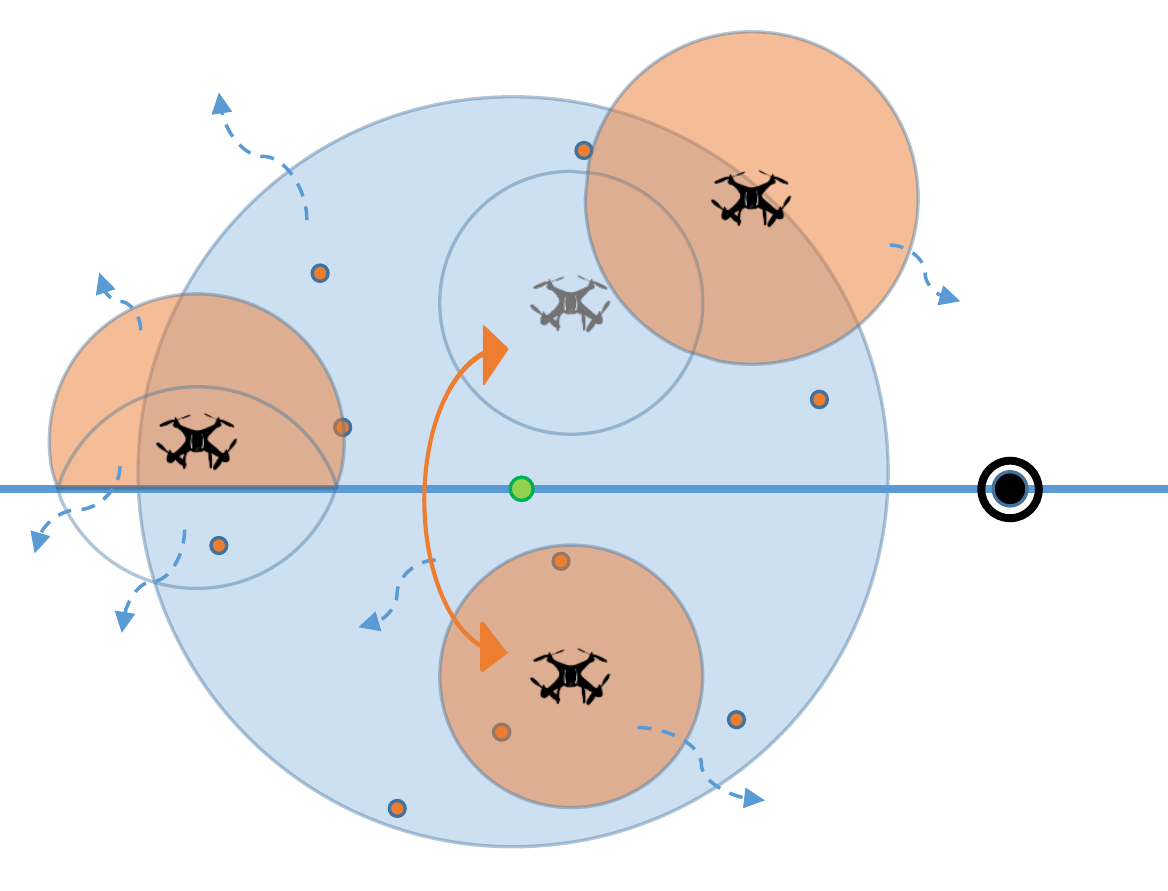}};	
			\draw (-3.7, 3.5) node {{$\mathcal{D}(\textbf{L}_o,R_o)$}};
			\draw (-4.9, 1.9) node {{$\bar{\mathcal{D} }\left( \textbf{L}_{n3},R_{(n3)}(h_u,\psi_u) \right)$}};
			\draw (3.0, -3.9) node {{$\bar{\mathcal{D} }\left( \textbf{L}_{n2},R_{(n2)}(h_u,\psi_u) \right)$}};
			\draw (3.9, 0.8) node {{$\bar{\mathcal{D} }\left( \textbf{L}_{n1},R_{(n1)}(h_u,\psi_u) \right)$}};
			\draw (-2.6, -2.2) node {{Symmetric}};
			\draw (-2.6, -2.6) node {{regions}};
			\draw (-4.6, -2.0) node {{Cropped}};
			\draw (-4.6, -2.4) node {{region}};
			\end{tikzpicture}
			\caption{Aerial view of the cone symmetry and cropped cone $\bar{\mathcal{M} }_n$ for a fixed T-UAV height.}  \label{fig:symmetry_around_x}
		\end{figure}
		
		\item 
		Let us define the angle and the distance between the ground projections of $\textbf{L}_n$ and $\textbf{L}_u$ as 
		\begin{align}
		\psi_u^{n} &= \angle ({\textbf{L}}_x^+,\textbf{L}_n',\textbf{L}_u') \text{ and}\\
		R_u^{n} &= \| \textbf{L}_n' - \textbf{L}_u' \|,
		\end{align}
		respectively. Accordingly, the spherical cone $\mathcal{M}_n$ can be expressed by the cylindrical coordinates as
		\begin{align}
		\mathcal{M}_n = \left\{
		R_u^{n}, \psi_u^{n}, h_u : \  h_u \in [h_n, h_n+T], \; \psi_u^{n} \in [0, 2\pi], \; R_u^{n} \leq R_n(h_u)
		\right\},
		\end{align}
		where 
		\begin{align}
		R_n(h_u)=	\begin{cases}
		(h_u - h_n) \tan(\phi) &\quad h_u < h_n + T \cos(\phi) ,\\
		\sqrt{T^2 - (h_u-h_n)^2 } &\quad  h_u \geq h_n + T \cos(\phi)
		\end{cases}
		\end{align}
		represents the cone bounds for given  T-UAV height $ h_u < h_n + T \cos(\phi)$ and the spherical bounds for $ h_u \geq h_n + T \cos(\phi)$. 
		\item 
		To define the cropped spherical cone $\bar{\mathcal{M}}_n$, we need to guarantee that the distance $R_u^{n}$ does not exceed the $x$-axis. Hence, $\bar{\mathcal{M}}_n$ is given by, 
		\begin{align}
		\bar{\mathcal{M} }_n =  \left\{ R_u^{n}, \psi_u^{n}, h_u : \  h_u \in [h_n, h_n+T], \;
		\psi_u^{n} \in [0, 2\pi], \; R_u^{n} \leq \bar{R}_n(h_u,\psi_u^{n})
		\right\},
		\end{align}
		where
		\begin{align}
		\bar{R}_n(h_u,\psi_u^{n}) = \begin{cases}
		R_n(h_u) &\quad \psi_u \in [0,\pi] , \\
		\min\left( R_n(h_u), \dfrac{-y_n}{\sin(\psi_u)} \right) &\quad \psi_u \in (\pi, 2\pi),
		\end{cases} 
		\end{align}
		is the truncated version of $R_n(h_u)$ as a result of the cropped spherical cone. 
	\end{itemize} 
	In the following theorem, we prove that the optimal T-UAV location belongs to a portion of the spherical cone surface. 
	
\begin{figure}
	\centering
	\begin{tikzpicture}[thick,scale=0.6, every node/.style={scale=0.6}]
	\draw (0, 0) node[inner sep=0] {	\includegraphics[trim={8cm 5cm  14cm 5cm},clip, width=.7 \linewidth]{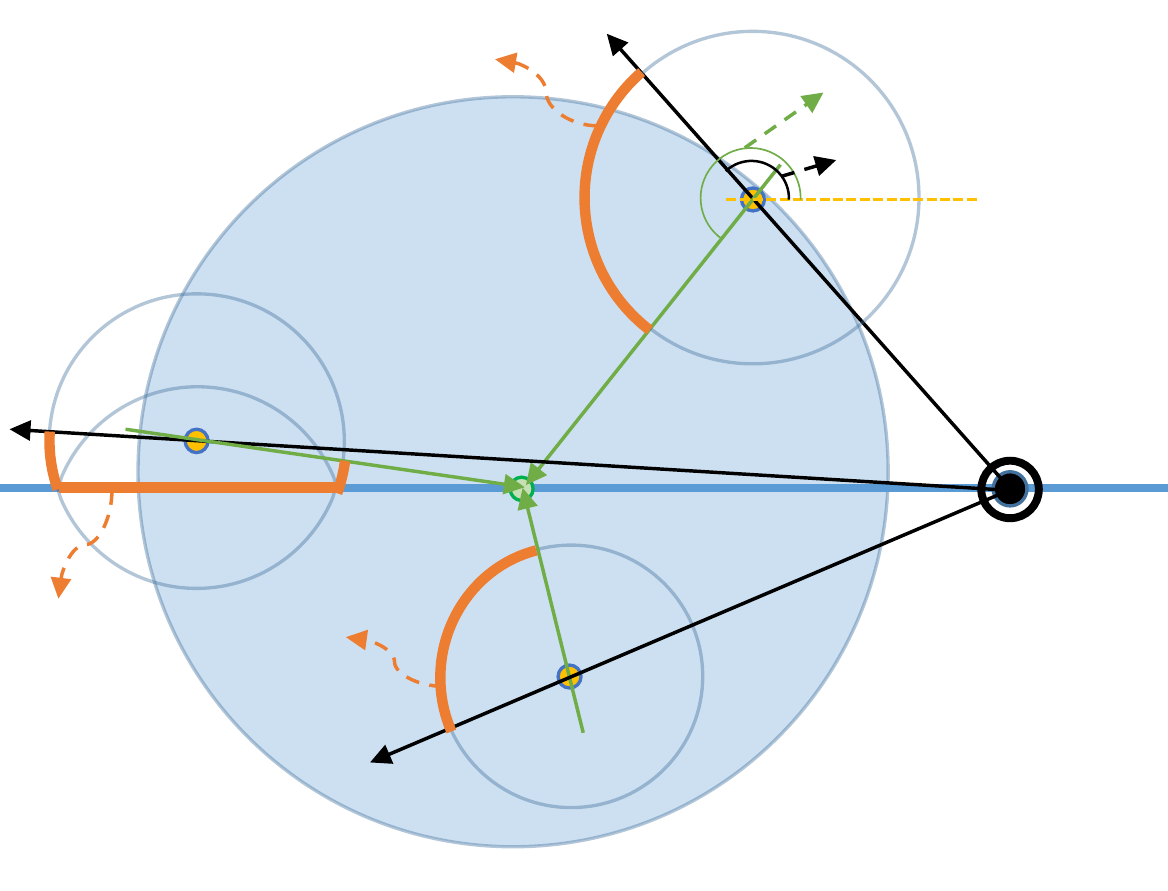}};	
	\draw (-4.9, -1.7) node {{$\mathcal{O}_{n3}(h_u)$}};
	\draw (-3.1, -1.9) node {{$\mathcal{O}_{n2}(h_u)$}};
	\draw (-1.7, 3.8) node {{$\mathcal{O}_{n1}(h_u)$}};
	\draw (2.9, 3.6) node {{$\psi_2^{(n1)}$}};
	\draw (3.0, 2.9) node {{$\psi_1^{(n1)}$}};
	\end{tikzpicture}
	\caption{The regions encompass the optimal T-UAV location at a given T-UAV altitude, $h_u$.} \label{fig:region3}
\end{figure}
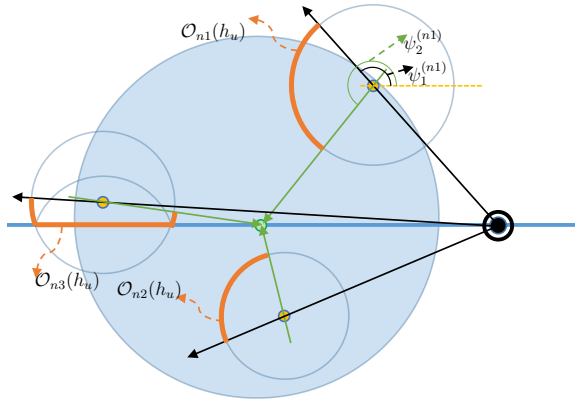

	\theorem For a given GS location $\textbf{L}_n = \{ x_n,y_n\geq 0, h_n\}$ and considered user association policy, the optimal T-UAV location, $\textbf{L}_u \in \bar{\mathcal{M} }_n$, that maximizes the overall coverage performance of the hot-spot, $P^t(\beta)$, falls within the following set of locations 
\begin{align}
\mathcal{O}_n =  \left\{
R_u^{n}, \psi_u^{n}, h_u : \
h_u \in [h_n, h_n+T], \;
\psi_u^{n} \in [\psi_1^{n}, \psi_2^{n}], \;
R_u^{n} = \bar{R}_n(h_u,\psi_u^{n})
\right\} \in \bar{\mathcal{M} }_n, 
\end{align}
where $\psi_1^{n} = \angle \left( {\textbf{L}}_x^+ , \textbf{L}_b', \textbf{L}_n'\right)$ and $\psi_2^{n} = \angle \left({\textbf{L}}_x^+, \textbf{L}_n',\textbf{L}_o \right)$. 
\proof  Please see Appendix \ref{app:optimal_region}. An illustration of the set $\mathcal{O}_n$ at a fixed $h_u$ is shown in Fig. \ref{fig:region3}.   \hfill $\blacksquare$
\label{lemma:optimal_region}

\corollary For a given GS location $\textbf{L}_n = \{ x_n,y_n\geq 0, h_n\}$ and considered user association policy, the optimal U-UAV location that maximizes the overall coverage performance within the hot-spot, $P^u(\beta)$ is located at $\textbf{L}_u = \{x_u,0,h_u \}$ such that  $x_u \leq x_b$ and $h_u \geq 0$. 
\proof  Please see Appendix \ref{app:optimal_region}.    \hfill $\blacksquare$

	Theorem 3 significantly reduces the 3D search space within the T-UAV spherical cone to a 2D search space within $\mathcal{O}_n$, which yields a significantly lower computational complexity for the optimal location search.

\section{Numerical Analysis}
\label{sec:results}

	In this section, we validate the mathematical analysis with independent Monte Carlo simulations and provide insightful performance results with respect to different system parameters and scenarios. Unless stated otherwise, we employ the default system parameters listed in Table \ref{table:parameters}, which mainly follow the parameters listed in \cite{Alzenad2018, Halim2016, kishk2019capacity}. Since T-UAVs and U-UAV systems have their own virtues and drawbacks, the limiting parameters are not the same for both. While tethered UAVs are limited by tether length and ground station building accessibility parameters, U-UAV is mainly limited by the wireless backhaul capacity and battery lifetime. To this end, we quantify the mobility limitation of tethered UAVs by considering different tether lengths and ground station building accessibility. To provide a fair comparison, we evaluate the performance of these systems in the same environment, with the same users, and given the same TBS and hot-spot locations.

	\begin{table}[H]
		\caption{Default System Parameters.}	\label{table:parameters} 
		\centering
		\begin{tabular}
			{| p{2cm} | p{2.1cm}  |  p{.9cm} | p{1.9cm} |  p{.9cm} | p{1.7cm} |} 
			\hline
			Par. & Value & Par. & Value & Par. & Value \\ [0.5ex] 
			\hline\hline

			$\rho_b$ / $\rho_u$ & $1$ dBm  &$\sigma_n^2$ & $-80$ dBm & $\beta$& $15$ \\ 
			\hline
			$\alpha_b$ & $3$ & $\alpha_u$ & $2.7$  & $m$ & $2$ \\
			\hline
			$ \eta_{LoS} / \eta_{NLoS}$ & $1.6/ 23$ dB  & $a_r/b_r$ & $ 13/ .22 $ &  $a_b/ b_b$   & $ 7/.2 $  \\
			\hline
			$R_o$ & $150 $ & $\textbf{L}_b$ & $\{170,0,10\}$  & $ T/\theta_T$ & $50 $m$/ 30^o$  \\
			\hline
		\end{tabular}
	\end{table}

\begin{figure}[htbp!]
	\centering
	\begin{subfigure}{.32\linewidth}
		\centering
		\begin{tikzpicture}[thick,scale=1, every node/.style={scale=1}]
		\draw (0, 0) node[inner sep=0] {	\includegraphics[clip, width=1 \linewidth]{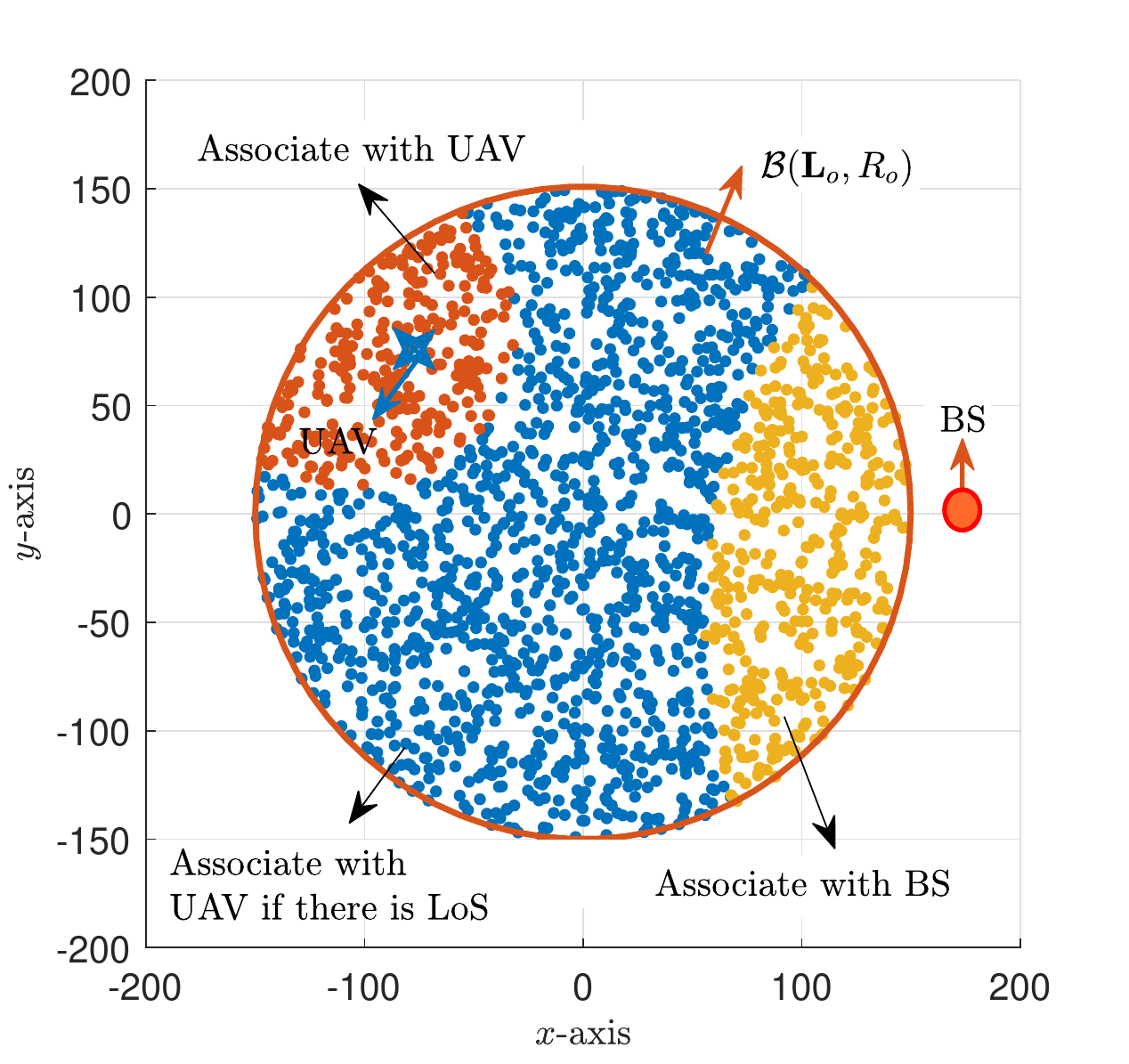}};	
		\end{tikzpicture}
		\caption{ } \label{fig:uni_asc}
	\end{subfigure}
	\begin{subfigure}{.32\linewidth}
		\centering
		\begin{tikzpicture}[thick,scale=1, every node/.style={scale=1}]
		\draw (0, 0) node[inner sep=0] {	\includegraphics[clip, width=1 \linewidth]{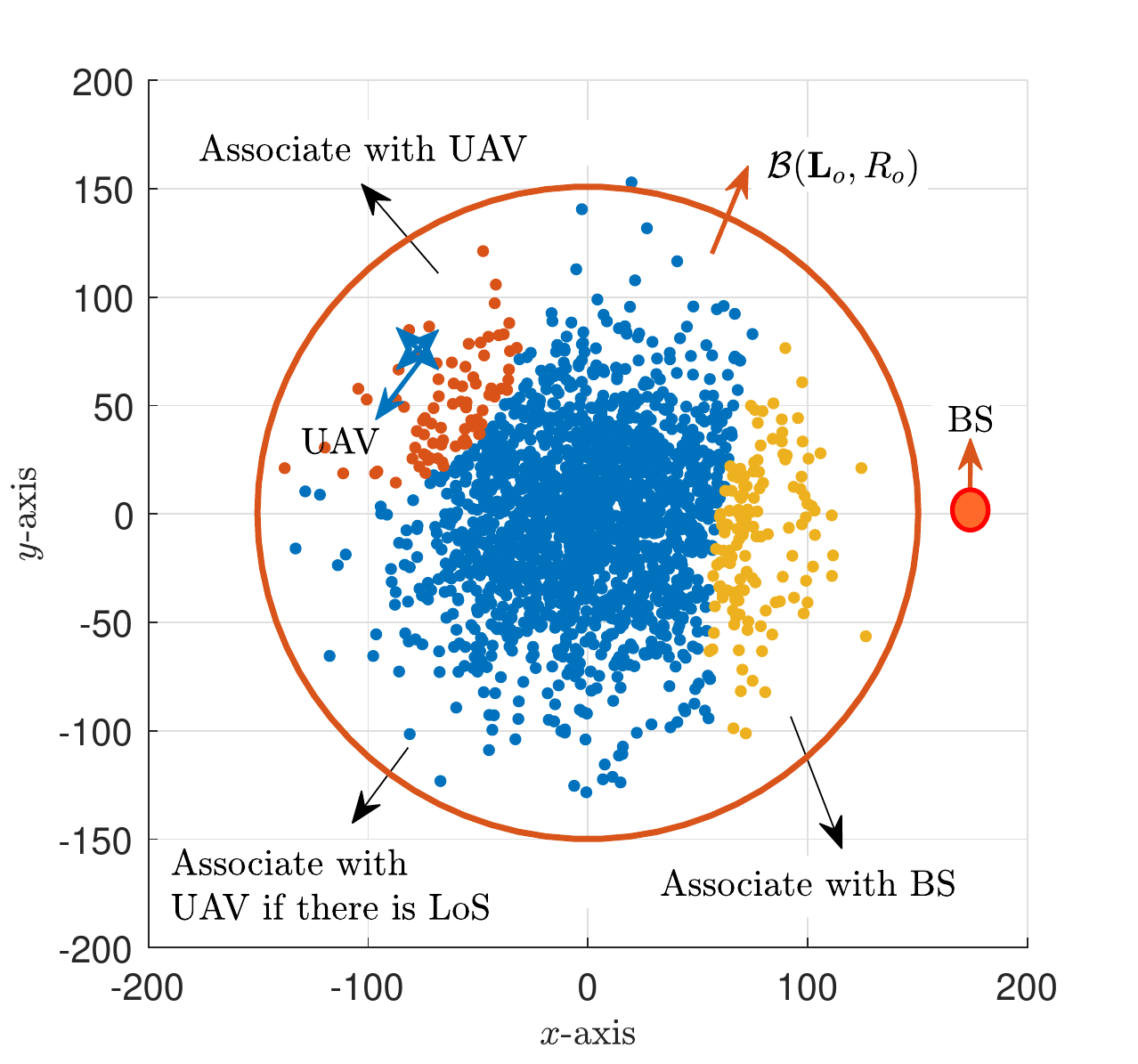}};	
		\end{tikzpicture}
		\caption{ }  \label{fig:normal_asc}
	\end{subfigure}
		\begin{subfigure}{.34\linewidth}
       \centering
	\begin{tikzpicture}[thick,scale=1, every node/.style={scale=1}]
	\draw (0, 0) node[inner sep=0] {	\includegraphics[clip, width=1 \linewidth]{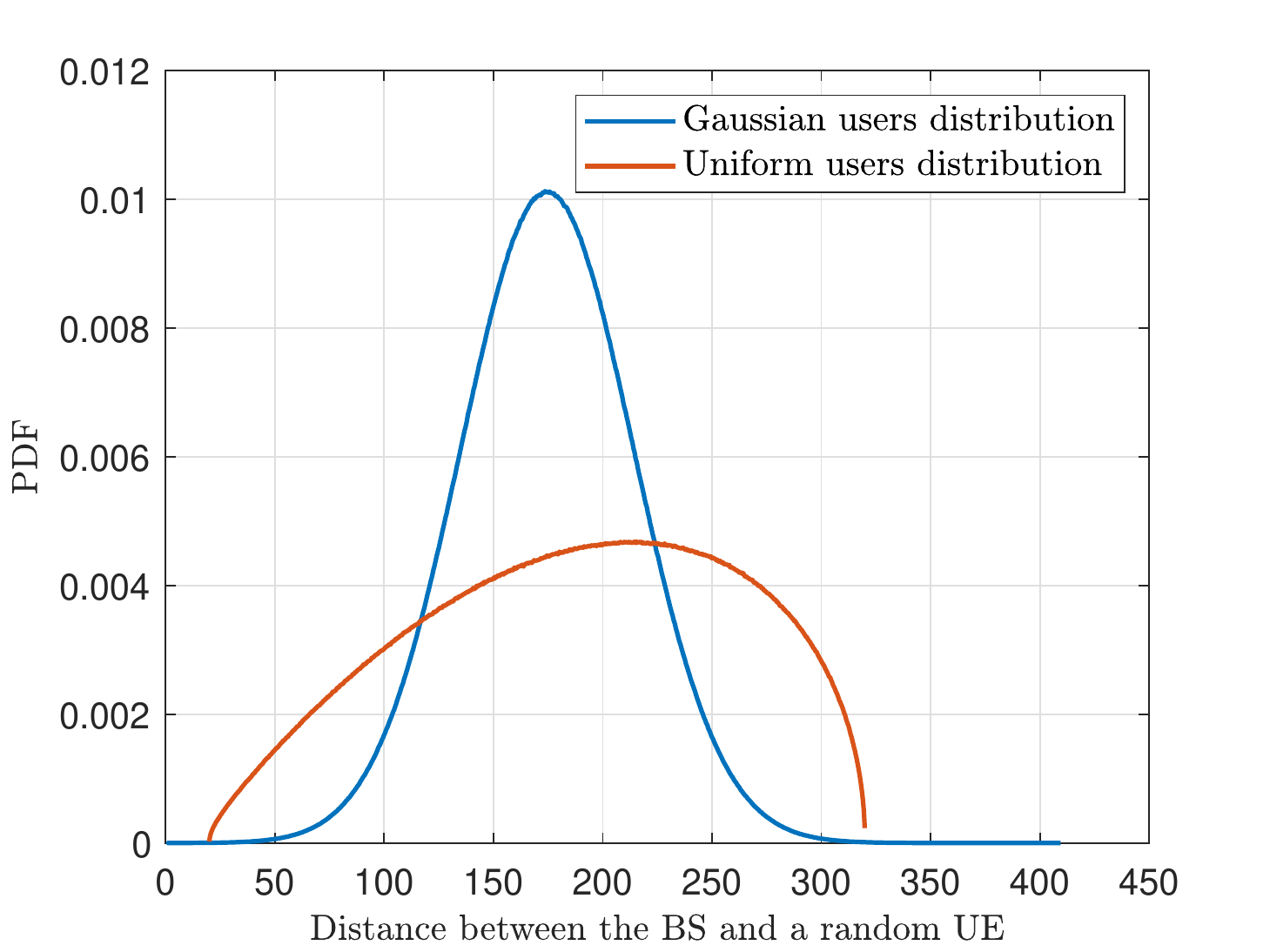}};	
	\end{tikzpicture}
		\caption{ }  \label{fig:uni_norm_dist}
	\end{subfigure}
	\caption{ The impact of UAV/BS location on user association and distance distribution: a) The user association regions of uniformly distributed UEs, a) The user association regions of Gaussian distributed UEs, and c) Corresponding distance PDF between a UE and the BS.}  \label{fig:asc_dist}
\end{figure}

To have a better insight into the relationship between coverage performance and UAV deployment, let us first explain how user association and distance distributions are commonly affected by deterministic UAV and BS locations. For a UAV located at $\textbf{L}_u = \{-75, 75, 50\}$, the association regions are shown in Fig. \ref{fig:uni_asc} where the users located within the orange and the yellow regions always associate with the UAV and the TBS, respectively. However, UEs which fall in the blue region associate with the UAV only if there is a LoS aerial access link. For the same UAV and BS locations, these regions do not change for other user distributions [c.f. Fig. \ref{fig:normal_asc} for Gaussian distribution]. However, it is worth noting that the shape, area, and orientation of these three regions vary with UAV and BS locations. We now focus on the distance distribution which describes the probability that a randomly selected user in the hot-spot is at a distance $D$ from a fixed/deterministic point, e.g., the TBS or the UAV. We illustrate in Figure \ref{fig:uni_norm_dist} the distance PDF between a BS at ${\textbf{L}}_b = \{170,0,10\}$ and a random user in the hot-spot $\mathcal{D}(\textbf{L}_o,150)$ for uniformly distributed and Gaussian distributed users. For example, the peak point in the Gaussian user distribution is around 170 as most users are very close to the hot-spot center. Therefore, one can expect that the distance PDF shape (mean, variance, skewness, kurtosis) will vary with both users' spatial distribution and location of the target point. Above discussions and results make it clear that 1) overall coverage and UAV deployment heavily depend on the TBS location and user distribution within the hot-spot region, and 2) the generalization of analytical results derived for a given spatial user distribution is not readily applicable to other distributions.  
	
\begin{figure}
		\centering
		\begin{tikzpicture}[thick,scale=0.8, every node/.style={scale=0.8}]
		\draw (0, 0) node[inner sep=0] {	\includegraphics[clip, width=.6 \linewidth]{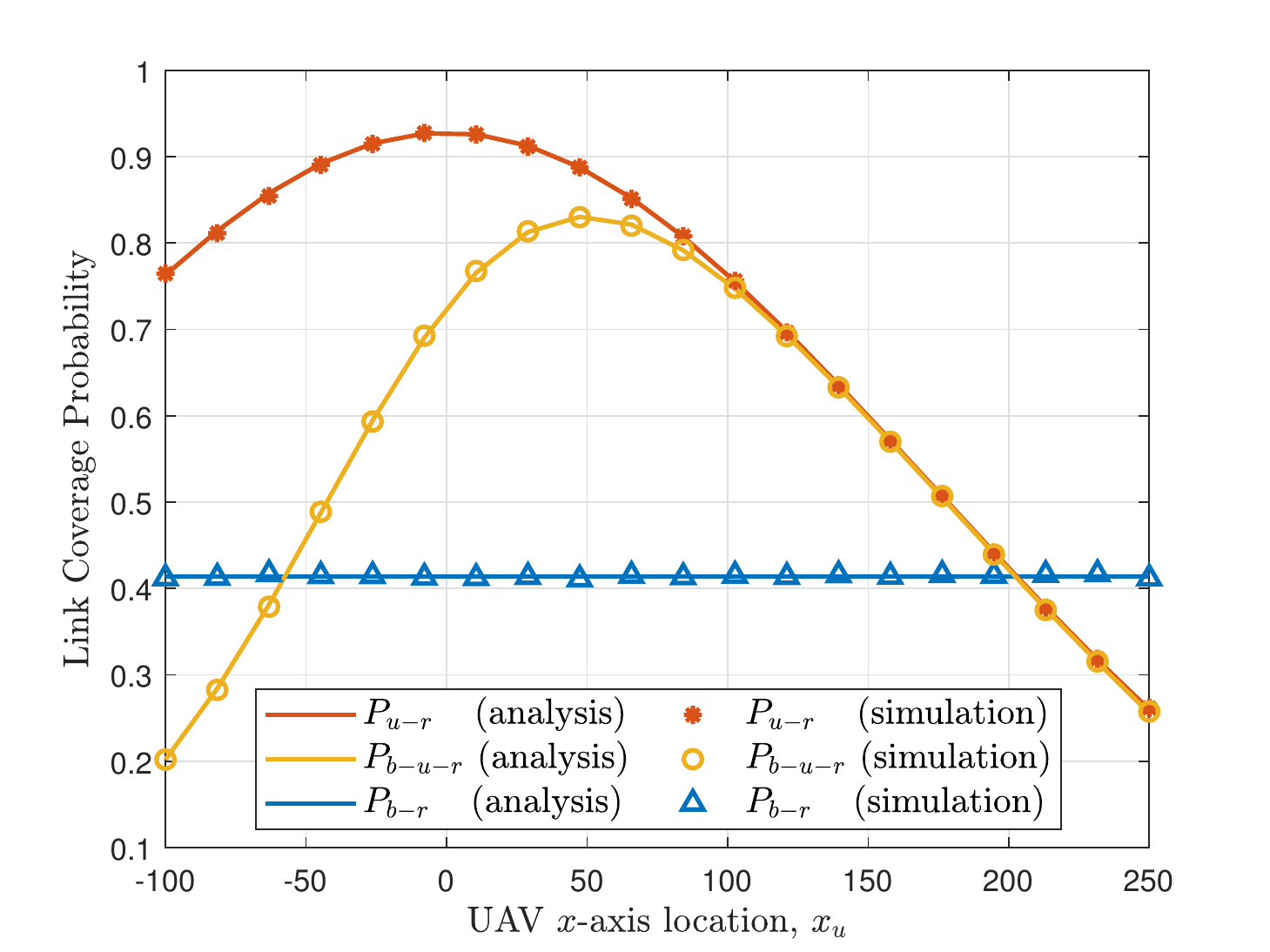}};	
		\end{tikzpicture}
		\caption{The coverage probability of access and backhaul links.} \label{fig:cov_p_links}
	\end{figure}

Let us now focus on the coverage performance of access and backhaul links for varying UAV locations $\textbf{L}_u = \{ x_u, 0, 100 \}$, $x_u \in [-100,175]$. Fig. \ref{fig:cov_p_links} shows the coverage probabilities for the terrestrial access link TBS--RUE,  $P_{b,r}$,  the aerial access link T-UAV--RUE, $P_{u,r}$, and the end-to-end TBS--U-UAV--RUE link,  $P_{b,u,r}$. As expected, the TBS link is not influenced by the UAV location. Given that T-UAV and U-UAV hover at the same location, T-UAV always outperforms the U-UAV thanks to the high capacity wired backhaul link.  For a clear comparison between T-UAVs and U-UAVs, let us focus on the locations where UAVs reach the maximum end-to-end coverage. The T-UAV reaches the maximum coverage when hovering over the hot-spot origin because it gives the maximum access link coverage to all users which are uniformly distributed over the area of interest. On the other hand, the U-UAV reaches the peak coverage at a point 50 m closer to the TBS, which is mainly because of the tradeoff between the backhaul and the access links. Since the end-to-end SNR is determined by the minimum of the access and the backhaul links, the maximum system coverage can be achieved in an equilibrium state which is obtained by getting closer to the TBS. 

\begin{figure}
		\centering
		\begin{tikzpicture}[thick,scale=0.8, every node/.style={scale=0.8}]
		\draw (0, 0) node[inner sep=0] {	\includegraphics[clip, width=.6 \linewidth]{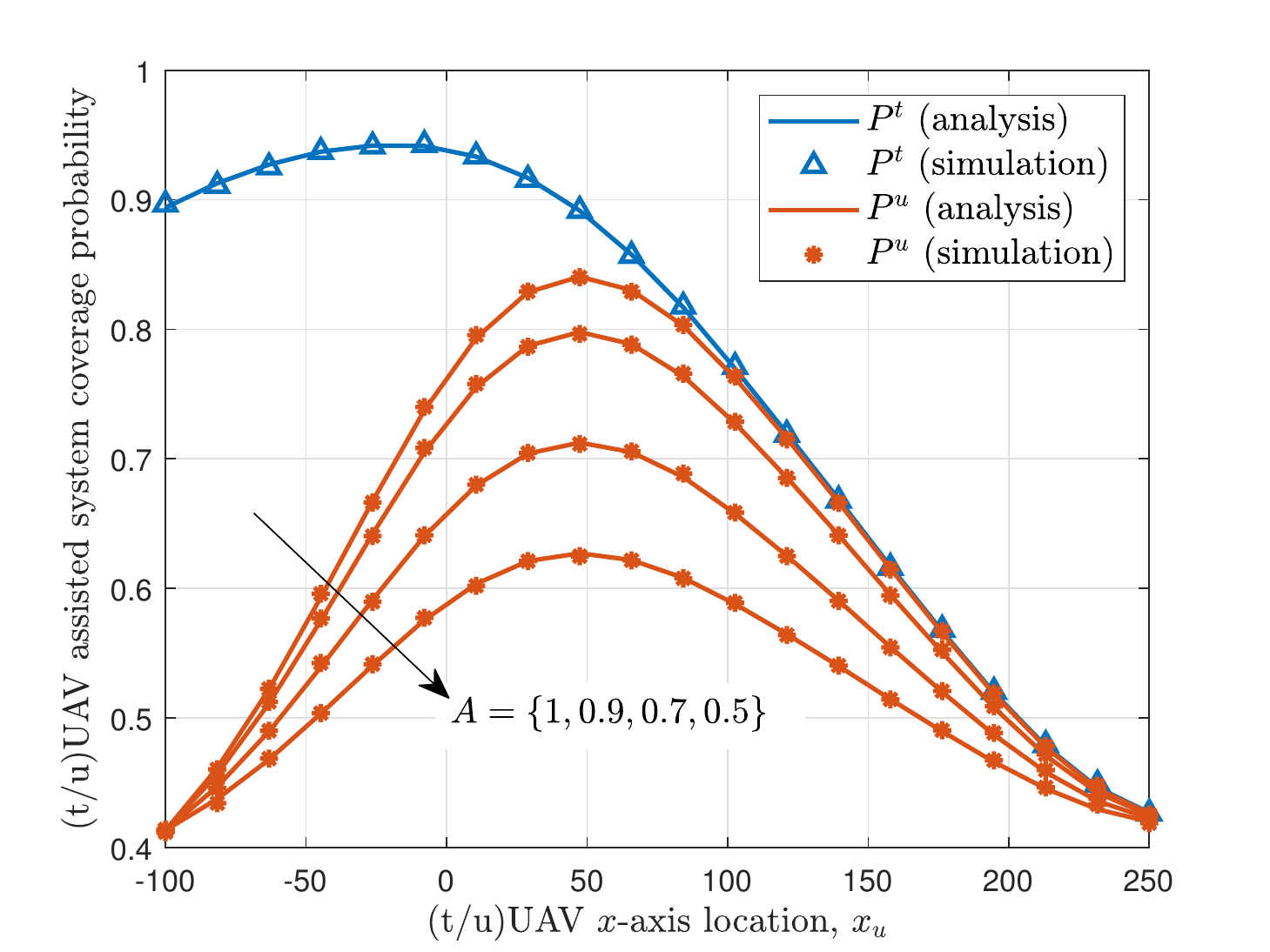}};	
		\end{tikzpicture}
		\caption{$P^u$ and $P^t$ for different U-UAV duty cycle values.} \label{fig:cov_p_sys_x}
	\end{figure}

	Furthermore, Fig. \ref{fig:cov_p_sys_x} shows the impact of U-UAV availability under the considered user association policy.  Intuitively, duty cycle of the U-UAV availability $A$ has a significant impact on the overall system coverage. One can observe that the maximum coverage point of the T-UAV is shifted towards the negative region because users closer to the TBS are associated with the TBS. On the other hand, the maximum coverage point of the U-UAV is still over the positive $x$-axis because the aforementioned tradeoff dominates the system behavior. Notice in  Fig. \ref{fig:cov_p_links} and Fig. \ref{fig:cov_p_sys_x} that the T-UAV and the U-UAV are assumed to be located at the same location. This assumption is made for the sake of a clear demonstration of the access and the backhaul link dynamics. However, in reality, the T-UAV is restricted by the tether length, inclination angle and the GS location. 
	
\begin{figure}
		\centering
		\begin{subfigure}{.45\linewidth}
			\centering
			\begin{tikzpicture}[thick,scale=0.9, every node/.style={scale=0.9}]
			\draw (0, 0) node[inner sep=0] {	\includegraphics[clip, width=1 \linewidth]{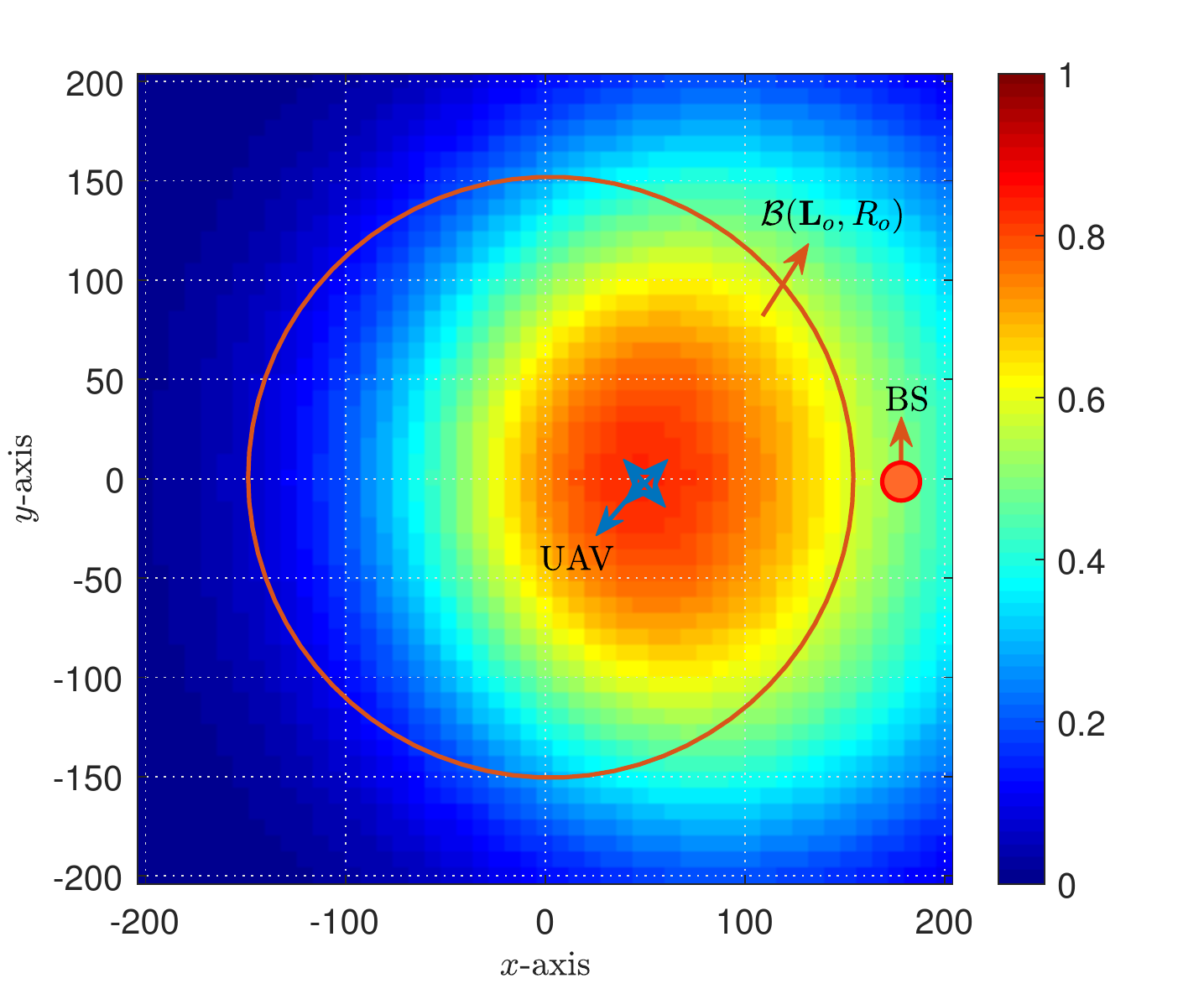}};	
			\end{tikzpicture}
			\caption{Optimal U-UAV location with maximum $P^u$.} \label{fig:cov_p_sys_uav}
		\end{subfigure}
		\begin{subfigure}{.45\linewidth}
			\centering
			\begin{tikzpicture}[thick,scale=0.9, every node/.style={scale=0.9}]
			\draw (0, 0) node[inner sep=0] {	\includegraphics[clip, width=1 \linewidth]{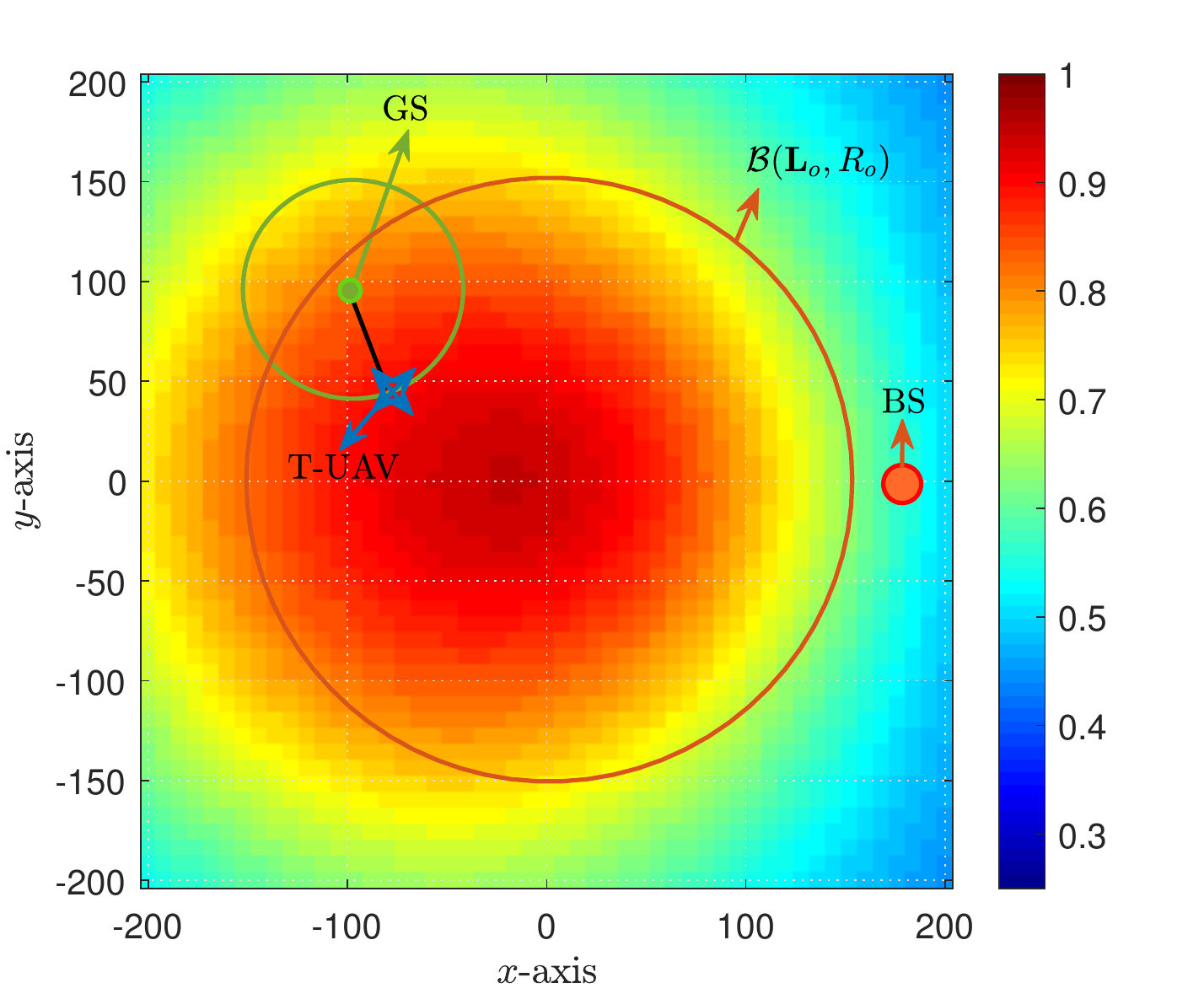}};	
			\end{tikzpicture}
			\caption{Optimal T-UAV location with maximum $P^t$.} \label{fig:cov_p_sys_tuav}
		\end{subfigure}
		\caption{Optimal U-UAV and T-UAV  locations that maximize $P^u$ and $P^t$.}  
	\end{figure}

	In order to consider a more realistic scenario, we present the overall system coverage probabilities for the U-UAV and the T-UAV in Fig. \ref{fig:cov_p_sys_uav} and \ref{fig:cov_p_sys_tuav}, respectively. To this aim, we first consider a discrete exhaustive solution approach by dividing the x-y plane into 8 m$^2$ grids at a fixed UAV height ($100$ m). Then, the coverage probability $P^u$ is calculated at the center of each grid and displayed by means of a color map. Intuitively, the best location for the U-UAV can be obtained by selecting the grid center with the maximum system coverage [c.f. Fig. \ref{fig:cov_p_sys_uav}]. In order to alleviate the computational complexity of the exhaustive approach, Fig. \ref{fig:cov_p_sys_uav} also shows the location calculated by the simulated annealing approach which can provide $10^{-3}$ coverage probability tolerance in only $20$ iterations. Likewise,  Fig. \ref{fig:cov_p_sys_tuav} shows the coverage probability $P^t$ of the T-UAV for a given GS location. It is obvious that the GS location and tether length poses a significant challenge to be located at the optimal location. Moreover, Theorem \ref{lemma:optimal_region} is numerically verified in Fig. \ref{fig:cov_p_sys_tuav}. By drawing any circle $\mathcal{C}( \textbf{L}_n',R_n(h_u) )$ with $\textbf{L}_n'$ and $R_n(h_u)$ representing the GS $x-y$ location and the radius within which the T-UAV can fly at the height $h_u$, the maximum $P^t$ in $\mathcal{D}( \textbf{L}_n',R_n(h_u) )$ belongs to the region described in the theorem.

\begin{figure}
		\centering
		\begin{tikzpicture}[thick,scale=.5, every node/.style={scale=.5}]
		\draw (0, 0) node[inner sep=0] {	\includegraphics[clip, width=1 \linewidth]{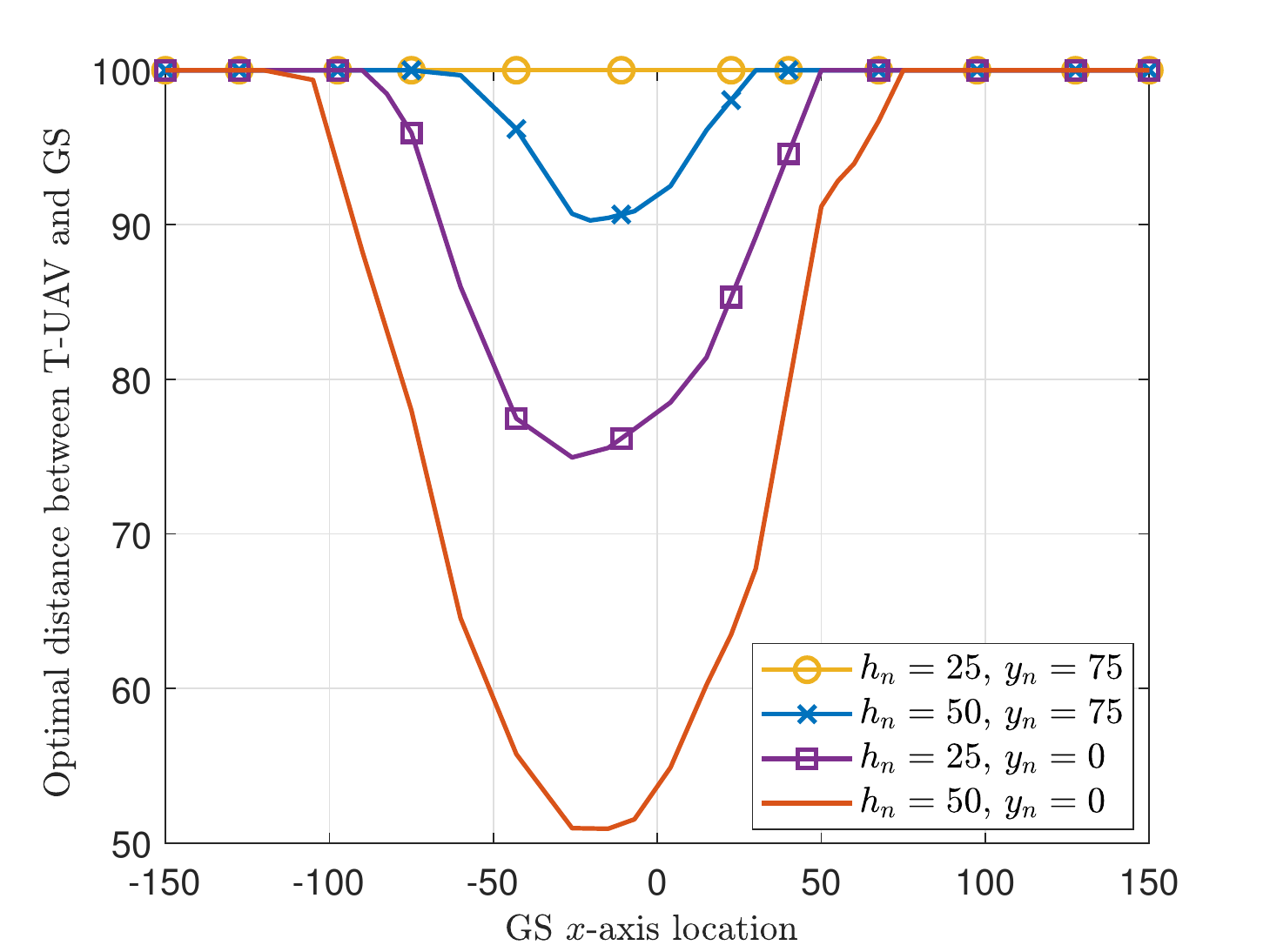}};
		\end{tikzpicture}
		\caption{Optimal distance between T-UAV and GS for several GS locations.}
		\label{fig:tether_length_e}
	\end{figure}

	Notice that T-UAV does not necessarily use the full tether length. In Fig. \ref{fig:tether_length_e}, we show the optimal distance, which maximizes the coverage probability, between the T-UAV and the GS for several GS locations. Given a maximum tether length $T=100$, a T-UAV connected to a GS at $\textbf{L}_n = \{x_n, 75, 25\} \, \forall x_n$ is located at a distance equal to $T$ from the GS, i.e., the T-UAV is located on the spherical edge of the tether spherical cone. As the GS gets closer to the T-UAV optimal location at $\textbf{L}_u = \{ -18.125, 0, 100\}$, the optimal distance between the T-UAV and the GS decreases.

\begin{figure}
		\begin{subfigure}{.5\linewidth}
			\centering
			\begin{tikzpicture}[thick,scale=1, every node/.style={scale=1}]
			\draw (0, 0) node[inner sep=0] {	\includegraphics[clip, width=1 \linewidth]{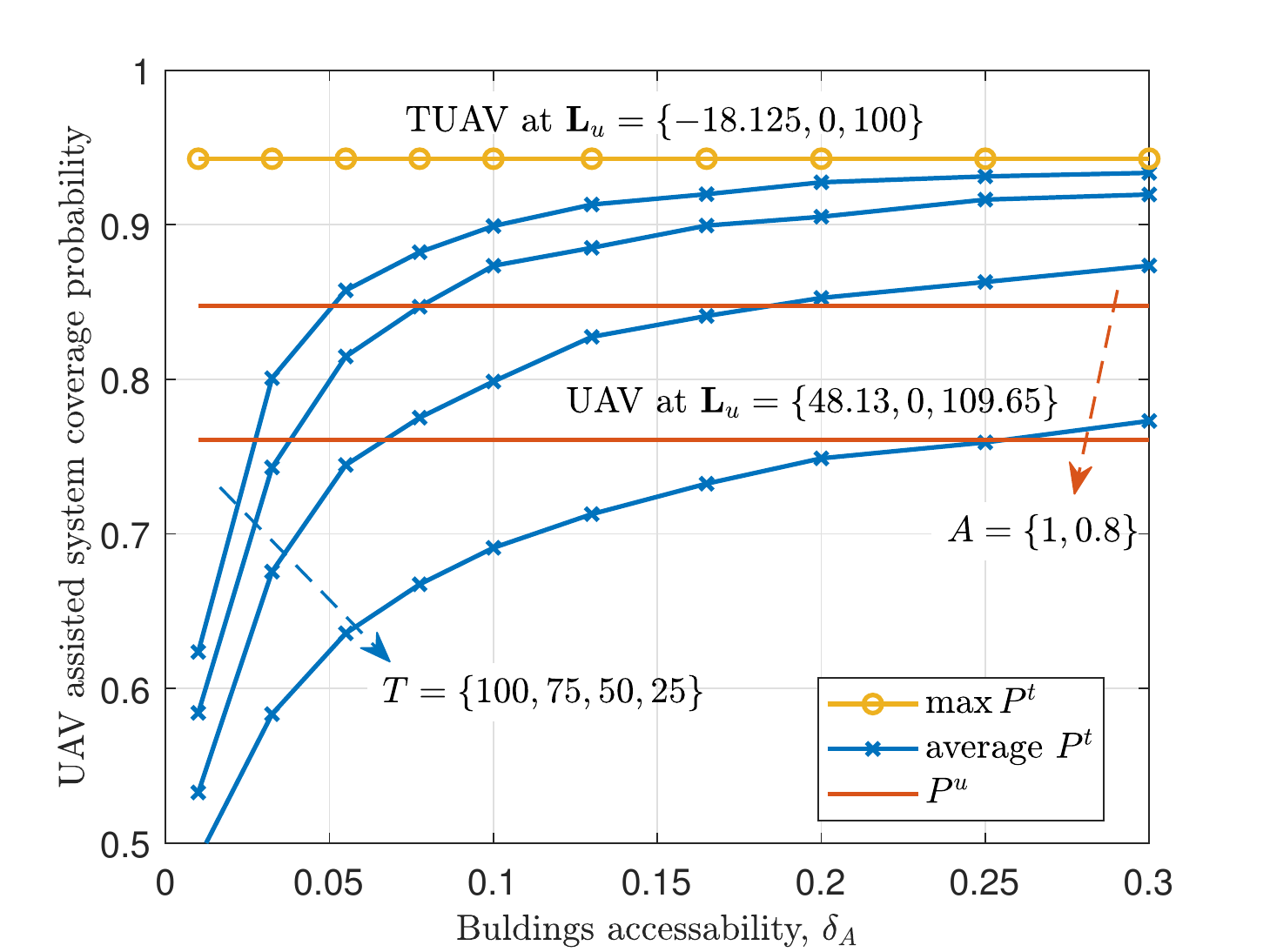}};	
			\end{tikzpicture}
			\caption{Dense urban environment}
			\label{fig:dense}
		\end{subfigure}
		\begin{subfigure}{.5\linewidth}
			\centering
			\begin{tikzpicture}[thick,scale=1, every node/.style={scale=1}]
			\draw (0, 0) node[inner sep=0] {	\includegraphics[clip, width=1 \linewidth]{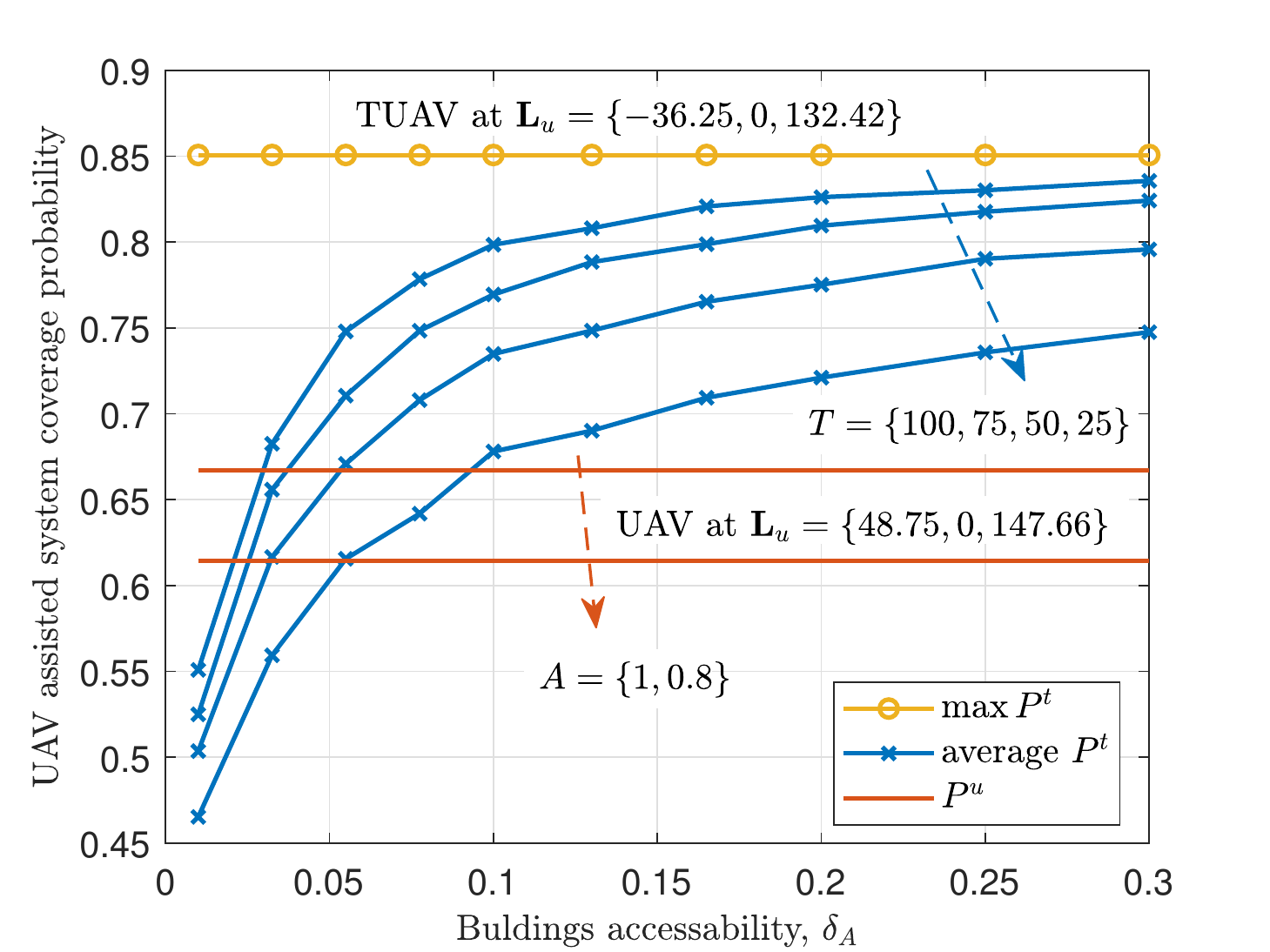}};	
			\end{tikzpicture}
			\caption{High-rise urban environment}
			\label{fig:highrise}
		\end{subfigure}
		\caption{The U-UAV and T-UAV-assisted systems performances against the percentage of accessible buildings. } 
		\label{fig:random_tether_base}
	\end{figure}

	Accordingly, we next compare the U-UAV and T-UAV coverage performance under different GS locations and availability scenarios. Based on the model developed by ITU, the average number of buildings per km$^2$ is given as $\gamma_3$ and the height of each building follows the Rayleigh distribution with the PDF expressed as \cite{Akram2014Dec2},
	\begin{align}
	f_{H_n}(h_n) = \dfrac{h_n}{\gamma_1^2} \exp\left(\dfrac{h_n^2}{2\gamma_1^2}\right) ,
	\end{align}
	where $\gamma_1$ is the Rayleigh distribution parameter. For dense urban environment, $\gamma_1=20$ and $\gamma_3 = 300$ while for high-rise urban environment $\gamma_1=50$ and $\gamma_3 = 300$. Given the tether length and the percentage of accessible GSs, $\delta_A$, the average system coverage probabilities are shown in Fig. \ref{fig:dense} and Fig. \ref{fig:highrise} for the dense and the high-rise urban environments, respectively. The average coverage probabilities are obtained by running a Monte Carlo simulation where the location and height of the GSs are random at each iteration. For the high-rise urban environment, we set the TBS height to $30$ m and approximate the environment parameters to $a_r=22$, $b_r=0.18$, $a_b=11$ and $b_b= 0.16$. The optimal T-UAV location is determined by using the simulated annealing search algorithm over the area described in Theorem \ref{lemma:optimal_region}. The optimal U-UAV location for the dense and high-rise urban environment scenarios are obtained as $\textbf{L}_u^* = \{48.13, 0, 109.65\}$ and $\textbf{L}_u^* = \{48.75, 0, 147.66\}$, respectively.  Fig. \ref{fig:random_tether_base} compares the U-UAV with the T-UAV for parameters $A \in \{0.8, 1\}$,  $T \in \{25,50,75,100\}$ m, and $\delta_A   \in [0,0.3]$. Thanks to increasing freedom of mobility, $P^t$ significantly improves with higher tether length and GS accessibility. Notice that increasing $T$ and $\delta_A$ eventually converges to the optimal case  (a freely moving T-UAV deployed at the optimal hovering location) where we achieve maximum $P^t$. 
	
	We also note that even for relatively high building accessibility, the coverage probability saturates at low values when the tether length is 25 and 50. This is because the average buildings height is $20$ m for dense urban environment and therefore only $1.11\%$ of the buildings are statistically higher than $60$ m. As a result, short tethers will prevent the T-UAV from reaching the optimal heights. Given a building accessibility of $\delta_A \geq 0.25$ and tether length of $100$ m, the coverage performance of the T-UAV-assisted system is very close to the maximum achievable coverage probability. Therefore, tether length of $100$ m is in general long enough to achieve near optimal coverage probability. Both the U-UAV and the T-UAV-assisted systems coverage probabilities are degraded for the high-rise urban environment as compared to the dense urban environment. We also note that the T-UAV system performs much better than the U-UAV system for the high-rise urban environment, while the systems performance is comparable for the dense urban environment. This is due to the lower LoS probability in the high-rise urban environment and because the U-UAV has to establish two A2G/G2A links to connect the RUE to the core network while the T-UAV only establish one link toward the RUE. 
	
\begin{figure}
	\centering
	\begin{tikzpicture}[thick,scale=.5, every node/.style={scale=.5}]
	\draw (0, 0) node[inner sep=0] {	\includegraphics[clip, width=1 \linewidth]{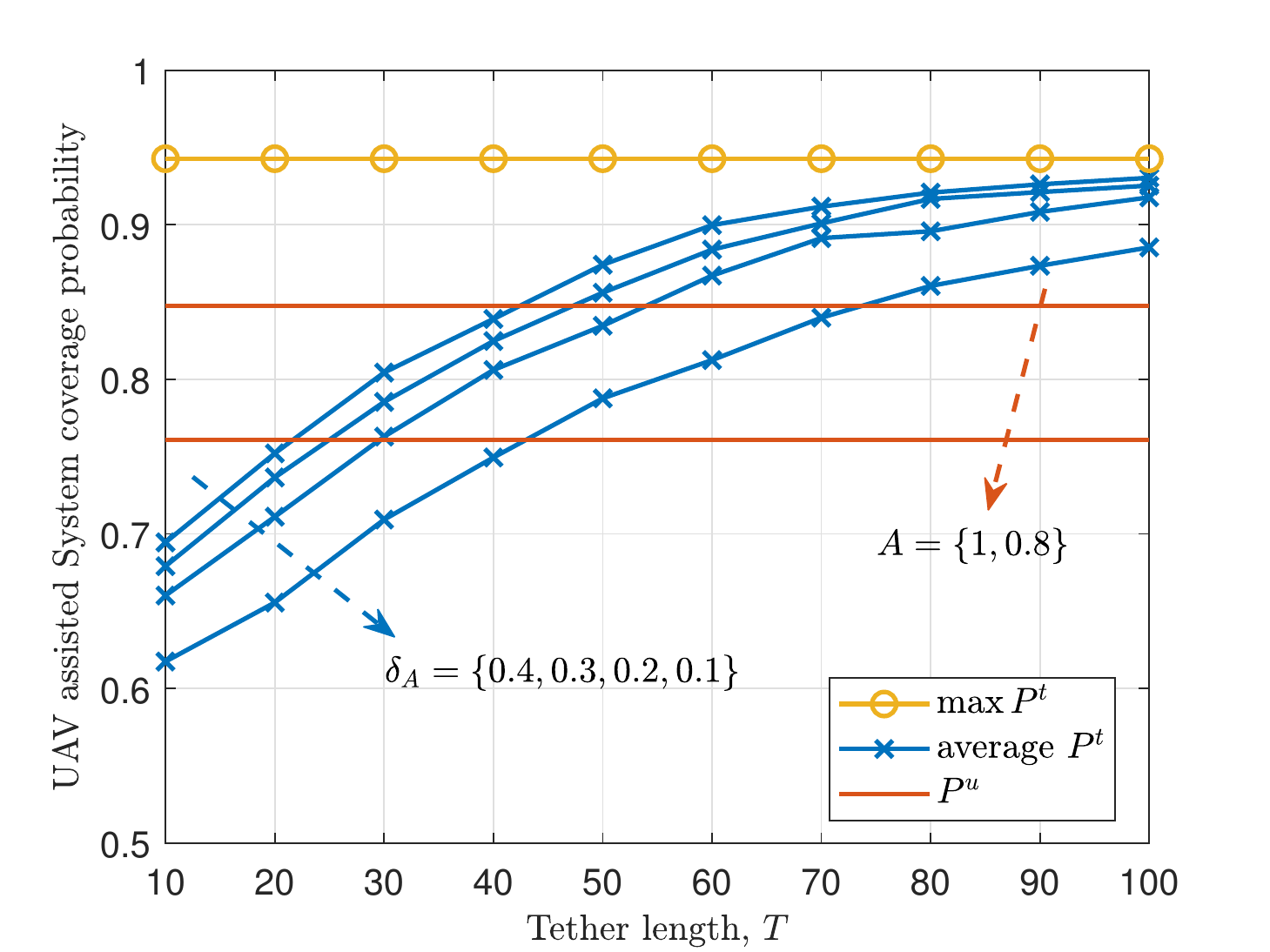}};
	\end{tikzpicture}
	\caption{$P^u$ and $P^t$ against maximum tether length, $T$. }
	\label{fig:max_tether_length_e}
\end{figure}

	In Fig. \ref{fig:max_tether_length_e}, T-UAV coverage performance is shown with respect to increasing tether length $T$, U-UAV availability $A$, and building accessibility $\delta_A$. It is clear that T-UAV cannot always outperform U-UAV if $T$ and $\delta_A$ is below a threshold. For $\delta_A=0.1$, $T>45$ and $T>75$ are required to outperform a U-UAV which is available $80\%$ and $100\%$ of the time, respectively. For $\delta_A=0.4$, $T>20$ and $T>40$ are required to outperform a U-UAV which is available $80\%$ and $100\%$ of the time, respectively. Indeed, this figure clearly and fairly compares the impact of UAV hardware specifications on the overall system performance.   

\begin{figure}
	\centering
	\begin{tikzpicture}[thick,scale=.5, every node/.style={scale=.5}]
	\draw (0, 0) node[inner sep=0] {	\includegraphics[clip, width=1 \linewidth]{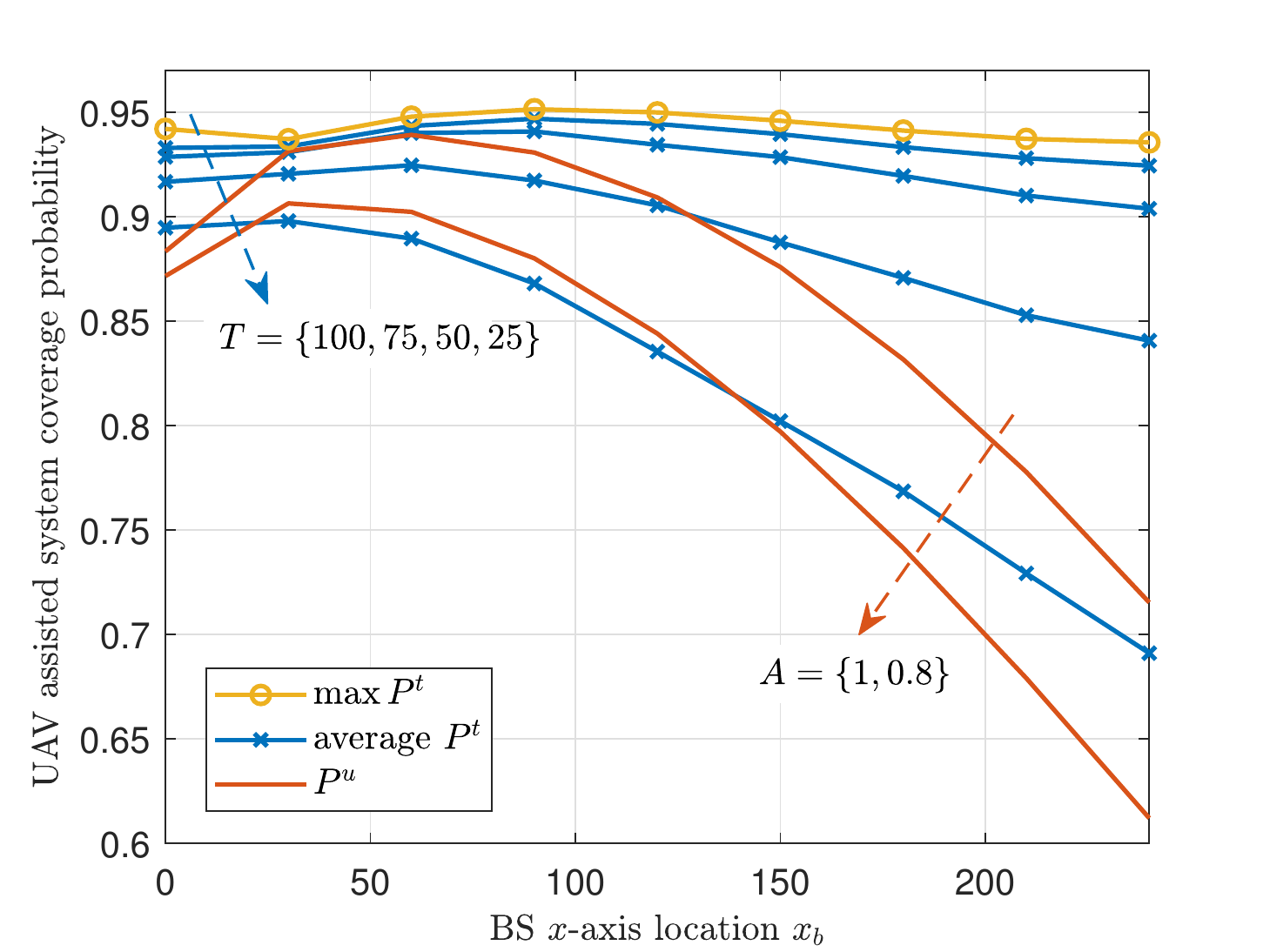}};	
	\end{tikzpicture}
	\caption{$P^t$ and $P^u$ for varying TBS distances from the hot-spot center.}
	\label{fig:bs_x}
\end{figure}

	In Fig. \ref{fig:bs_x}, $P^t$ and $P^u$ are shown with respect to the TBS $x$-axis location where we consider a  dense urban environment with random GS locations and an accessibility factor $\delta_A=0.3$. The coverage probability for the U-UAV and the T-UAV-assisted systems are high and comparable when the TBS is near to hot-spot center. This is because the TBS has a good coverage over the hot-spot and the U-UAV can cover close to the hot-spot center with desirable backhaul link conditions. As the distance between the TBS and the hot-spot center increases, the T-UAV-assisted system starts significantly outperforming the U-UAV-assisted system. Interestingly, the coverage probability is not maximum when the TBS is at the hot-spot center. Optimally, the TBS location is at one side of the hot-spot to serve the nearby users while the users on the other side are served by the UAV.

\begin{figure}
			\centering
			\begin{tikzpicture}[thick,scale=.5, every node/.style={scale=.5}]
			\draw (0, 0) node[inner sep=0] {	\includegraphics[clip, width=1 \linewidth]{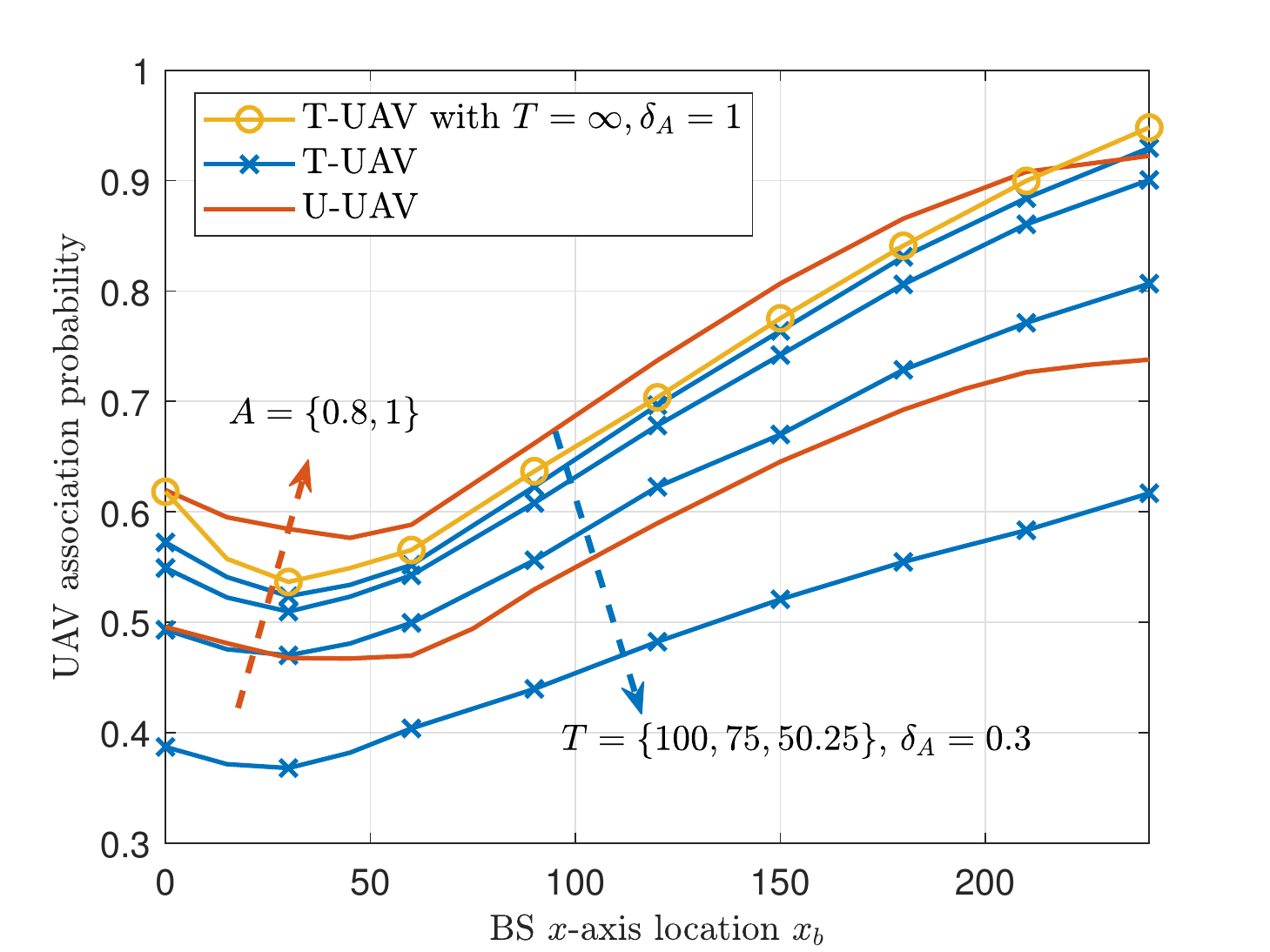}};
			\end{tikzpicture}
			\caption{T-UAV and U-UAV association probabilities for varying TBS distances from the hot-spot center. T-UAV and U-UAV are placed such that $P^t$ and $P^u$ are maximized.}
			\label{fig:association_probability_e}
		\end{figure}

		Traffic offloading can be assessed by evaluating the RUE association probability with the U-UAV and T-UAV. The association probability depends on the UAV hovering location and the U-UAV availability. In Fig. \ref{fig:association_probability_e}, we show the association probabilities for different TBS locations over the $x$-axis such that the UAV is deployed to maximize coverage probability. From the figure, we note that, in general, the UAV association probability increases as the T-UAV mobility is less restricted, the U-UAV availability is increased, and the TBS is farther away from the hot-spot center. Since the aerial access link is generally stronger than the terrestrial link, the U-UAV with $100$\% availability has a higher association probability than the T-UAV, as the backhaul link quality restricts the U-UAV distance from the TBS. For practical T-UAV tether length and U-UAV availability duty cycle, the T-UAV association probability is higher than the U-UAV association probability.

\section{Conclusions}
\label{sec:conc}
In this paper, we provide a comparative performance analysis of U-UAV and T-UAV-assisted cellular traffic offloading from a geographical region that experiences heavy traffic conditions. To achieve this, we exploit stochastic geometry tools and derive joint distance distributions between users, the terrestrial base station (TBS), and UAV. To maximize the end-to-end signal-to-noise ratio, a user association policy is proposed and the corresponding association regions are analytically identified. Thereafter, the overall coverage probability of U-UAV/T-UAV-assisted systems is obtained for given locations of the TBS and the U-UAV/T-UAV. Furthermore, the set of optimal UAV locations is shown to belong to the surface of the spherical cone centered at the GS. Extensive simulation results are presented to validate analytical results and compare the performance of U-UAV and T-UAV-assisted systems. Numerical results show that T-UAV outperforms U-UAV given a sufficient number of GS locations accessibility and long enough tether are provided.

\appendices
\section{Lemma \ref{lemma:f_D} Proof} \label{app:f_D}

	Unlike the PDF derivations in \cite{Afshang_2017,Dhillon}, we consider a more general case where $\textbf{L}_i$ can be inside or outside $\mathcal{D}(\textbf{L}_o,R_o)$. The cumulative distribution function (CDF) of the distance between a point $\textbf{L}_i'$ and a uniformly distributed RUE location within $\mathcal{D}(\textbf{L}_o,R_o)$ is given by
	\begin{align}
	F_{D_{i,j}'} \left(r_i \right) = \mathbb{P}\left( D_{i,j}' \leq r_i \right) = \dfrac{ \left| \mathcal{D} \left( \textbf{L}_i',r_i \right) \cap \mathcal{D}(\textbf{L}_o,R_o) \right| } { |\mathcal{D}(\textbf{L}_o,R_o)| }.
	\end{align}
	In order to find $ \left| \mathcal{D} \left( \textbf{L}_i',r_i \right) \cap \mathcal{D}(\textbf{L}_o,R_o) \right| $, we consider the following two cases:

	\begin{figure}
		\centering
		\begin{subfigure}{.3\linewidth}
			\centering
			\begin{tikzpicture}[thick,scale=.8, every node/.style={scale=.8}]
			\draw (0, 0) node[inner sep=0] {	\includegraphics[trim={7cm 5cm  14cm 5cm},clip, width=1.3 \linewidth]{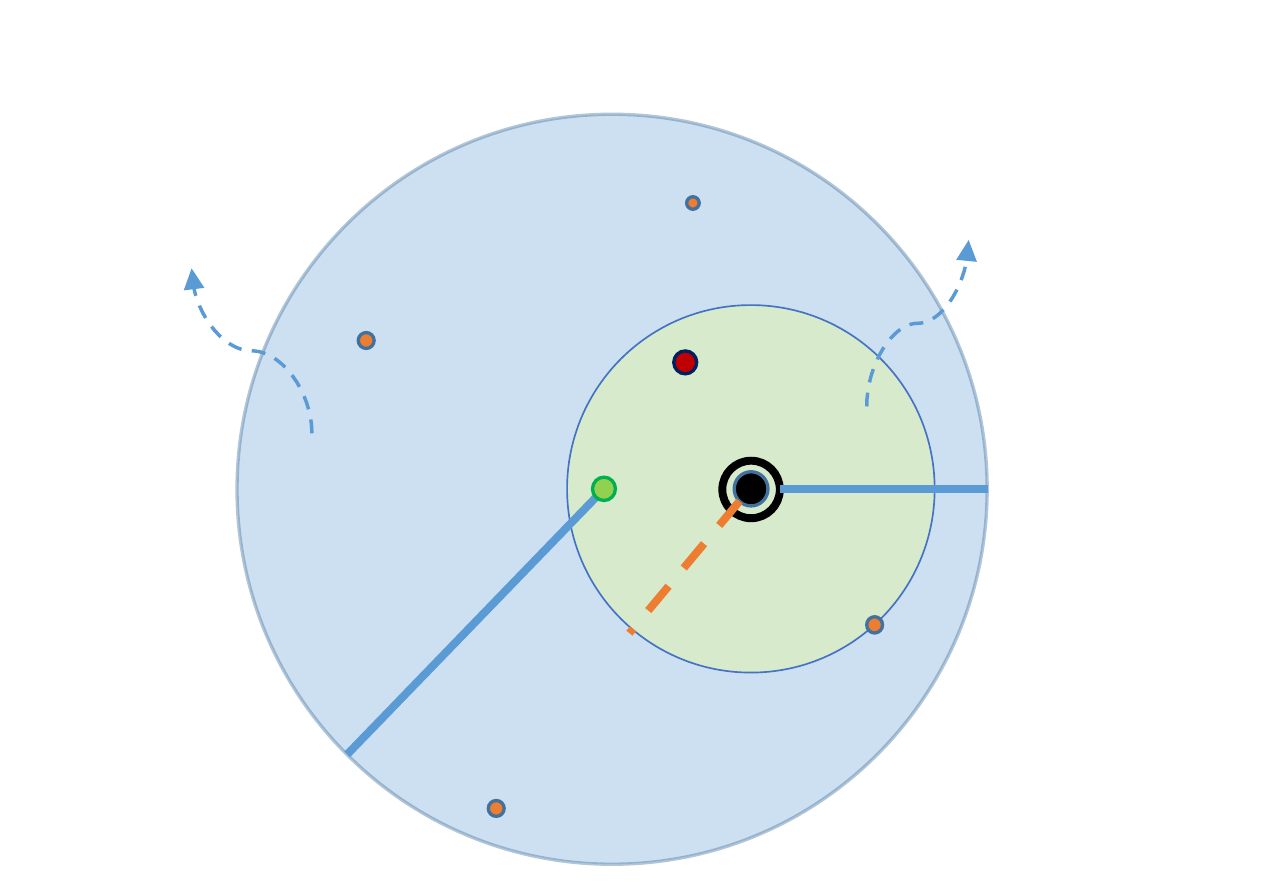}};	
			\draw (2.0, 1.5) node {\small{$\mathcal{D}(\textbf{L}_i',r_i)$}};	
			\draw (-2.5, 1.5) node {\small{$\mathcal{D}(\textbf{L}_o,R_o)$}};	
			\draw (-1.0, -0.6) node {\small{$R_o$}};
			\draw (0.1, -0.5) node {\small{$r_i$}};
			\draw (1.5, -0.5) node {\small{$|R_o-D_{i,o}'|$}};
			\end{tikzpicture}
			\caption{} \label{fig:limma2a}
		\end{subfigure}
		\begin{subfigure}{.3\linewidth}
			\centering
			\begin{tikzpicture}[thick,scale=.8, every node/.style={scale=.8}]
			\draw (0, 0) node[inner sep=0] {	\includegraphics[trim={7cm 5cm  14cm 5cm},clip, width=1.3 \linewidth]{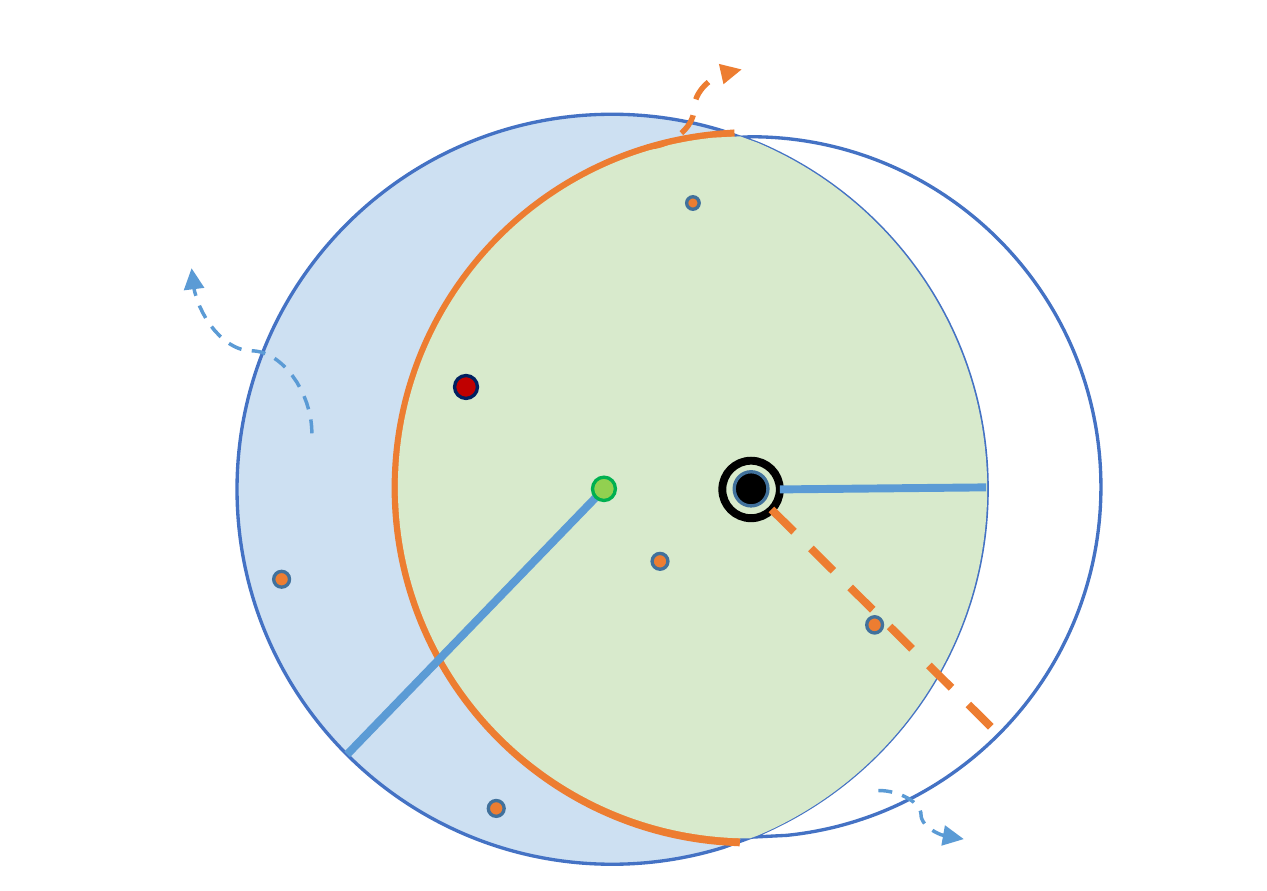}};	
			\draw (2.4, -2) node {\small{$\mathcal{D}(\textbf{L}_i',r_i)$}};	
			\draw (-2.3, 1.2) node {\small{$\mathcal{D}(\textbf{L}_o,R_o)$}};	
			\draw (-1.0, -0.6) node {\small{$R_o$}};
			\draw (1.4, -0.6) node {\small{$r_i$}};
			\draw (1.2, 0.1) node {\small{$|R_o-D_{i,o}'|$}};
			\draw (1.8, 1.9) node {\small{	$   \mathcal{A}(\textbf{L}_i',r_i, \textbf{L}_o,R_o) $}};
		
			\end{tikzpicture}
			\caption{ } \label{fig:limma2b}
		\end{subfigure}
		\begin{subfigure}{.30\linewidth}
			\centering
			\begin{tikzpicture}[thick,scale=.8, every node/.style={scale=.8}]
			\draw (0, 0) node[inner sep=0] {	\includegraphics[trim={7cm 5cm  14cm 5cm},clip, width=1.3 \linewidth]{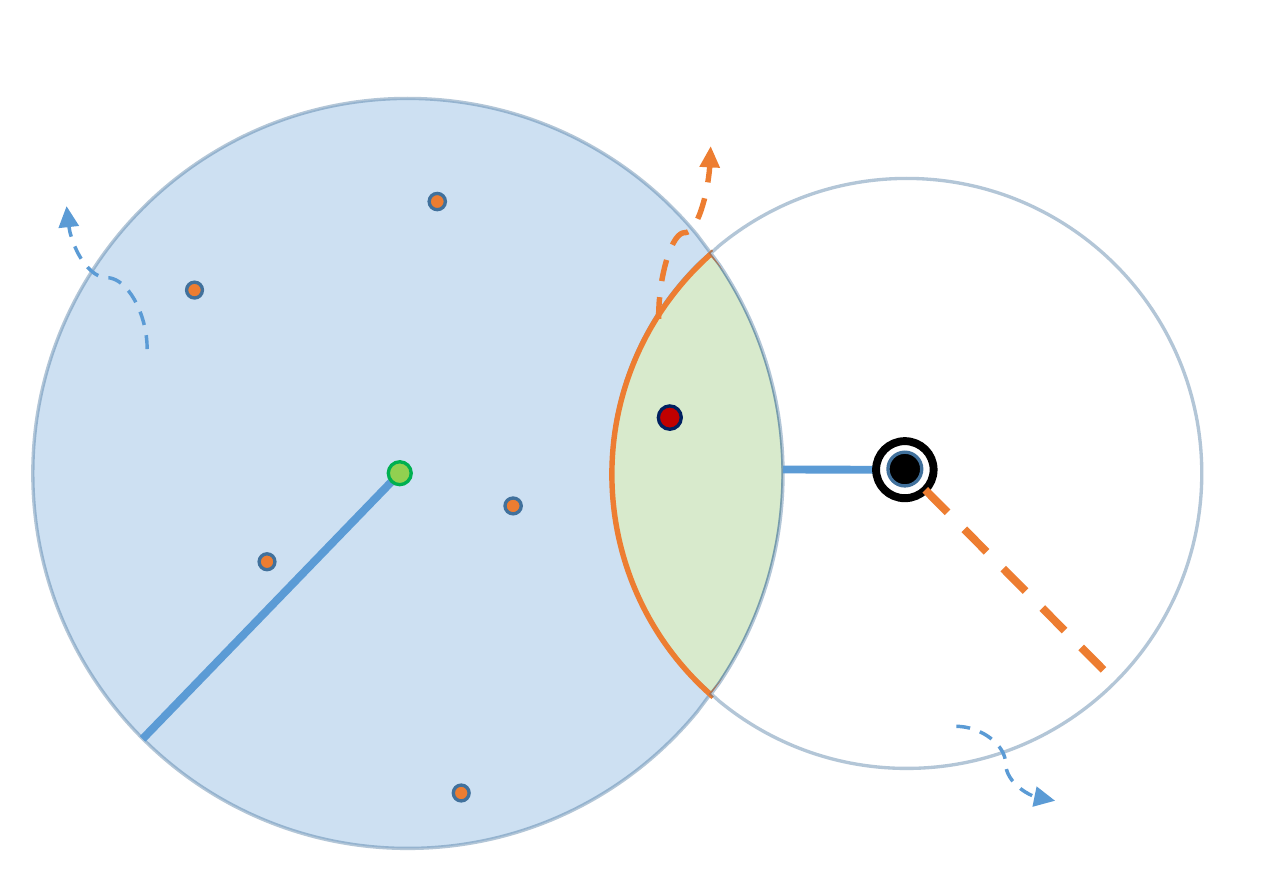}};	
			\draw (2.9, -2) node {\small{$\mathcal{D}(\textbf{L}_i',r_i)$}};	
			\draw (-3.2, 1.4) node {\small{$\mathcal{D}(\textbf{L}_o,R_o)$}};	
			\draw (-2.2, -0.5) node {\small{$R_o$}};
			\draw (2.4, -0.6) node {\small{$r_i$}};
			\draw (1.3, 0.3) node {\small{$|R_o-D_{i,o}'|$}};
			\draw (1.4, 1.7) node {\small{	$  \mathcal{A}(\textbf{L}_i',r_i, \textbf{L}_o,R_o) $}};
			\end{tikzpicture}
			\caption{ } \label{fig:limma2c}
		\end{subfigure}
		\caption{ Different cases for Lemma 2: a) $\mathcal{D} \left( \textbf{L}_i',r_i \right) \cap \mathcal{D}(\textbf{L}_o,R_o) \equiv \mathcal{D} \left( \textbf{L}_i',r_i \right)$, b) $ \left\{\mathcal{D} \left( \textbf{L}_i',r_i \right) \cap \mathcal{D}(\textbf{L}_o,R_o) \right\}   \subset \mathcal{D} \left( \textbf{L}_i',r_i \right)$ and $\textbf{L}_i' \in  \mathcal{D}(\textbf{L}_o,R_o)$, and c) $ \left\{\mathcal{D} \left( \textbf{L}_i',r_i \right) \cap \mathcal{D}(\textbf{L}_o,R_o) \right\}   \subset \mathcal{D} \left( \textbf{L}_i',r_i \right)$ and $\textbf{L}_i' \notin \mathcal{D}(\textbf{L}_o,R_o)$ }  \label{fig:limma2}
	\end{figure}

	\begin{enumerate}
		\item $\mathcal{D} \left( \textbf{L}_i',r_i \right) \cap \mathcal{D}(\textbf{L}_o,R_o) \equiv \mathcal{D} \left( \textbf{L}_i',r_i \right)$: In this case, $\mathcal{D} \left( \textbf{L}_i',r_i \right)$ is completely inside $\mathcal{D}(\textbf{L}_o,R_o)$ such that $0 \leq r_i \leq \max(0,R_o-D_{i,o}')$ where $D_{i,o}'=\|\textbf{L}_i'\|$. This case is illustrated in Fig. \ref{fig:limma2a} where the intersection region is highlighted by green color. Accordingly, the CDF of  this case is given by the area ratio of the disks, i.e.,
		\begin{align}\label{eq:F_R_r1}
		F_{D_{i,j}'} \left(r_i \right) = \dfrac{\pi r_i^2}{\pi R_o^2}.
		\end{align}
		
		\item $ \left \{\mathcal{D} \left( \textbf{L}_i',r_i \right) \cap \mathcal{D}(\textbf{L}_o,R_o) \right\}   \subset \mathcal{D} \left( \textbf{L}_i',r_i \right)$: In this case,  a part of $\mathcal{D} \left( \textbf{L}_i',r_i \right)$ is outside of  $\mathcal{D}(\textbf{L}_o,R_o)$ such that $|R_o-D_{i,o}'| \leq r_i \leq R_o+D_{i,o}'$. The intersection regions of this case is illustrated  in Fig. \ref{fig:limma2b} and \ref{fig:limma2c} for situations where $\textbf{L}_i' \in \mathcal{D}(\textbf{L}_o,R_o)$ and $\textbf{L}_i' \notin \mathcal{D}(\textbf{L}_o,R_o)$, respectively. Accordingly, the CDF of this case is given by
		\begin{align}\label{eq:F_R_r2}
		F_{D_{i,j}'} \left(r_i \right) = \dfrac{\max(0,R_o-D_{i,o}')^2}{ R_o^2} + \int_{|R_o-D_{i,o}'|}^{ r_i} \dfrac{ |  \mathcal{A}(\textbf{L}_i',r_i, \textbf{L}_o,R_o) | }{\pi R_o^2} dr_i,
		\end{align}
		where $ |  \mathcal{A}(\textbf{L}_i',r_i, \textbf{L}_o,R_o) |$ is the arc length as shown in Fig. \ref{fig:limma2}. 
		
		The arc length can be derived as follows: Let us consider two generic intersecting circles $\mathcal{C}(\textbf{L}_i',R_i')$ and  $\mathcal{C}(\textbf{L}_j',R_j')$. Because $|  \mathcal{A}(\textbf{L}_i',r_i, \textbf{L}_j',R_j')| $ is independent from the circles' absolute locations given a fixed distance $D_{i,j}'$ from their centers, we assume $\textbf{L}_i'=\{D_{i,j}',0,0\}$ and $\textbf{L}_j'=\{0,0,0\}$. Following from the mathematical definition of a circle, these circles intersect at,
		${\bf{\check{L}_{ij}}}=\{\check{x}_{ij},\check{y}_{ij},0\}$ and ${\bf{\hat{L}_{ij}}}=\{\hat{x}_{ij},\hat{y}_{ij},0\}$ where,
		\begin{align}
		\check{x}_{ij} =\check{x}_{ij} &= \dfrac{(R_i')^2  - (R_j')^2 +(D_{i,j}')^2}{2D_{i,j}'}, \\
		\check{y}_{ij} =- \hat{y}_{ij} &= \sqrt{ (R_i')^2 - (\check{x}_{ij})^2}. 
		\end{align}
		The angle at $\textbf{L}_i'$ enclosed by the lines $\mathcal{L}(\textbf{L}_i',\textbf{L}_j')$ on one side and $\mathcal{L}(\textbf{L}_i',{\bf{\check{L}_{ij}}})$ or $\mathcal{L}(\textbf{L}_i',{\bf{\hat{L}_{ij}}})$ on the other side is expressed as,
		\begin{align}\label{eq:phi}
		\phi_j = \arccos \left( \dfrac{\check{x}_{ij} }{R_j'} \right) =\arccos \left( \dfrac{R_j'^2+(D_{i,j}')^2-D_{i,j}'^2}{2D_{i,j}'R_j'} \right).
		\end{align}
		By use of \eqref{eq:phi}, the arc length $|  \mathcal{A}(\textbf{L}_i',r_i, \textbf{L}_j',R_j'|)$ is given as, 
		\begin{align}\label{eq:arc_length}
		|  \mathcal{A}(\textbf{L}_i',R_j', \textbf{L}_j',R_j')| = 2\phi_i R_j' = 2 R_j' \arccos\left( \dfrac{(D_{i,j}')^2 +(R_i')^2-(R_j')^2}{2 D_{i,j}' R_j'} \right) .
		\end{align}
		Therefore, the arc length $ |  \mathcal{A}(\textbf{L}_i',r_i, \textbf{L}_o,R_o)| $ in \eqref{eq:F_R_r2} is expressed as, 
		\begin{align}
		|  \mathcal{A}(\textbf{L}_i',r_i, \textbf{L}_o,R_o)| = 2\phi_i r_i = 2 r_i \arccos\left( \dfrac{(D_{i,o}')^2 +r_i^2-R_o^2}{2 D_{i,o}' r_i} \right) .
		\end{align}	 
	\end{enumerate}
	By taking the derivate of \eqref{eq:F_R_r1} and \eqref{eq:F_R_r2} w.r.t. $r_i$, the PDFs can be derived from the CDFs.

\section{Lemma \ref{lemma:f_Du|Db_general} Proof}
\label{app:f_Du|Db}

	Denoting the conditional distance between the RUE and  TBS  by $r_b=D_{b,r}'$, we consider two cases as depicted  in Fig. \ref{fig:limma3}:
	\begin{figure}
		\begin{subfigure}{.5\linewidth}
			\centering
			\begin{tikzpicture}[thick,scale=.8, every node/.style={scale=.8}]
			\draw (0, 0) node[inner sep=0] {	\includegraphics[trim={7cm 5cm  16cm 3cm},clip, width=1 \linewidth]{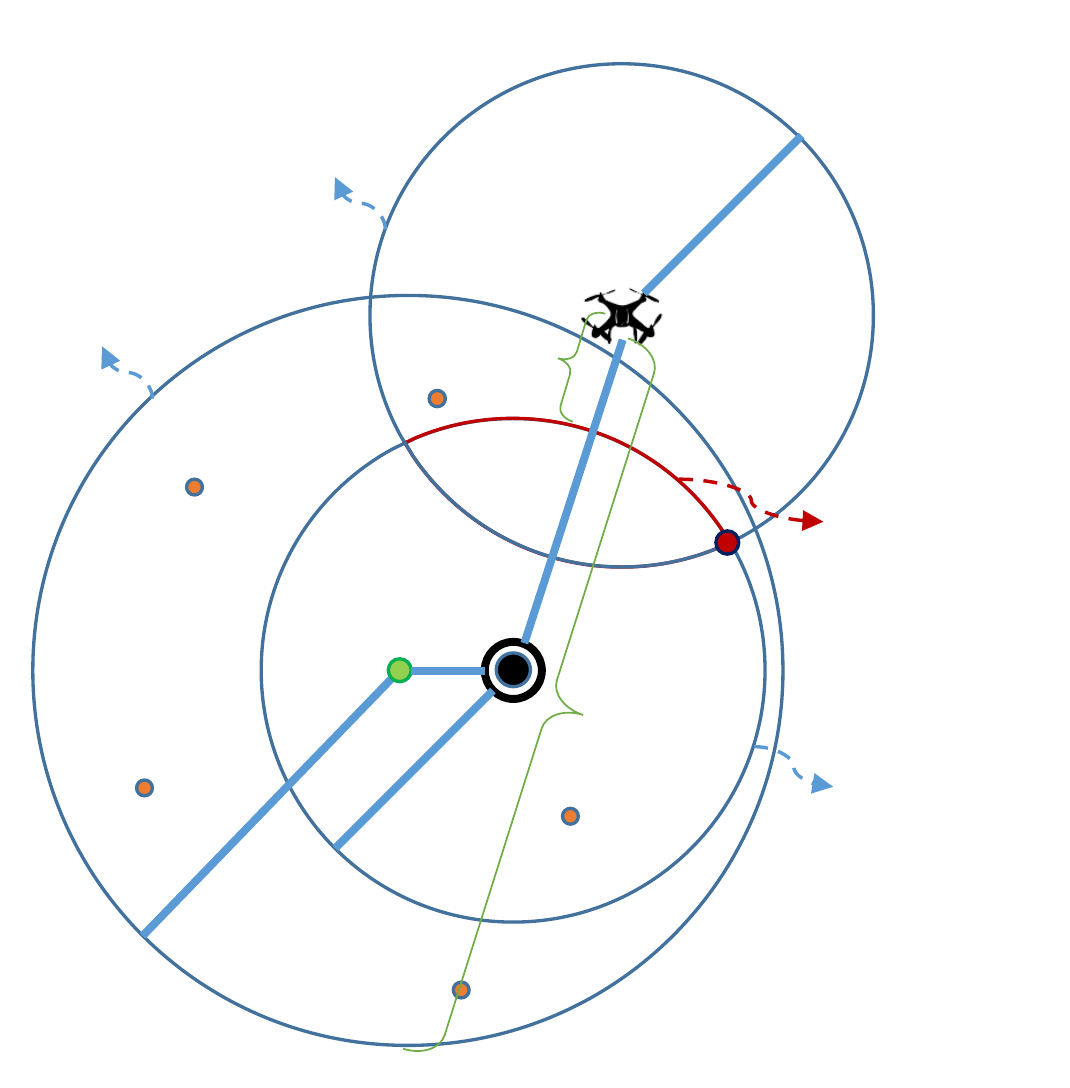}};	
			\draw (1.2, 2.7) node {\small{$r_u$}};	
			\draw (-1.8, -1.4) node {\small{$R_o$}};
			\draw (-0.9, -1.4) node {\small{$r_b$}};
			\draw (-.5, -0.3) node {\small{$D_{b,u}'$}};
			\draw (-.7, 1.4) node {\small{$| D_{b,u}' - r_b |$}};
			\draw (1.0, -1.3) node {\small{$ D_{b,u}' + r_b$}};
			\draw (-1.6, 3.1) node {\small{$\mathcal{C}(\textbf{L}_u',r_u)$}};
			\draw (-3.3, 1.8) node {\small{$\mathcal{C}(\textbf{L}_o,R_o)$}};
			\draw (3.1, -1.9) node {\small{$\mathcal{C}(\textbf{L}_b',r_b)$}};
			\draw (2.9, 0.4) node {\small{$\mathcal{A}(\textbf{L}_b',r_b,\textbf{L}_u',r_u)$}};
			\end{tikzpicture}
			\caption{ }   \label{fig:limma3a}
		\end{subfigure}
		\begin{subfigure}{.5\linewidth}
			\centering
			\begin{tikzpicture}[thick,scale=.8, every node/.style={scale=.8}]
			\draw (0, 0) node[inner sep=0] {	\includegraphics[trim={7cm 5cm  16cm 3cm},clip, width=1 \linewidth]{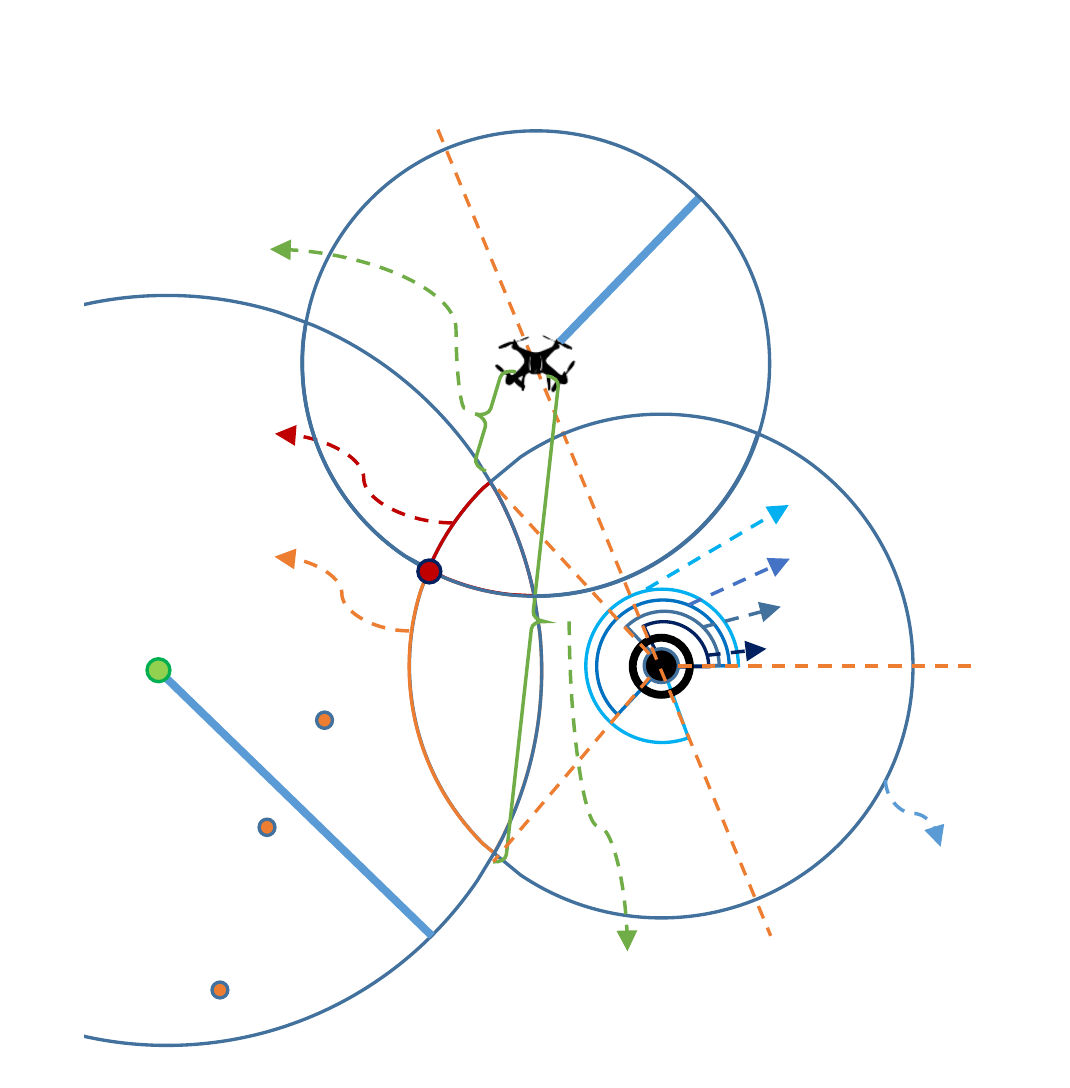}};	
			\draw (0.5, 2.3) node {\small{$r_u$}};
			\draw (-1.1, -2.3) node {\small{$R_o$}};
			\draw (-2.7, 2.3) node {\small{$\| {\bf L}_u' - {\bf \check{L} }_b \|$}};
			\draw (0.5, -3.3) node {\small{$\| {\bf L}_u' - {\bf \hat{L} }_b \|$}};
			\draw (3.1, -2.5) node {\small{$\mathcal{C}(\textbf{L}_b',r_b)$}};
			\draw (-2.7, 0.8) node {\small{$\mathcal{A}_{\rm{int}}(r_u)$}};	
			\draw (-2.9, 0.2) node {\small{$\mathcal{A}(\textbf{L}_b',r_b,\textbf{L}_o,R_o)$}};	
			\draw (2.0, -0.9) node {\small{$ \check{\theta}_u$}};
			\draw (2.1, -0.5) node {\small{$ \check{\theta}_b$}};
			\draw (2.2, -0.1) node {\small{$ \hat{\theta}_b$}};
			\draw (2.2, +0.3) node {\small{$ \hat{\theta}_u$}};		
			\end{tikzpicture}
			\caption{  }   \label{fig:limma3b}
		\end{subfigure}
		\caption{ Illustration of cases in Lemma \ref{lemma:f_Du|Db_general}: a) $\mathcal{D} \left( \textbf{L}_b',r_b \right) \cap \mathcal{D}(\textbf{L}_o,R_o) \equiv \mathcal{D} \left( \textbf{L}_b',r_b \right)$, and b)  $ \left \{\mathcal{D} \left( \textbf{L}_b',r_b \right) \cap \mathcal{D}(\textbf{L}_o,R_o) \right\}   \subset \mathcal{D} \left( \textbf{L}_b',r_b \right)$. 
		}  \label{fig:limma3}
	\end{figure}

\begin{enumerate}
	\item $r_b \leq \max(0,R_o-D_{b,o}')$: In this case, $\mathcal{D}(\textbf{L}_b',r_b)$ is completely inside $\mathcal{D}(\textbf{L}_o,R_o)$ as shown in Fig. \ref{fig:limma3a}. Therefore, the RUE location $\textbf{L}_r$ is uniformly distributed over $\mathcal{C} (\textbf{L}_b', r_b)$. The distance between the RUE and the UAV is bounded by $r_u \in \left[ | D_{b,u}' - r_b |, D_{b,u}' + r_b \right]$. By defining the circle $\mathcal{C}(\textbf{L}_u',r_u)$ with $r_u \in \left[ | D_{b,u}' - r_b |, D_{b,u}' + r_b \right]$, the conditional CDF of the distance $D_{u,r}'$ given $D_{b,r}'$ is expressed as,
	\begin{align}
	F_{D_{u,r}'|D_{b,r}'}(r_u|r_b) &= \mathbb{P}(D_{u,r}' <r_u|r_b) = \dfrac{ |\mathcal{A}(\textbf{L}_b',r_b,\textbf{L}_u',r_u)|} {|\mathcal{C}(\textbf{L}_b',r_b) | } \\
	& =\dfrac{1}{\pi} \arccos \left( \dfrac{(D_{b,u}')^2 + r_b^2 - r_u^2}{2r_b D_{b,u}'} \right).
	\end{align}
	By taking the derivative w.r.t. $r_u$, the conditional PDF is obtained as,
	\begin{align}
	f_{D_{u,r}'|D_{b,r}'}(r_u|r_b) = \dfrac{ r_u}{D_{b,u}'\pi r_b} \dfrac{1}{ \sqrt{ 1 - \left(  \dfrac{(D_{b,u}')^2 +r_b^2 - r_u^2}{2D_{b,u}'r_b }  \right)^2 } }.
	\end{align}
	\item $r_b \in \left[ |R_o-D_{b,o}'| , R_o +D_{b,o} \right]$: In this case, only an arc, $ \mathcal{A}(\textbf{L}_b',r_b,\textbf{L}_o,r_o ) \subseteq \mathcal{C}(\textbf{L}_b',r_b) $, is inside $\mathcal{D}(\textbf{L}_o,r_o )$ as shown in Fig. \ref{fig:limma3}b. Therefore, $\textbf{L}_r$ is uniformly distributed over $\mathcal{A}(\textbf{L}_b',r_b,\textbf{L}_o,r_o )$. The conditional CDF $F_{D_{u,r}'|D_{b,r}'}(r_u|r_b) $ is given as,
	\begin{align}\label{eq:F_r_u|r_b,case2}
	F_{D_{u,r}'|D_{b,r}'}(r_u|r_b) &= \mathbb{P}(D_{u,r}' <r_u|r_b) = \dfrac{  |\mathcal{A}_{\rm{int}}(r_u)|} {| \mathcal{A}(\textbf{L}_b',r_b,\textbf{L}_o,R_o | },
	\end{align}	  
	
	where $ |\mathcal{A}_{\rm{int}}(r_u)|= |\mathcal{A}(\textbf{L}_b',r_b,\textbf{L}_u',r_u) \cap \mathcal{A}(\textbf{L}_b',r_b,\textbf{L}_o,R_o)  | $. To find $ |\mathcal{A}_{\rm{int}}(r_u)|$, we first define the angles
	$\check{\theta}_b = \angle ({\textbf{L}}_x^+,\textbf{L}_b',{\bf \check{L} }_b) $, $\hat{\theta}_b= \angle ({\textbf{L}}_x^+,\textbf{L}_b',{\bf \hat{L} }_b)$ with ${\bf \check{L} }_b$ and ${\bf \hat{L} }_b$ being the points of intersection between $\mathcal{C}(\textbf{L}_o,R_o)$ and $\mathcal{C}(\textbf{L}_b',r_b)$, and the angles $\check{\theta}_u = \angle ({\textbf{L}}_x^+,\textbf{L}_b',{\bf{L} }_u') $ and $\hat{\theta}_u=( \pi + \check{\theta}_u) \mod 2\pi$ as shown in Fig. \ref{fig:limma3}b. Now we consider the following three cases for $r_u$,

	\begin{itemize}
		\item[a)] $ r_u \in \left[ | D_{b,u}' - r_b | ,  \| {\bf L}_u' - {\bf \check{L} }_b \| \right] $: if $\check{\theta}_b \leq \check{\theta}_u \leq \hat{\theta}_b$, then, $\mathcal{A}(\textbf{L}_b',r_b,\textbf{L}_u',r_u) $ and $ \mathcal{A}(\textbf{L}_b',r_b,\textbf{L}_o,R_o) $ completely intersect over $\mathcal{A}(\textbf{L}_b',r_b,\textbf{L}_u',r_u) $. Otherwise, $\mathcal{A}_{\rm{int}}(r_u)=\emptyset$. Hence, 
		\begin{align}\label{eq:A_int1}
		|\mathcal{A}_{\rm{int}}(r_u)| = |\mathcal{A}_{\rm{int}}^{(1)}(r_u)| = |\mathcal{A}(\textbf{L}_b',r_b,\textbf{L}_u',r_u) | \mathbbm{1}_{ \{ \check{\theta}_b \leq \check{\theta}_u \leq \hat{\theta}_b \} }
		\end{align}
		By substituting \eqref{eq:A_int1} in \eqref{eq:F_r_u|r_b,case2} and taking the derivative w.r.t. $r_u$, 	$f_{D_{u,r}'|D_{b,r}'}(r_u|r_b)$ is obtained as,
		\begin{align}
		f_{D_{u,r}'|D_{b,r}'}(r_u|r_b) = \dfrac{2 r_u}{D_{b,u}' | \mathcal{A}(\textbf{L}_b',r_b,\textbf{L}_o,R_o) |} \dfrac{\mathbbm{1}_{ \{ \check{\theta}_b \leq \check{\theta}_u \leq \hat{\theta}_b \} } }{ \sqrt{ 1 - \left(  \dfrac{(D_{b,u}')^2 +r_b^2 - r_u^2}{2D_{b,u}'r_b }  \right)^2 } }.
		\end{align} 
		
		\item[b)] $r_u \in \left[ \| {\bf L}_u' - {\bf \check{L} }_b \|,  \| {\bf L}_u' - {\bf \hat{L} }_b \| \right]$: The arc $\mathcal{A}(\textbf{L}_b',r_b,\textbf{L}_u',r_u)$ is symmetric around the line connecting $\textbf{L}_b'$ and $\textbf{L}_u'$ and can be split into two sides. When $r_u \in \left[ \| {\bf L}_u' - {\bf \check{L} }_b \|,  \| {\bf L}_u' - {\bf \hat{L} }_b \| \right]$, the arcs $\mathcal{A}(\textbf{L}_b',r_b,\textbf{L}_u',r_u)$ and $\mathcal{A}(\textbf{L}_b',r_b,\textbf{L}_o,R_o)$ intersect only from one side. Therefore, $|\mathcal{A}_{\rm{int}}(r_u)|$ is equal to $|\mathcal{A}_{\rm{int}}^{(1)}(\| {\bf L}_u' - {\bf \check{L} }_b \|)|$ plus half the difference between $|\mathcal{A}(\textbf{L}_b',r_b,\textbf{L}_u',r_u)|$ and $|\mathcal{A}(\textbf{L}_b',r_b,\textbf{L}_u', \| {\bf L}_u' - {\bf \check{L} }_b \|)|$. Hence, 
		\begin{align}
		|\mathcal{A}_{\rm{int}}(r_u)| &= |\mathcal{A}_{\rm{int}}^{(2)}(r_u)| =  |\mathcal{A}_{\rm{int}}^{(1)}(\| {\bf L}_u' - {\bf \check{L} }_b \|)|  \nonumber \\ \label{eq:A_int2}
		&+ \dfrac{1}{2} |\mathcal{A}(\textbf{L}_b',r_b,\textbf{L}_u',r_u) | - \dfrac{1}{2} |\mathcal{A}\left(\textbf{L}_b',r_b,\textbf{L}_u',\| {\bf L}_u' - {\bf \check{L} }_b \| \right) |.
		\end{align}
		By substituting \eqref{eq:A_int2} in \eqref{eq:F_r_u|r_b,case2} and taking the derivative w.r.t. $r_u$, 	$f_{D_{u,r}'|D_{b,r}'}(r_u|r_b)$ is obtained as,
		\begin{align}
		f_{D_{u,r}'|D_{b,r}'}(r_u|r_b) = \dfrac{r_u}{D_{b,u}' | \mathcal{A}(\textbf{L}_b',r_b,\textbf{L}_o,R_o) |} \dfrac{1 }{ \sqrt{ 1 - \left(  \dfrac{(D_{b,u}')^2 +r_b^2 - r_u^2}{2D_{b,u}'r_b }  \right)^2 } }.
		\end{align} 
		
		\item[c)] $r_u \in \left[ \| {\bf L}_u' - {\bf \hat{L} }_b \|, D_{b,u}' + r_b \right]$: when $r_u \in \left[ \| {\bf L}_u' - {\bf \hat{L} }_b \|, D_{b,u}' + r_b \right]$, the arcs $\mathcal{A}(\textbf{L}_b',r_b,\textbf{L}_u',r_u)$ and $\mathcal{A}(\textbf{L}_b',r_b,\textbf{L}_o,R_o)$ intersect only if $\{ \check{\theta}_b \leq \hat{\theta}_u \leq \hat{\theta}_b \}$. Therefore, 
		\begin{align}
		|\mathcal{A}_{\rm{int}}(r_u)| &= |\mathcal{A}_{\rm{int}}^{(3)}(r_u)| =  |\mathcal{A}_{\rm{int}}^{(2)}(\| {\bf L}_u' - {\bf \hat{L}}_b \|)|  \nonumber \\ \label{eq:A_int3}
		&+ \left( |\mathcal{A}(\textbf{L}_b',r_b,\textbf{L}_u',r_u) | - |\mathcal{A}\left(\textbf{L}_b',r_b,\textbf{L}_u',(\| {\bf L}_u' - {\bf \hat{L} }_b \|) \right) | \right) \mathbbm{1}_{ \{ \check{\theta}_b \leq \hat{\theta}_u \leq \hat{\theta}_b \} }.
		\end{align}
		By substituting \eqref{eq:A_int3} in \eqref{eq:F_r_u|r_b,case2} and taking the derivative w.r.t. $r_u$, 	$f_{D_{u,r}'|D_{b,r}'}(r_u|r_b)$ is obtained as,
		\begin{align}
		f_{D_{u,r}'|D_{b,r}'}(r_u|r_b) = \dfrac{2r_u}{D_{b,u}' | \mathcal{A}(\textbf{L}_b',r_b,\textbf{L}_o,R_o) |} \dfrac{\mathbbm{1}_{ \{ \check{\theta}_b \leq \hat{\theta}_u \leq \hat{\theta}_b \} } }{ \sqrt{ 1 - \left(  \dfrac{(D_{b,u}')^2 +r_b^2 - r_u^2}{2D_{b,u}'r_b }  \right)^2 } }.
		\end{align}
		
	\end{itemize} 
\end{enumerate}
Combining all cases, Lemma \ref{lemma:f_Du|Db_general} is proved.

\section{Lemma \ref{theorem:p_c_b} Proof}
\label{app:p_c_b}
The Rayleigh fading channel coverage probability from the TBS is derived as follows,
{\small
	\begin{align*}
	P_{b,r}(\beta) & \overset{(a)}{=} \mathbb{P} \left( \text{SNR}_{b,r} > \beta \right) ,\\
	&\overset{(b)}{=}  \mathbb{E}_{D_{b,r}'} \left[ \mathbb{P} \left( G_b > \bar{\beta}_b \left( (D_{b,r}')^2+h_b^2 \right)^{\alpha_b/2}  | D_{b,r}' \right) \right] ,\\
	&\overset{(c)}{=}  \mathbb{E}_{D_{b,r}'} \left[ \exp \left(-\bar{\beta}_b \left( (D_{b,r}')^2+h_b^2 \right)^{\alpha_b/2} \right)  \right] ,\\
	&\overset{(d)}{=}  \int_{-\infty}^{\infty}  \exp \left( - \bar{\beta}_b \left( r_b^2+h_b^2 \right)^{\alpha_b/2} \right) f_{D_{b,r}'}(r_b) \: d r_b.
	\end{align*}
}
where ($a$) follows from the coverage probability definition, ($b$) follows by substituting $SNR_{b,r}$ from \eqref{eq:SNR_b-r} with $\bar{\beta}_b = \dfrac{\sigma_n^2 \beta}{\rho_b}$, (c) follows from the CCDF of $G_b$, and (d) follows from the expectation over $D_{b,r}'$.

\section{Lemma \ref{theorem:p_c_u} Proof}
\label{app:p_c_u}
The Nakagami-$m$ fading channel coverage probability from the UAV to the RUE is derived as follows,
{\small
	\begin{align*}
	P_{u,r}(\beta) &= \mathbb{P} \left( \text{SNR} > \beta \right) = \mathbb{E}_{D_{u,r}',\eta_i} \left[ \mathbb{P} \left( G_{u,r} > \bar{\beta}_u \left( (D_{u,r}')^2+h_u^2 \right)^{\alpha_u/2} \eta_i | D_{u,r}', \eta_i \right) \right] ,\\
	&\overset{(a)}{=} \mathbb{E}_{D_{u,r}',\eta_i} \left[ \dfrac{\Gamma(m,m  \bar{\beta}_u \left( (D_{u,r}')^2+h_u^2 \right)^{\alpha_u/2} \eta_i) }{\Gamma(m)}  \right] ,\\
	&\overset{(b)}{=} \mathbb{E}_{D_{u,r}',\eta_i} \left[ \sum_{k=0}^{m-1} \dfrac{ \left(m  \bar{\beta}_u \left( (D_{u,r}')^2+h_u^2 \right)^{\alpha_u/2} \eta_i \right)^k }{k!} \exp \left(-m  \bar{\beta}_u \left( (D_{u,r}')^2+h_u^2 \right)^{\alpha_u/2} \eta_i \right) \right] ,\\
	&= \mathbb{E}_{D_{u,r}'} \left[\sum_{i\in \{\text{LoS},\text{NLoS}\}} \kappa_{u,r}^{i} \sum_{k=0}^{m-1} \dfrac{ \left(m  \bar{\beta}_u \left( (D_{u,r}')^2+h_u^2 \right)^{\alpha_u/2} {\eta_i} \right)^k }{k!} \exp \left(-m  \bar{\beta}_u \left( (D_{u,r}')^2+h_u^2 \right)^{\alpha_u/2} {\eta_i} \right)  \right] ,\\
	&\overset{(c)}{=} \int_{-\infty}^{\infty}  \sum_{i\in \{\text{LoS},\text{NLoS}\}} \kappa_{u,r}^{i} \sum_{k=0}^{m-1} \dfrac{ \left(m  \bar{\beta}_u (r_u^2+h_u^2)^{\alpha_u/2} {\eta_i} \right)^k }{k!} \exp \left(-m  \bar{\beta}_u (r_u^2+h_u^2)^{\alpha_u/2} {\eta_i} \right) f_{D_{u,r}'}(r_u) \; d r_u.
	\end{align*}
}
where ($a$) follows from the CCDF of $G_{u,r}$, ($b$) follows from the incomplete gamma function definition for $m \in \mathbb{Z}^+$, and ($c$) follows from the expectation over $D_{u,r}'$.

\section{Theorem \ref{lemma:optimal_region} Proof  }
\label{app:optimal_region}
For a given GS, the maximum coverage probability is obtained by placing the T-UAV at the optimal location on the (cropped) spherical cone, $\bar{\mathcal{M} }_n$. By fixing the T-UAV hovering height at $h_u \in [h_n,h_n+T]$, the T-UAV can fly within the (cropped) desk, $\bar{\mathcal{D}}\left( \textbf{L}_n, \bar{R}_n(h_u,\psi_u^{n}) \right)$.

To prove Theorem \ref{lemma:optimal_region}, we prove the following \textit{two claims}: (1) As the T-UAV moves far from the TBS with a constant distance from $\textbf{L}_o$, the coverage probability $P^t$ is improved. Hence, the optimal T-UAV location belongs to $\mathcal{A}_1$ as shown in Fig. \ref{fig:part1_proof_optimal_arc}. (2) As the T-UAV moves closer to $\textbf{L}_o$ with a constant distance from the TBS, $P^t$ is also improved. As a result, the optimal T-UAV location belongs to $\mathcal{A}_2$ as shown in Fig. \ref{fig:part2_proof_optimal_arc}.  The intersection region, $\mathcal{A}_1\cap \mathcal{A}_2, \; \forall h_u \in [h_n,h_n+T]$ is $\mathcal{O}_n$ as described in Theorem \ref{lemma:optimal_region}. Therefore, by proving these two claims, Theorem \ref{lemma:optimal_region} is proved. The \textit{two claims} are proved as follows:
\begin{enumerate}

	\item Compare $P^t$ at two T-UAV locations, $\textbf{L}_{u1}$ and $\textbf{L}_{u2}$, with same distances from $\textbf{L}_o$, $D_{u1,o} = D_{u2,o}$ but different distances from the TBS, $D_{b,u1} < D_{b,u2}$. We divide $\mathcal{D}(\textbf{L},o,R_o)$ into two halves, $\mathcal{H}_1$ and $\mathcal{H}_2$, by a hypothetical line where $\overline{\text{SNR} }_{u1,r} =\overline{\text{SNR} }_{u2,r}$ for any RUE on the line, this line is denoted as $\mathcal{L}_{H1}$ (see Fig. \ref{fig:part1_proof_optimal_arc}). For any user location $\textbf{L}_{p1} \in \mathcal{H}_1$, there is another user location (with same probability) such that $\textbf{L}_{p2} \in \mathcal{H}_2$ and, $\textbf{L}_{p1}$ and $\textbf{L}_{p2}$ are symmetric around $\mathcal{L}_{H1}$. We note that $P_{b,r}(p_1) \geq P_{b,r}(p_2)$, $P_{u1,r}(p_2) = P_{u2,r}(p_1)$, $ P_{u1,r}(p_1)= P_{u2,r}(p_2)$ and $P_{u1,r}(p_2) \leq P_{u1,r}(p_1)$ where $P_{i,j}(p_k)$ is the $i$--$j$ link coverage probability for a point located at $\textbf{L}_{p_k}$. Now, compare $P^t$ for the T-UAV locations $\textbf{L}_{u1}$ and $\textbf{L}_{u2}$ for all the possible six cases:		
	
	\begin{figure}
		\centering
		\begin{subfigure}{.49\linewidth}
			\centering
			\begin{tikzpicture}[thick,scale=0.8, every node/.style={scale=0.8}]
			\draw (0, 0) node[inner sep=0] {	\includegraphics[trim={7cm 8cm  14cm 1cm},clip, width=1 \linewidth]{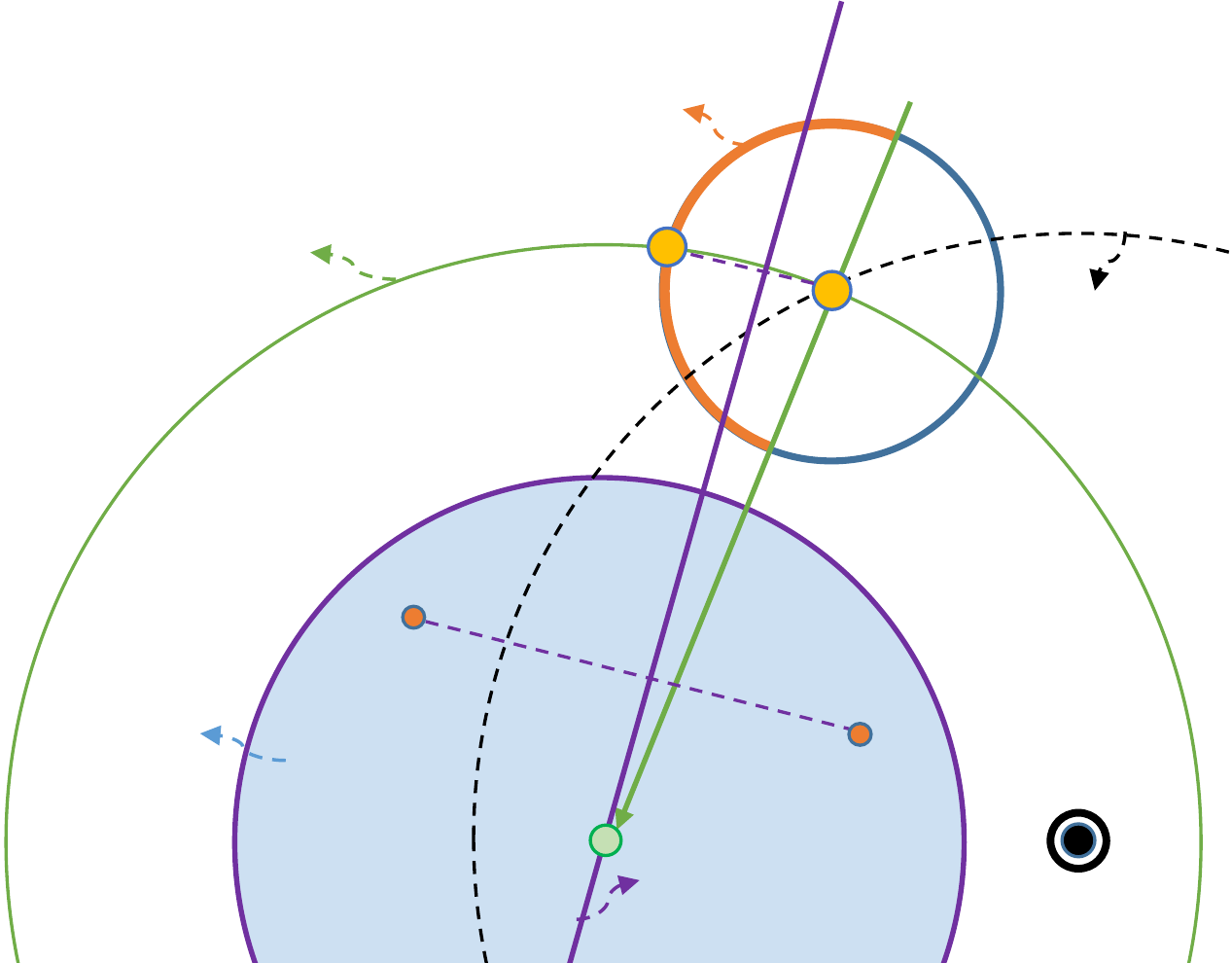}};
			\draw (-.1, 1.8) node {\small{$\textbf{L}_{u2}$}};
			\draw (1.8, 1.6) node {\small{$\textbf{L}_{u1}$}};	
			\draw (.3, 2.4) node {\small{$\mathcal{A}_1$}};
			\draw (.4, -2.8) node {\small{$\mathcal{L}_{H1}$}};
			\draw (3.2, -2) node {\small{$\textbf{L}_b$}};
			\draw (-1.6, -1.1) node {\small{$\textbf{L}_{p2}$}};
			\draw (1.6, -1.9) node {\small{$\textbf{L}_{p1}$}};
			\draw (-3.5, -1.7) node {\small{$\mathcal{D}(\textbf{L}_o,R_o)$}};	
			\draw (-3.1, 1.6) node {\small{$\mathcal{C}(\textbf{L}_o,D_{u1,o})$}};	
			\draw (3.9, 1.) node {\small{$\mathcal{C}(\textbf{L}_b',D'_{b,u1})$}};	
			\draw (1.5, -2.9) node {\small{$\mathcal{H}_1$}};
			\draw (-1.6, -2.7) node {\small{$\mathcal{H}_2$}};
			\end{tikzpicture}
			\caption{Comparison between two T-UAV locations at the same distance from $\textbf{L}_o$ but different distances from $\textbf{L}_b$.} \label{fig:part1_proof_optimal_arc}
		\end{subfigure}
		\begin{subfigure}{.49\linewidth}
			\centering
			\begin{tikzpicture}[thick,scale=0.8, every node/.style={scale=0.8}]
			\draw (0, 0.3) node[inner sep=0] {	\includegraphics[trim={9cm 8cm  15cm 3.3cm},clip, width=1 \linewidth]{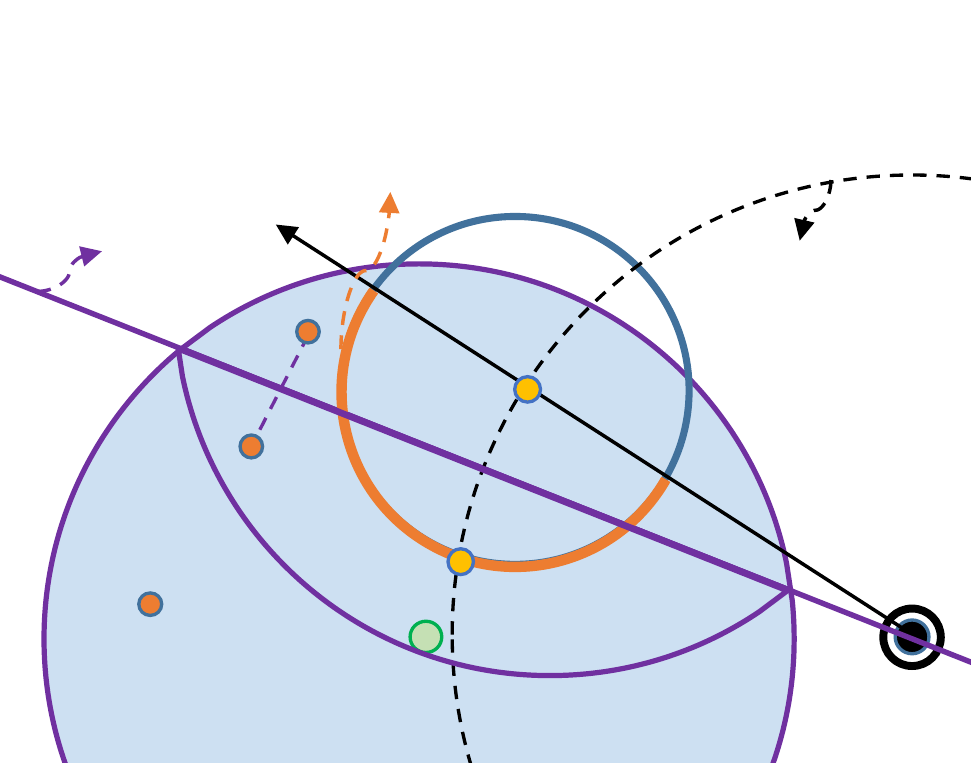}};
			\draw (.2, -1.) node {\small{$\textbf{L}_{u2}$}};
			\draw (.2, .6) node {\small{$\textbf{L}_{u1}$}};	
			\draw (-.9, 2.1) node {\small{$\mathcal{A}_2$}};
			\draw (-2.9, 1.6) node {\small{$\mathcal{L}_{H2}$}};
			\draw (3.6, -1.3) node {\small{$\textbf{L}_b$}};
			\draw (-1.9, .6) node {\small{$\textbf{L}_{p1}$}};
			\draw (-1.6, -.6) node {\small{$\textbf{L}_{p2}$}};
			\draw (-2.6, -1.3) node {\small{$\textbf{L}_{p3}$}};
			\draw (1.8, -.9) node {\small{$\mathcal{H}_1$}};
			\draw (1.6, -1.7) node {\small{$\mathcal{H}_2$}};
			\draw (1.8, -2.4) node {\small{$\mathcal{H}_3$}};		
			\draw (2.8, 1.2) node {\small{$\mathcal{C}(\textbf{L}_b',D'_{b,u1})$}};	
			\end{tikzpicture}
			\caption{Comparison between two T-UAV locations at the same distance from $\textbf{L}_b$ but different distances from $\textbf{L}_o$.}  \label{fig:part2_proof_optimal_arc}
		\end{subfigure}
		\caption{Optimal T-UAV location at a given height. }  
	\end{figure}
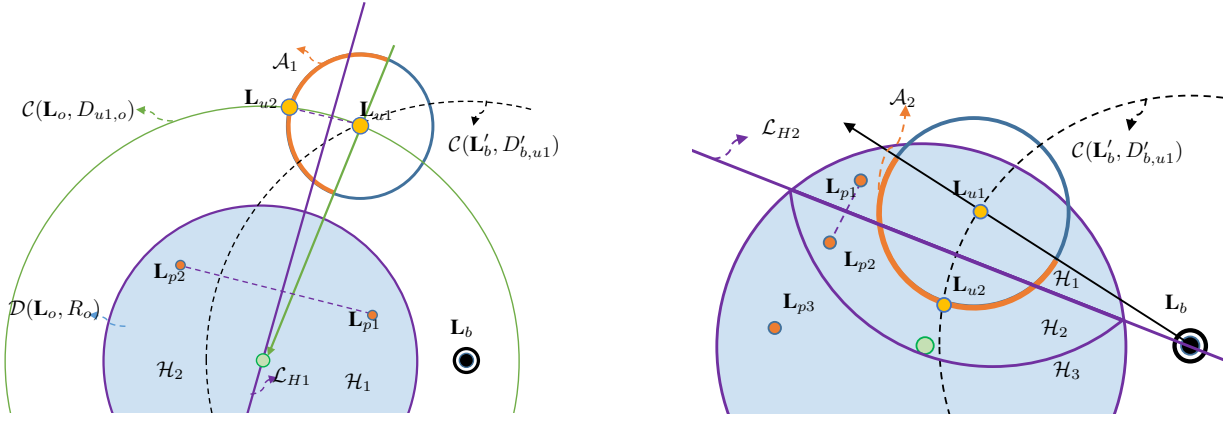

	\begin{enumerate}
		\item $P_{b,r}(p_2) < P_{u1,r}(p_1)$: The users at $p_1$ and $p_2$ are served by the TBS whether the T-UAV is at $\textbf{L}_{u1}$ and $\textbf{L}_{u2}$. Therefore, $P^t(p_1)=P_{b,r}(p_1)$ and $P^t(p_2) =P_{b,r}(p_2)$. 
		
		\item $P_{b,r}(p_1) \geq P_{u1,r}(p_1) \geq   P_{b,r}(p_2) \geq P_{u1,r}(p_2)$: 
		\begin{itemize}
			\item T-UAV at $\textbf{L}_{u1}$: $P^t(p_1) = P_{b,r}(p_1)$, and,
			$P^t(p_2) = P_{b,r}(p_2)$.
			\item T-UAV at $\textbf{L}_{u2}$: $P^t(p_1) = P_{b,r}(p_1)$, and,
			$P^t(p_2) = P_{u2,r}(p_2) > P_{b,r}(p_2)$.
		\end{itemize}
		
		\item $P_{b,r}(p_1) \geq P_{u1,r}(p_1) \geq   P_{u1,r}(p_2) \geq P_{b,r}(p_2)$: 
		\begin{itemize}
			\item T-UAV at $\textbf{L}_{u1}$: $P^t(p_1) = P_{b,r}(p_1)$, and,
			$P^t(p_2) = P_{u1,r}(p_2)$.
			\item T-UAV at $\textbf{L}_{u2}$: $P^t(p_1) = P_{b,r}(p_1)$, and,
			$P^t(p_2) = P_{u2,r}(p_2) > P_{u1,r}(p_2)$.
		\end{itemize}
		
		\item $P_{u1,r}(p_1) \geq P_{b,r}(p_1) \geq   P_{u1,r}(p_2) \geq P_{b,r}(p_2)$: 
		\begin{itemize}
			\item T-UAV at $\textbf{L}_{u1}$: $P^t(p_1) = P_{u1,r}(p_1)$, and,
			$P^t(p_2) = P_{u1,r}(p_2)$.
			\item T-UAV at $\textbf{L}_{u2}$: $P^t(p_1) = P_{b,r}(p_1) \geq P_{u1,r}(p_2)$. And,
			$P^t(p_2) = P_{u2,r}(p_2) = P_{u1,r}(p_1)$.
		\end{itemize}
		
		\item $P_{u1,r}(p_1) \geq P_{b,r}(p_1) \geq   P_{b,r}(p_2) \geq P_{u1,r}(p_2)$: 
		\begin{itemize}
			\item T-UAV at $\textbf{L}_{u1}$: $P^t(p_1) = P_{u1,r}(p_1)$, and,
			$P^t(p_2) = P_{b,r}(p_2)$.
			\item T-UAV at $\textbf{L}_{u2}$: $P^t(p_1) = P_{b,r}(p_1) \geq P_{b,r}(p_2) $, and,
			$P^t(p_2) = P_{u2,r}(p_2)= P_{u1,r}(p_1)$.
		\end{itemize}
		
		\item $P_{u1,r}(p_1) \geq P_{u2,r}(p_2) \geq   P_{b,r}(p_1) \geq P_{b,r}(p_2)$: 
		\begin{itemize}
			\item T-UAV at $\textbf{L}_{u1}$: $P^t(p_1) = P_{u1,r}(p_1)$, and,
			$P^t(p_2) = P_{u1,r}(p_2)$.
			\item T-UAV at $\textbf{L}_{u2}$: $P^t(p_1) = P_{u2,r}(p_1)= P_{u1,r}(p_2)$, and,
			$P^t(p_2) = P_{u2,r}(p_2) = P_{u1,r}(p_1)$.
		\end{itemize}

	\end{enumerate}
	In all the cases, the overall coverage probability $P^t$ is enhanced or unchanged when the T-UAV is located at $\textbf{L}_{u2}$ as compared with $\textbf{L}_{u1}$, which proves the first claim.

	\item Compare $P^t$ at two T-UAV locations, $\textbf{L}_{u1}$ and $\textbf{L}_{u2}$, with same distances from the TBS, $D_{b,u1} = D_{b,u2}$ but different distances from $\textbf{L}_o$, $D_{u1,o} > D_{u2,o}$. We draw a hypothetical line, denoted as $\mathcal{L}_{H2}$, where $\overline{\text{SNR} }_{u1,r} =\overline{\text{SNR} }_{u2,r}$ at any RUE on the line (see Fig. \ref{fig:part2_proof_optimal_arc}). Since $D_{b,u1} = D_{b,u2}$, the line passes through the TBS as Fig. \ref{fig:part2_proof_optimal_arc} shows. Let $\mathcal{D}(\textbf{L}_o,R_o) = \mathcal{H}_1 \cup \mathcal{H}_2 \cup \mathcal{H}_3  $ where $ \mathcal{H}_1$ is the smaller part of $\mathcal{D}(\textbf{L}_o,R_o)$ that is on one side of $\mathcal{L}_{H2}$, $ \mathcal{H}_2$ is symmetric to $ \mathcal{H}_1$ around $\mathcal{L}_{H2}$ and $ \mathcal{H}_3 = \mathcal{D}(\textbf{L}_o,R_o) \setminus ( \mathcal{H}_1 \cup  \mathcal{H}_2)$ (see Fig. \ref{fig:part2_proof_optimal_arc}). If $\mathcal{L}_{H2}$ does not intersect with $\mathcal{D}(\textbf{L}_o,R_o)$, then $ \mathcal{H}_1 = \emptyset$, $ \mathcal{H}_2 = \emptyset$ and $ \mathcal{H}_3 = \mathcal{D}(\textbf{L}_o,R_o)$.
	
	For any user location $\textbf{L}_{p1} \in \mathcal{H}_1$, there is another user location (with same probability) such that $\textbf{L}_{p2} \in \mathcal{H}_2$ where $\textbf{L}_{p1}$ and $\textbf{L}_{p2}$ are symmetric around $\mathcal{L}_{H2}$. We note that $P_{b,r}(p_1) = P_{b,r}(p_2)$, and $P_{u1,r}(p_1) = P_{u2,r}(p_2) $ and $ P_{u1,r}(p_2)= P_{u2,r}(p_1)$. In each of the following cases,
	\begin{itemize}
		\item $P_{b,r}(p_1) > P_{u1,r}(p_1) $,
		\item $P_{b,r}(p_1) <P_{u1,r}(p_2) $,
		\item $P_{u1,r}(p_2) <P_{b,r}(p_1) < P_{u1,r}(p_1)$,
	\end{itemize}
	the same coverage probability over the region $\mathcal{H}_1 \cup \mathcal{H}_2$ is obtained whether the T-UAV is placed at $\textbf{L}_{u1}$ or $\textbf{L}_{u2}$. However, if the RUE is located at $\textbf{L}_{p3} \in \mathcal{H}_3$, the coverage probability at $\textbf{L}_{p_3}$ can be computed as follows:

	\begin{enumerate}
		\item If $P_{b,r}(p_3) > P_{u2,r}(p_3) $: The user associates with the TBS in all cases. Therefore, $P^t(p_3) =P_{b,r}(p_3)$ whether the T-UAV located at $\textbf{L}_{u1}$ or $\textbf{L}_{u2}$.
		
		\item If $P_{b,r}(p_3) <P_{u1,r}(p_3) $: The user associates with the T-UAV in all cases. Therefore,
		\begin{itemize}
			\item T-UAV at $\textbf{L}_{u1}$: $P^t(p_3) = P_{u1,r}(p_3)$. 
			\item T-UAV at $\textbf{L}_{u2}$: $P^t(p_3) = P_{u2,r}(p_3) > P_{u1,r}(p_3)$. 
		\end{itemize}
		
		\item If $P_{u1,r}(p_3) <P_{b,r}(p_3) < P_{u2,r}(p_3)$: Consider the T-UAV locations $\textbf{L}_{u1}$ and $\textbf{L}_{u2}$ as follows,
		\begin{itemize}
			\item T-UAV at $\textbf{L}_{u1}$: $P^t(p_3) = P_{b,r}(p_3)$. 
			\item T-UAV at $\textbf{L}_{u2}$: $P^t(p_3) = P_{u2,r}(p_1) \geq P_{b,r}(p_3)$. 
		\end{itemize}
	\end{enumerate}
	Therefore, the overall coverage probability $P^t$ is enhanced or unchanged when the T-UAV is located at $\textbf{L}_{u2}$ as compared with $\textbf{L}_{u1}$ which proves the second claim. In case of U-UAV, $P^u$ is a function of the link quality between the U-UAV and the TBS. Therefore, the first part of the above proof cannot be used. Based on the second part, where the distance between the UAV and the TBS is fixed at $D_{b,u}$, we note that $P^u$ is maximized when the U-UAV is located at $\textbf{L}_u = \{ x_u,0,h_u\}$ where $x_u = x_b - D_{b,u}, \; \forall D_{b,u} \geq 0$ and $\forall h_u \geq 0$.
	
\end{enumerate}

%

\bibliographystyle{IEEEbib}
\bibliography{Ref}

\end{document}